\documentclass[twocolumn]{aastex63}
\usepackage{amssymb,amsmath}
\usepackage{graphicx}
\usepackage{xcolor,natbib}
\usepackage{multirow}
\usepackage[T1]{fontenc}
\usepackage{ae,aecompl}
\usepackage{newtxtext, newtxmath}
\usepackage{float}
\usepackage{subfigure}
\usepackage{longtable}
\usepackage{rotating}
\usepackage{enumitem}
\usepackage{longtable,listings}
\usepackage[flushleft]{threeparttable}
\usepackage{parcolumns}
\usepackage{lineno}


\submitjournal{ApJS}

\shorttitle{Type-2 AGN with apparent variabilities}
\shortauthors{Zhang XueGuang}

\begin{document}

\title{Systematic research of low redshift optically selected SDSS Type-2 AGN but with apparent 
long-term optical variabilities from Catalina Sky Survey, I: data sample and basic results}

\correspondingauthor{XueGuang Zhang}
\email{xgzhang@gxu.edu.cn}
\author{XueGuang Zhang$^{*}$}
\affiliation{Guangxi Key Laboratory for Relativistic Astrophysics, School of Physical Science and Technology, GuangXi 
University, No. 100, Daxue Road, Nanning, 530004, P. R. China}

\begin{abstract}
	The main objective of the paper, the first paper in a dedicated series, is to report basic results on 
systematic research of low-redshift optically selected SDSS Type-2 AGN but with apparent optical variabilities. 
For all the pipeline classified Type-2 AGN in SDSS DR16 with $z<0.3$ and $SN>10$, long-term optical V-band 
light curves are collected from Catalina Sky Survey. Through all light curves described by Damped Random Walk 
process with process parameters of $\sigma/(mag/days^{0.5})$ and $\tau/days$, 156 Type-2 AGN have apparent 
variabilities with process parameters at least three times larger than corresponding uncertainties and with 
$\ln(\sigma/(mag/days^{0.5}))>-4$, indicating central AGN activity regions directly in line-of-sight, leading 
the 156 Type-2 AGN as mis-classified Type-2 AGN. Furthermore, based on spectroscopic emission features around 
H$\alpha$, 31 out of the 156 AGN have broad H$\alpha$, indicating the 31 Type-2 AGN are actually Type-1.8/1.9 
AGN. Meanwhile, 14 out of the 156 AGN have multi-epoch SDSS spectra. After checking multi-epoch spectra of 
the 14 objects, no clues for appearance/disappearance of broad lines indicate true Type-2 AGN rather than 
changing-look AGN are preferred in the collected Type-2 AGN with long-term variabilities. Moreover, a small 
sample of Type-2 AGN have long-term variabilities with features roughly described by theoretical TDEs expected 
$t^{-5/3}$, indicating probable central TDEs as further and strong evidence to support true Type-2 AGN. 
\end{abstract}


\section{Introduction}

	Different observational spectroscopic phenomena between broad emission line AGN (Active Galactic Nuclei) 
(Type-1 AGN) and narrow emission line AGN (Type-2 AGN) can be well explained by the well-known constantly being 
improved Unified Model (UM) of AGN, after considering effects of different orientation angles of central accretion 
disk \citep{an93}, combining with different central activities and different properties of inner dust torus etc. 
\citep{bm12, mb12, Oh15, ma16, aa17, bb18, bn19, kw21}. More recent reviews on the UM can be found in \citet{nh15}. 
Accepted the UM, Type-2 AGN are intrinsically like Type-1 AGN, but Type-2 AGN have their central accretion disk 
around black hole (BH) and broad line regions (BLRs) seriously obscured by central dust torus, leading to no 
optical broad line emission features in Type-2 AGN. The simple UM has been strongly supported by clearly detected 
polarized broad emission lines and/or clearly detected broad infrared emission lines for some Type-2 AGN \citep{mg90, 
hl97, tr03, nk04, zs05, or17, sg18, mb20}, and the strong resonance of silicate dust at 10${\rm \mu m}$ seen in 
absorption towards many Type-2 AGN but in emission in Type-1 AGN \citep{sh05}.

	However, even after considering necessary modifications to the UM, some challenges have been reported. 
Different evolutionary patterns in Type-1 and Type-2 AGN have been reported in \citet{fb02}. Higher average star 
formation rates in Type-2 AGN than in Type-1 AGN have been reported in \citet{hi09}. Different neighbours around 
Type-1 AGN and Type-2 AGN can be found in \citet{vk14} and then followed in \citet{jw16}. Lower stellar masses 
of host galaxies in Type-1 AGN than in Type-2 AGN have been reported in \citet{zy19}, through 2463 X-ray selected 
AGN in the COSMOS field. More recently, different host galaxy properties have been discussed in \citet{bg20}, 
to favour an evolutionary scenario rather than a strict unified model in obscured and unobscured AGN. Probably 
higher stellar velocity dispersions have been reported in our recent paper \citet{zh22}. As detailed discussions 
in \citet{nh15}, the UM has been successfully applied to explain different features between Type-1 and Type-2 
AGN in many different ways, however, there are many other features of structures/environments proved to be far 
from homogeneous among the AGN family.

	In order to provide further clues to support and/or to modify the current UM of AGN, better classifying 
Type-1 AGN and Type-2 AGN is the first thing which should be well done in optically selected AGN, in order to 
ignore effects of mis-classified AGN on discussions on UM of AGN. \citet{ba14} have reported g-band variabilities 
in 17 out of 173 Type-2 quasars covered in the Stripe 82 region, indicating part of Type-2 objects with their 
central AGN activity regions are directly being observed, and the part of Type-2 quasars are not the UM expected 
Type-2 quasars obscured by central dust torus. In other words, based on only optical spectroscopic broad emission 
line features, it is not efficient enough to classify optically selected Type-1 AGN and Type-2 AGN.

	Among optically selected Type-2 AGN with none detected broad emission lines, there are at least two 
interesting kinds of mis-classified AGN (AGN have their central AGN activity regions been directly observed, but 
been classified as Type-2 AGN through single-epoch spectroscopic results): the called True Type-2 AGN (TT2 AGN) 
without hidden central BLRs (broad emission line regions), the called changing-look AGN (CLAGN) with transitioned 
types between Type-1 with broad emission lines and Type-2 without apparent broad emission lines. 

	The first kind of mis-classified AGN, the True Type-2 AGN (TT2 AGN) without hidden central BLRs, have been 
firstly reported in \citet{tr03} that there are no hidden central BLRs in some Type-2 AGN through studying 
polarized broad emission lines. And then, \citet{sg10, ba14, zh14, ly15, pw16} have confirmed the existence of 
the rare kind of TT2 AGN. More recently, We \citet{zh21d} have reported the composite galaxy SDSS J1039 to be 
classified as a TT2 AGN, due to its apparent long-term optical variabilities and none existence of broad emission 
lines. Moreover, considering study of TT2 AGN should provide further clues on formation and/or suppression of 
central BLRs in AGN, \citet{eh09, cao10, nm13, ip15, en16} have discussed and proposed theoretical models and/or 
explanations on probable disappearance of BLRs in TT2 AGN, either depending on physical properties of central 
AGN activities and/or depending on properties of central dust obscurations. Although the very existence of TT2 
AGN is still an open question, as the well known case of NGC 3147 previously classified as a TT2 AGN but now 
discussed in \citet{ba19} reported with detected broad H$\alpha$ in high quality HST spectrum. Probably, either 
the very existence of TT2 AGN with none hidden central BLRs or the quite weak broad emission lines in TT2 AGN 
can lead some Type-1 AGN (central regions directly observed) to be mis-classified as Type-2 AGN, indicating that 
there are some mis-classified Type-1 AGN in the optically selected Type-2 AGN, if only single-epoch spectroscopic 
emission line features are considered.

	The second kind of mis-classified AGN, the changing-look AGN (CLAGN), have been firstly reported in NGC 
7603\ in \citet{to76} with its broad H$\beta$ becoming much weaker in one year. After studying on CLAGN for more 
than four decades, there are more than 40 CLAGN reported in the literature, according to the basic properties 
that spectral types of AGN are transitioned between Type-1 AGN (apparent broad Balmer emission lines and/or 
Balmer decrements near to the theoretical values) and Type-2 AGN (no apparent broad Balmer emission lines and/or 
Balmer decrements much different from the theoretical values).

	There are several well known individual CLAGN, such as the CLAGN Mrk1086 which has firstly reported in 
\citet{cr86} with its type changed from Type-1.9 to Type-1\ in 4 years, and then followed in \citet{mh16}, and 
as the CLAGN in NGC 1097 reported in \citet{sb93} with the detected Seyfert 1 nucleus which has previously shown 
only LINER characteristics, and as the CLAGN NGC 7582 reported in \citet{aj99} with the transition toward a Type-1 
Seyfert experienced by the classical Type-2 Seyfert nucleus, and as the CLAGN NGC 3065 reported in \citet{eh01} 
with the new detected broad Balmer emission lines, and as the CLAGN Mrk 590 reported in \citet{dd14} with its 
type changed from Seyfert 1\ in 1970s to Seyfert 1.9\ in 2010s, and as the CLAGN NGC 2617 reported in \citet{sp14} 
classified as a Seyfert 1.8 galaxy in 2003 but as a Seyfert 1 galaxy in 2013, and as the CLAGN SDSS J0159 
reported in \citet{la15} with its type transitioned from Type-1\ in 2000 to Type-1.9\ in 2010, and as the CLAGN 
SDSS J1554 reported in \citet{gh17} with its type changed from Type-2 to Type-1\ in 12 years.

	More recently, we \citet{zh21b} have reported the bluest CLAGN SDSS J2241 with its flux ratio of broad 
H$\alpha$ to broad H$\beta$ changed from 7\ in 2011 to 2.7\ in 2017. Moreover, besides the individual CLAGN, 
\citet{mr16} have reported ten CLAGN with variable and/or changing-look broad emission line features, and 
\citet{yw18} have reported a sample of 21 CLAGN with the appearance or the disappearance of broad Balmer 
emission lines within a few years, and \citet{pv21} have reported six CLAGN within the redshift range $0.1<z<0.3$ 
using SDSS difference spectra, and \citet{gp22} have reported 61 newly discovered CLAGN candidates through 
multi-epoch spectroscopic properties in the known Time Domain Spectroscopic Survey in SDSS-IV. Different 
models have been proposed to explain the nature of CLAGN, such as the dynamical movement of dust clouds well 
discussed in \citet{em12}, the common variations in accretion rates well discussed in \citet{em14}, the 
variations in accretion rates due to transient events as discussed in \citet{er95, bn17}. Although theoretical 
explanations to CLAGN are still unclear, spectroscopic properties in quiet state of CLAGN could lead to 
CLAGN with central activity regions directly observed but mis-classified as Type-2 AGN.

	Besides the spectroscopic features well applied to classify Type-1 AGN and Type-2 AGN, long-term 
variabilities, one of fundamental intrinsic characteristics of AGN \citep{mr84, um97, ms16, bv20} tightly related 
to central BH accreting processes, can also be well applied to classify AGN: Type-1 AGN with apparent 
long-term optical variabilities, but Type-2 AGN with no apparent long-term optical variabilities because of 
serious obscurations of central AGN activities. And the long-term optical variabilities of AGN can be be 
well modeled by the well-applied Continuous AutoRegressive process (CAR process) firstly proposed by 
\citet{kbs09} and then the improved damped random walk process (DRW process) in \citet{koz10, zk13, kb14, sh16, 
zk16}. Therefore, combining variability properties and spectroscopic properties can lead to more confident 
classifications of AGN, which is the main objective of the paper, in order to check how many Type-1 AGN 
(central activity regions directly observed) mis-classified as Type-2 AGN through optical spectroscopic 
emission features.

	The manuscript is organized as follows. Section 2 presents the data sample of the pipeline classified 
Type-2 AGN in SDSS DR16 (Sloan Digital Sky Survey, Data Release 16, \citet{ap20}). Section 3 shows the method 
to describe the long-term variabilities from CSS (Catalina Sky Survey) \citep{dr09, dg14, gr17, sb21} for the 
collected SDSS pipeline classified Type-2 AGN and the basic results on a small sample of optically selected 
SDSS Type-2 AGN with apparent long-term variabilities. Section 4 shows the main spectroscopic results for the 
collected Type-2 AGN but with apparent long-term optical variabilities. Section 5 shows the necessary 
discussions. Section 6 gives the final summaries and conclusions. And in the manuscript, the cosmological 
parameters of $H_{0}~=~70{\rm km\cdot s}^{-1}{\rm Mpc}^{-1}$, $\Omega_{\Lambda}~=~0.7$ and $\Omega_{m}~=~0.3$ 
have been adopted.

\section{Parent Samples of SDSS pipeline classified Type-2 AGN}

	As well described in \url{https://www.sdss.org/dr16/spectro/catalogs/}, each spectroscopic object has 
a main classification\footnote{Not considering the 'UNKNOWN' classification in SDSS databases.} in SDSS: GALAXY, 
QSO or STAR. And, for objects classified as GALAXY, there are at least three subclasses\footnote{Not considering 
the 'UNKNOWM' subclass in SDSS databases}, STARBURST, STARFORMING, AGN. Here, the main classification of GALAXY 
(class='galaxy') and subclass of AGN (subclass = 'AGN') are mainly considered to collect the SDSS pipeline 
classified low redshift Type-2 AGN. Here, only one main criterion of redshift smaller than 0.3 ($z~<~0.3$) is 
applied to collect all the low redshift Type-2 AGN from SDSS pipeline classified main galaxies in DR16, through 
the SDSS provided SQL (Structured Query Language) Search tool 
(\url{http://skyserver.sdss.org/dr16/en/tools/search/sql.aspx}) by the following query
\begin{lstlisting}
SELECT plate, fiberid, mjd, ra, dec
FROM SpecObjall
WHERE
   class='galaxy' and subclass = 'AGN' 
   and (z between 0 and 0.30) and zwarning=0 
   and snmedian > 10
\end{lstlisting}
In the query above, 'SpecObjall' is the SDSS pipeline provided database including basic properties of spectroscopic 
emission features of emission line galaxies in SDSS DR16, 'snmedian' means the median signal-to-noise (SN) of SDSS 
spectra, class='galaxy' and subclass='AGN' mean the SDSS spectrum can be well identified with a galaxy template 
and the galaxy has detectable emission lines that are consistent with being a Seyfert or LINER by the dividing line 
applied in the known BPT diagram \citep{bpt, kb01, ka03a, kb06, kb19, zh20} through flux ratios of 
[O~{\sc iii}]$\lambda5007$\AA~ to narrow H$\beta$ (O3HB) and of [N~{\sc ii}]$\lambda6583$\AA~ to narrow H$\alpha$ 
(N2HA). More detailed information of the database 'SpecObjall' can be found in the SDSS webpage 
\url{http://skyserver.sdss.org/dr16/en/help/docs/tabledesc.aspx}. The SQL query leads 14354 emission line main 
galaxies collected as Type-2 AGN in DR16.

	Before proceeding further, two points are noted. On the one hand, the criteria of  $z<0.3$ and $S/N>10$ 
are applied, mainly in order to ensure reliable narrow emission lines of [N~{\sc ii}] doublet and narrow H$\alpha$ 
totally covered in SDSS spectra, to well apply the narrow emission line ratios to classify Type-2 AGN in the BPT 
diagram. On the other hand, due to few contaminations of host galaxy starlight to narrow emission lines, different 
techniques to determine starlight in SDSS spectrum can lead to totally similar measurements of emission lines. 
Therefore, there are no further discussions on emission line properties of the collected Type-2 AGN, but the 
reliable line parameters measured by the MPA-JHU group have been well collected from the database of 'GalSpecLine'. 
And more detailed descriptions on measurements of emission line properties in main galaxies can be found in 
\url{https://www.sdss.org/dr16/spectro/galaxy_mpajhu/} and in \citet{bc04, ka03a, th04}. Certainly, when the sample 
is created for the SDSS Type-2 AGN but with apparent optical variabilities, there are detailed discussions in the 
following sections in the manuscript on procedure to determine host galaxy contributions in SDSS spectra and 
procedure to describe emission lines after subtractions of starlight.

\begin{figure*}
\centering\includegraphics[width = 18cm,height=5cm]{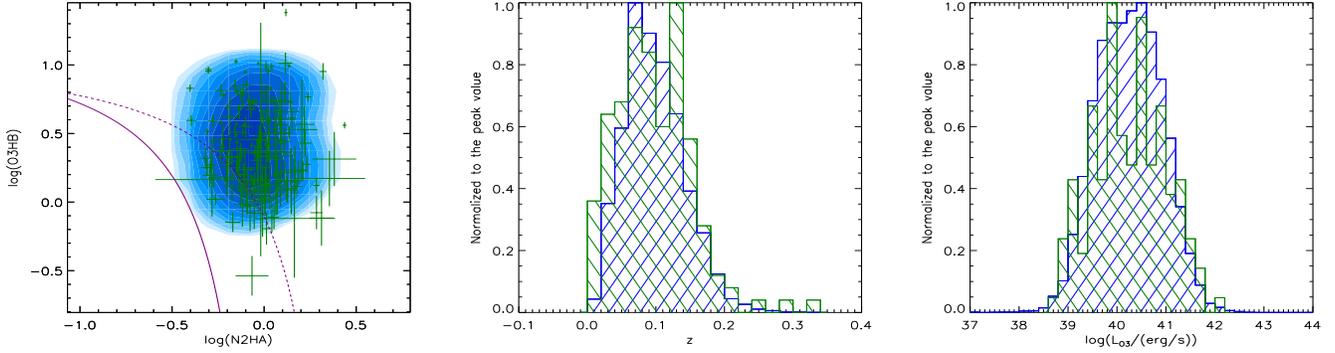}
\caption{Left panel shows the collected Type-2 AGN in the BPT diagram (contour filled by bluish colors) of O3HB 
versus N2HA. Solid and dashed purple lines show the dividing lines between different kinds of narrow emission 
line galaxies in \citet{kb06, ka03a}, HII galaxies, composite galaxies and AGN. In left panel, small circles plus 
error bars in dark green show the collected 156 Type-2 AGN with apparent long-term variabilities. Middle panel 
and right panel show the distributions of redshift and [O~{\sc iii}] line luminosity ($L_{O3}$) of the 14353 
collected Type-2 AGN (histogram filled by blue lines) and of the 155 Type-2 AGN with apparent long-term 
variabilities (histogram filled by dark green lines). The individual Type-2 AGN SDSS J075736.47+532557.1 
(plate-mjd-fiberid=1870-53383-0466) has measured redshift to be zero in SDSS, therefore, the SDSS J075736.47+532557.1 
is not included in the middle panel and right panel.
}
\label{bpt}
\end{figure*}

	Based on the collected line parameters and the other necessary information of the Type-2 AGN from the SDSS 
databases, Fig.~\ref{bpt} shows properties of BPT diagram of $\log(O3HB)$ (mean value about 0.451, standard deviation 
about 0.298) versus N2HA (mean value about -0.058, standard deviation about 0.141), distributions of redshift $z$ 
(mean value about 0.098, standard deviation about 0.044) and luminosity of [O~{\sc iii}] line $\log(L_{O3}/{\rm erg/s})$ 
(mean value about 40.255, standard deviation about 0.658).

\section{Long-term CSS photometric variabilities of the SDSS Type-2 AGN}
	As well described in the homepage of the Catalina Sky Survey (CSS) 
(\url{https://catalina.lpl.arizona.edu/about/facilities/telescopes}) and in \citet{dr09}, the three CSS telescopes 
are located in the Santa Catalina Mountains just north of Tucson, Arizona. The 1.5-meter Cassegrain telescope 
(field of view about 5.0 deg$^2$) is located on the 9157-foot summit of Mt. Lemmon, along-side the 1.0-meter 
telescope (field of view about 0.3 deg$^2$). The 0.7-meter Schmidt telescope (field of view about 19.4 deg$^2$) 
is located on Mt. Bigelow, just east of Mt. Lemmon. The CSS telescopes operate 24 nights per lunation with a 4-5 
day break surrounding the full moon. The CSS database encompasses about 10-years long photometry for more than 
500 million objects with V magnitudes between 11.5 and 21.5 from an area of 33,000 square degrees.

\begin{figure*}
\centering\includegraphics[width = 18cm,height=6cm]{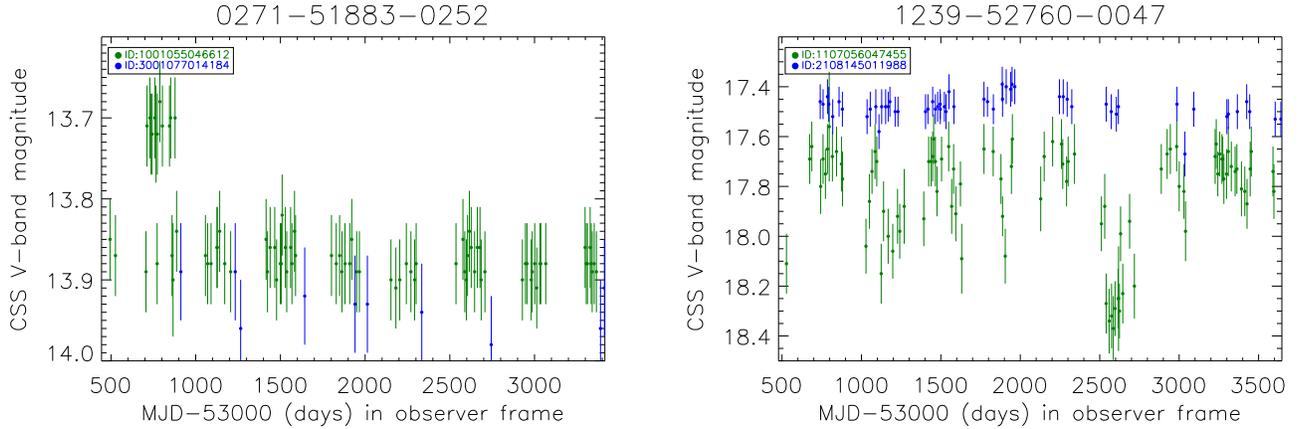}
\caption{CSS V-band light curves of SDSS 0271-51883-0252 and 1239-52760-0047. In each panel, symbols in 
different colors show the light curves with different CSS IDs as shown in legends. And the light curve shown in dark 
green is the final accepted CSS light curve.}
\label{Lex}
\end{figure*}

	Based on RA and DEC of the collected 14354 SDSS Type-2 AGN, long-term light curves can be collected and 
downloaded from \url{http://nesssi.cacr.caltech.edu/DataRelease/} for 12881 out of the collected 14354 SDSS Type-2 
AGN, with searching radius of 3 arcseconds. Meanwhile, as discussed in \citet{dr09} and descriptions in 
CSS homepage, there is a photometry blending flag for CSS light curves, probably causing by different photometric 
detections in different seeing conditions and/or by causing by two and more sources detected within the searching 
radius. Fortunately, among the download light curves of the 12881 Type-2 AGN, all the blending flags are zero, 
indicating that the light curves can be used reliably. Moreover, when cross matching database between CSS and SDSS, 
3520 of the 12881 Type-2 AGN have their collected light curves with more than one CSS IDs. Therefore, two simple 
criteria are applied to determine the final accepted CSS light curves for the 3520 Type-2 AGN. First, if the multiple 
CSS light curves with different IDs for one Type-2 AGN have the same mean photometric magnitudes within the same 
time durations, the CSS light curve with more data points (corresponding longer time durations) is accepted as the 
CSS light curve of the Type-2 AGN. Second, if the multiple CSS light curves with different IDs for one Type-2 AGN 
have quiet different mean photometric magnitudes, the CSS light curve with mean photometric V-band magnitude nearer 
to SDSS photometric g-band Petrosian magnitude is accepted as the CSS light curve of the Type-2 AGN, due to the CSS V-band 
and the SDSS g-band covering similar wavelength ranges. Fig.~\ref{Lex} shows the light curves of SDSS 0271-51883-0252 
(PLATE-MJD-FIBERID) and 1239-52760-0047 as examples to show applications of the two criteria above to determine 
the final accepted CSS light curve. For the SDSS 0271-51883-0252, there are two CSS light curves with IDs of 
1001055046612 and 3001077014184. For the same time durations of the two light curves with MJD-53000 from 912 to 
3365, there are similar mean photometric magnitudes about 13.9, therefore, the CSS light curve with ID:1001055046612 
having more data points and also longer time duration is accepted as the final light curve of SDSS 0271-51883-0252. 
For the SDSS 1239-52760-0047, the collected CSS light curves with two IDs of 1107056047455 and 2108145011988 have 
mean magnitude difference to be about 0.5mag. Considering the SDSS photometric g-band magnitude about 18.36, therefore, 
the CSS light curve (ID:1107056047455) with mean V-band magnitude 17.85 is the final accepted light curve of SDSS 
1239-52760-0047. Actually, the final accepted light curves of the SDSS 0271-51883-0252 and 1239-52760-0047, two 
candidates of Type-2 AGN with apparent variabilities in the manuscript, are also shown in the following Fig.~\ref{lmc}.

	Then, in order to check whether are there apparent variabilities, the commonly accepted DRW process 
\citep{kbs09, koz10, zk13} is applied to describe the collected light curves. There are many other reported 
studies on the AGN variabilities through the DRW process. \citet{mi10} have modeled the variabilities of about 
9000 spectroscopically confirmed quasars covered in the SDSS Stripe82 region, and found correlations between 
the AGN parameters and the DRW process determined parameters. \citet{bj12} proposed an another fully probabilistic 
method for modeling AGN variabilities by the DRW process. \citet{ak13} have shown that the DRW process is preferred 
to model AGN variabilities, rather than several other stochastic and deterministic models, by fitted results of 
long-term variabilities of 6304 quasars. \citet{zk13} have checked that the DRW process provided an adequate 
description of AGN optical variabilities across all timescales. \citet{zh17a} have checked long-term variability 
properties of AGN with double-peaked broad emission lines, and found the difference in intrinsic variability 
timescales between normal broad line AGN and the AGN with double-peaked broad emission lines. More recently, 
in our previous paper, \citet{zh21d} have shown apparent long-term variabilities well described by the DRW 
process to report a composite galaxy as a better candidate of true Type-2 AGN. Therefore, the DRW process 
determined parameters from the long-term variabilities can be well used to predict whether are there apparent 
clues to support central AGN activities.

	In the manuscript, the public code of JAVELIN (Just Another Vehicle for Estimating Lags In Nuclei) 
described in \citet{koz10} and provided by \citet{zk13} has been applied here to describe the long-term CSS 
variabilities of the 12881 Type-2 AGN. When the JAVELIN code is applied, through the MCMC (Markov Chain Monte 
Carlo) \citep{fh13} analysis with the uniform logarithmic priors of the DRW process parameters of $\tau$ and 
$\sigma$ covering every possible corner of the parameter space ($0~<~\tau/days~<~1e+5$ and 
$0~<~\sigma/(mag/days^{0.5})~<~1e+2$), the posterior distributions of the DRW process parameters can be well 
determined and provide the final accepted parameters and the corresponding statistical confidence limits. 
Meanwhile, when the JAVELIN is applied, each long-term CSS light curve has been firstly re-sampled with one 
mean value for multiple observations per day. Then, the best descriptions to each re-sampled light curve can 
be well determined.

	Based on the JAVELIN code determined process parameters of $\sigma$ and $\tau$ at least three times 
larger than their corresponding uncertainties, there are 156 Type-2 AGN which have apparent long-term 
variabilities. The long-term light curves and corresponding best-fitting results of the 156 Type-2 AGN are 
shown in Fig.~\ref{lmc}, and the corresponding two dimensional posterior distributions in contour of the 
parameters of $\sigma$ and $\tau$ are shown in Fig.~\ref{mcmc}. And the basic information of the 156 
Type-2 AGN are listed in Table~1, including information of plate-mjd-fiberid, SDSS coordinate based name 
(Jhhmmss.s$\pm$ddmmss.s), redshift $z$, averge photometric CSS V-band magnitude, SDSS provided g-band Petrosian 
magnitude, $O3HB$, $N2HA$, the parameters and uncertainties of $\ln(\sigma)$ and $\ln(\tau)$. Properties of 
$O3HB$ and $N2HA$ of the 156 Type-2 AGN are also shown as small circles in dark green in left panel of 
Fig.~\ref{bpt}, with mean $\log(O3HB)$ and standard deviation about 0.421 and 0.316, and with mean $\log(N2HA)$ 
and standard deviation about -0.036 and 0.160, which are roughly similar as the values of all the 14354 
Type-2 AGN. Meanwhile, distribution of redshift $z$ of the 155 Type-2 AGN (the SDSS J075736.47+532557.1 not 
included due to its zero redshift in SDSS) is also shown in the middle panel of Fig.~\ref{bpt}, with mean 
value about 0.097 and standard deviation about 0.056, roughly similar as the values of all the 14353 Type-2 
AGN (SDSS J075736.47+532557.1 not included). Distribution of luminosity of [O~{\sc iii}] line 
$\log(L_{O3}/{\rm erg/s})$ of the 155 Type-2 AGN (the SDSS J075736.47+532557.1 not included due to its zero 
redshift) is also shown in the right panel of Fig.~\ref{bpt}, with mean value about 40.257 and standard 
deviation about 0.730, roughly similar as the values of all the 14353 Type-2 AGN (SDSS J075736.47+532557.1 not included).

	Before proceeding further, one additional point is noted. As discussed in \citet{gr17, lv20}, 
uncertainties of CSS photometric magnitudes are overestimated at the brighter magnitudes, but underestimated 
at fainter magnitudes. Therefore, it is necessary to check effects of corrections of photometric magnitude 
uncertainties on the variability properties. Fig.~\ref{db2} shows photometric magnitude distributions of 
the collected 156 Type-2 AGN with apparent variabilities and the 12881 Type-2 AGN in the parent sample. 
The mean photometric magnitudes are 16.17 (standard deviation 1.51) and 16.58 (standard deviation 0.92) 
for the 156 Type-2 AGN with apparent variabilities and for the 12881 Type-2 AGN, respectively. Therefore, 
the collected 156 Type-2 AGN with apparent variabilities do not tend to be chosen from brighter or fainter 
sources. In other words, there are tiny effects of corrections of uncertainties of photometric magnitudes 
on the collected 156 Type-2 AGN through variability properties.

	Besides properties of $\log(N2HA)$, $\log(O3HB)$, $z$ and $\log(L_{O3}/{\rm erg/s})$, left panel 
of Fig.~\ref{dis} shows properties of $\ln(\sigma/(mag/days^{0.5}))$ (mean value about -2.45 and standard 
deviation about 0.54) and $\ln(\tau/days)$ (mean value about 5.06 and standard deviation about 1.09). 
Before proceeding further, as well discussed on variability properties of OGLE-III quasars in \citet{koz10} 
(see their Figure~9) by the same method as the JAVELIN code, the DRW process parameters of 
$\sigma/(mag/days^{0.5})$ and $\tau/days$ have values about 0.005-0.52\footnote{In \citet{koz10}, the unit 
of $\sigma$ is ${\rm mag/years^{0.5}}$. Here, the values 0.005-0.52 are calculated by $\sigma/{\rm mag/years^{0.5}}$ 
larger than 0.1 and smaller than 10 as shown in Figure~9\ in \citet{koz10}.} and about 10-1000days, strongly 
indicating that the measured parameters of $\ln(\sigma/(mag/days^{0.5}))$ and $\ln(\tau/days)$ of the 156 
Type-2 AGN are reliable enough. Moreover, right panel of Fig.~\ref{dis} shows distributions of $NP$, with 
mean values of 81 and 84 for the 156 Type-2 AGN and the other Type-2 AGN, respectively, indicating there 
are no effects of different numbers of data points $NP$ on the measured DRW process parameters. Besides the 
156 Type-2 AGN with $\ln(\sigma/(mag/days^{0.5}))$ larger than -4, the other Type-2 AGN have their measured 
$\ln(\sigma/(mag/days^{0.5}))$ smaller than -10. Therefore, in left panel of Fig.~\ref{dis}, there are no 
plots on the other Type-2 AGN with $\ln(\sigma/(mag/days^{0.5}))$ smaller than -10.

	Finally, based on the DRW process well applied to describe intrinsic AGN activities, there are 
1.2\% (156/12881) optically selected Type-2 AGN which have apparent long-term optical variabilities, 
indicating the 156 Type-2 AGN of which central AGN activity regions directly in the line-of-sight.

\begin{figure*}
\centering\includegraphics[width = 18cm,height=20cm]{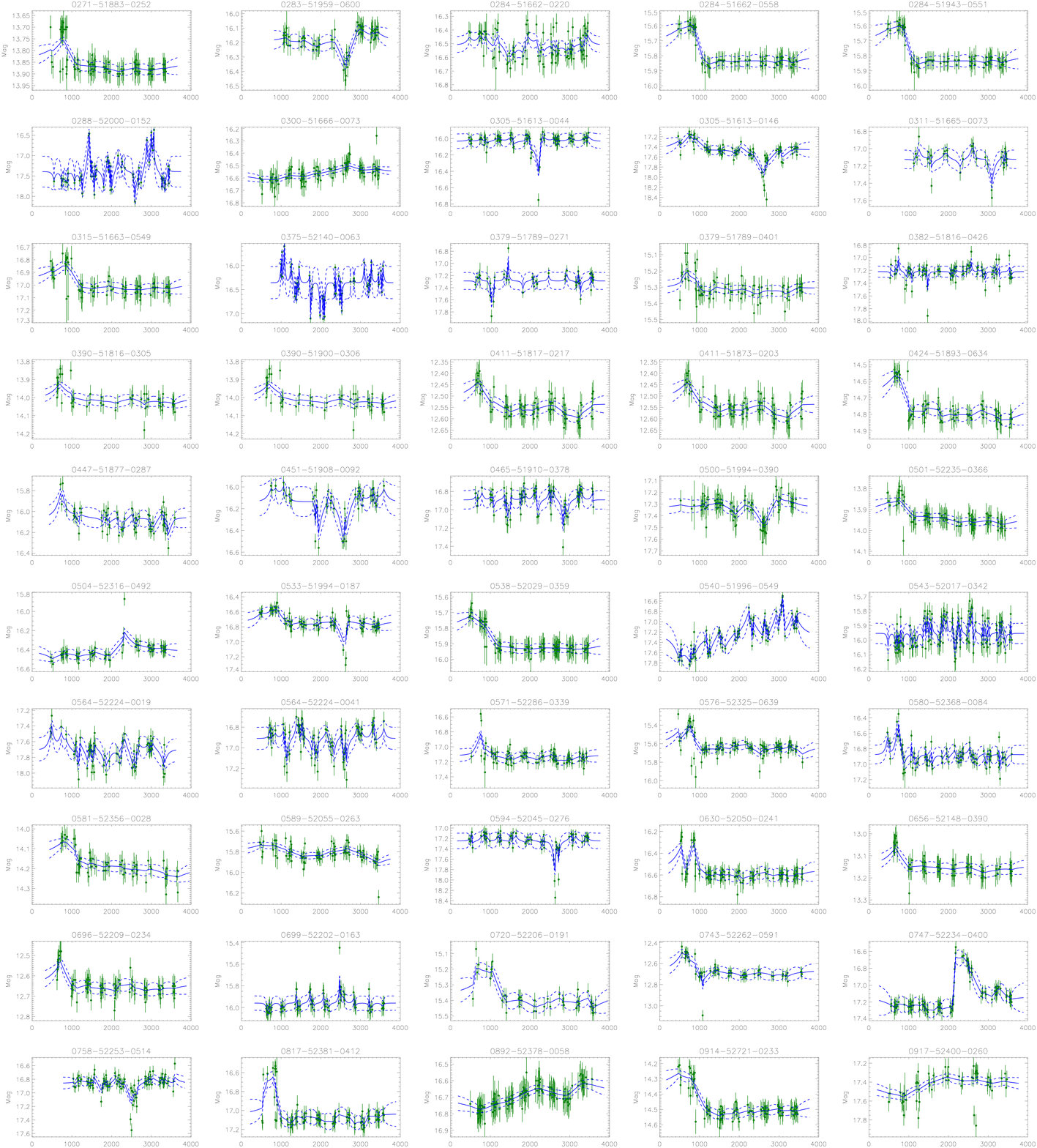}
\caption{CSS long-term light curves (solid circles plus error bars in dark green, with MJD-53000 (days) in 
x-axis and CSS apparent magnitude in y-axis) of the 156 Type-2 AGN with apparent variabilities well 
described by the DRW process. In each panel, solid blue line and dashed blue lines show the JAVELIN code 
determined best descriptions to the light curve and the corresponding 1sigma confidence bands, respectively. 
The information of plate-mjd-fiberid is marked as title of each panel.
}
\label{lmc}
\end{figure*}

\setcounter{figure}{2}

\begin{figure*}
\centering\includegraphics[width = 18cm,height=22cm]{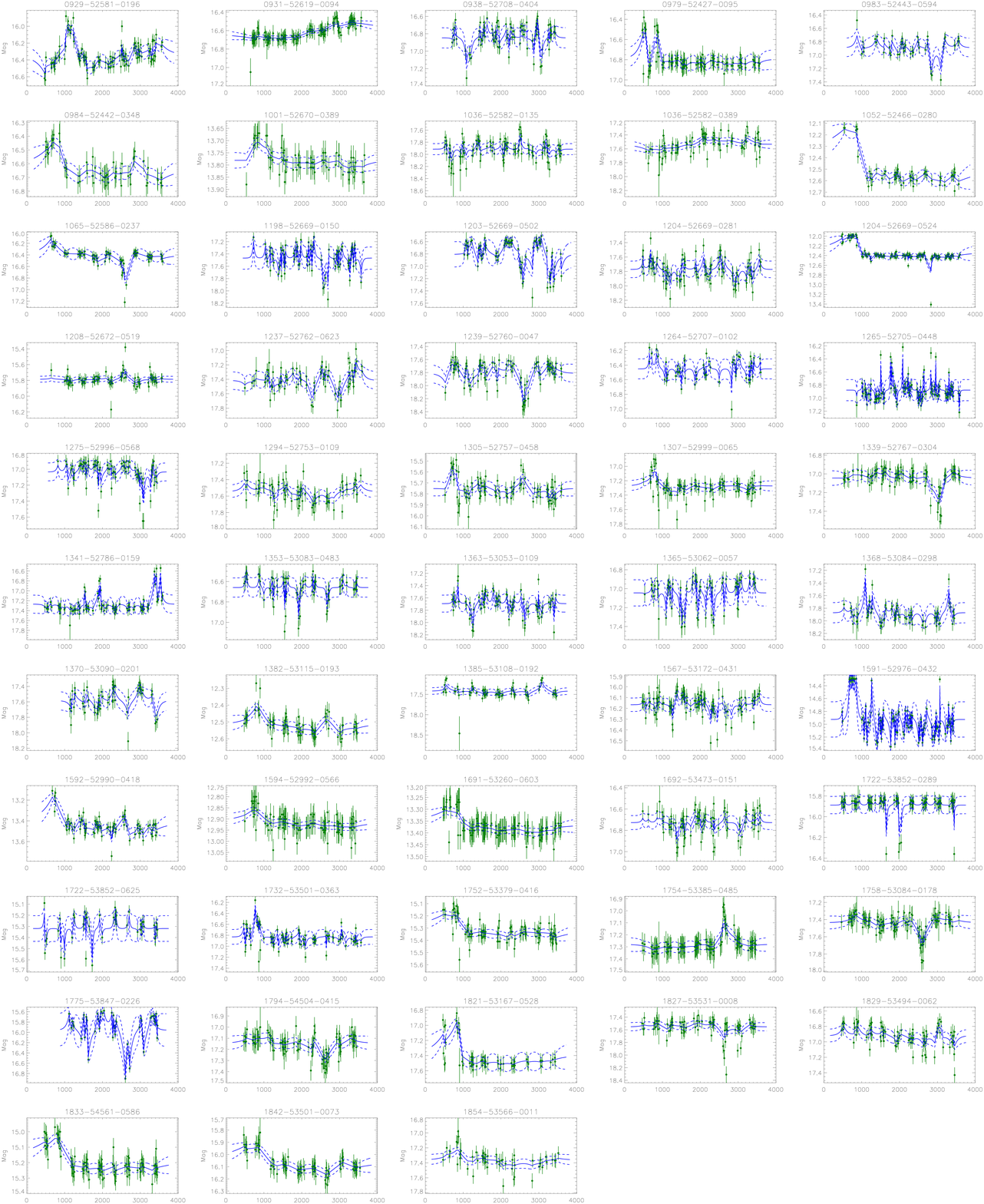}
\caption{--to be continued}
\end{figure*}

\setcounter{figure}{2}

\begin{figure*}
\centering\includegraphics[width = 18cm,height=22cm]{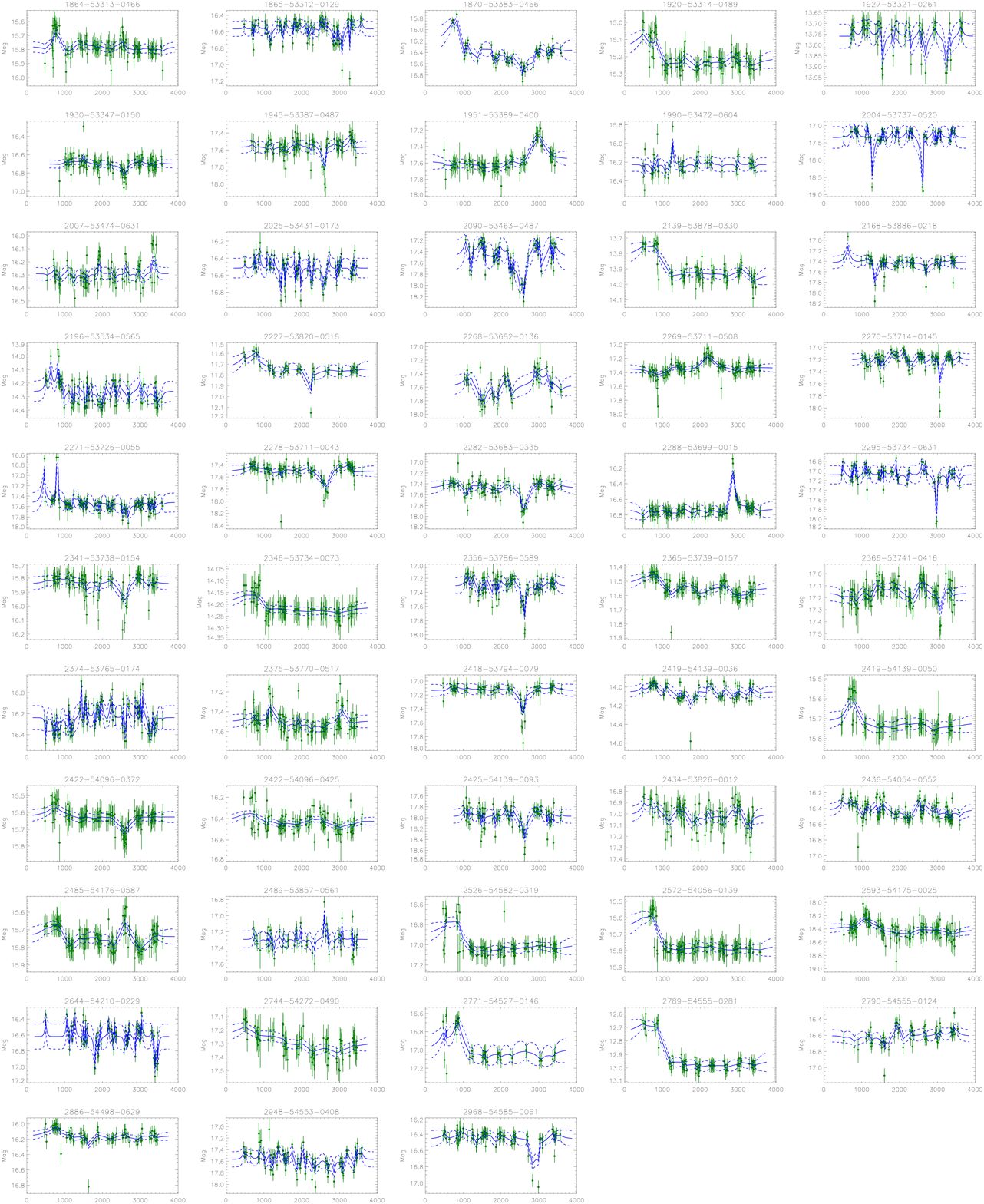}
\caption{--to be continued}
\end{figure*}

\begin{figure*}
\centering\includegraphics[width = 18cm,height=20cm]{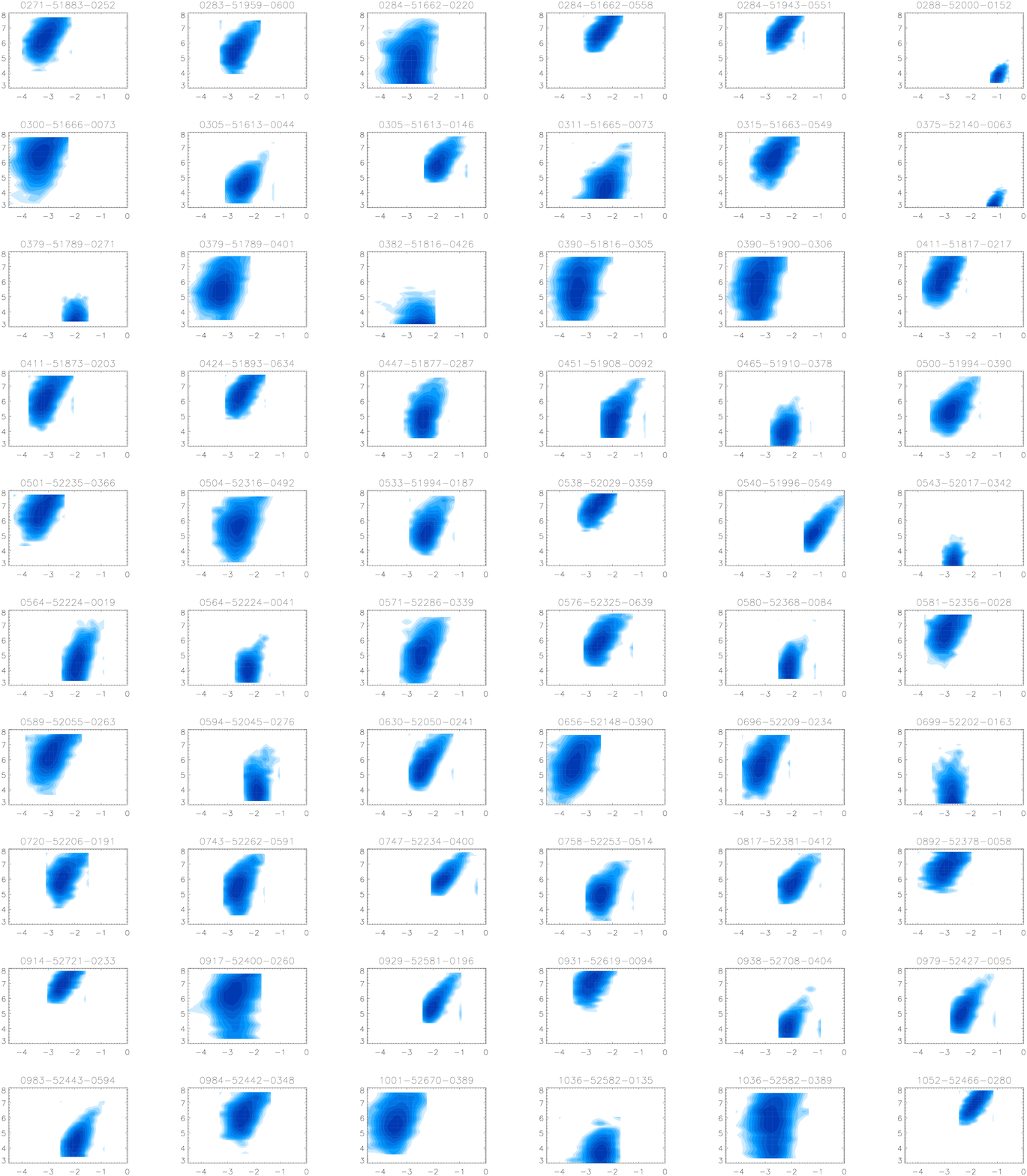}
\caption{The corresponding two dimensional posterior distributions shown in contour of the DRW process 
parameters of $\ln(\sigma/(mag/days^{0.5}))$ (x-axis) and $\ln(\tau/days)$ (y-axis) of the light curves 
shown in Fig.~\ref{lmc} of the 156 Type-2 AGN. The information of plate-mjd-fiberid is marked as title 
of each panel. In order to show clear parameter difference between different objects, the same 
limit from -4.5 to 0 has been applied to x-axis, and the same limit from 3 to 8 has been applied to y-axis.
}
\label{mcmc}
\end{figure*}

\setcounter{figure}{3}

\begin{figure*}
\centering\includegraphics[width = 18cm,height=22cm]{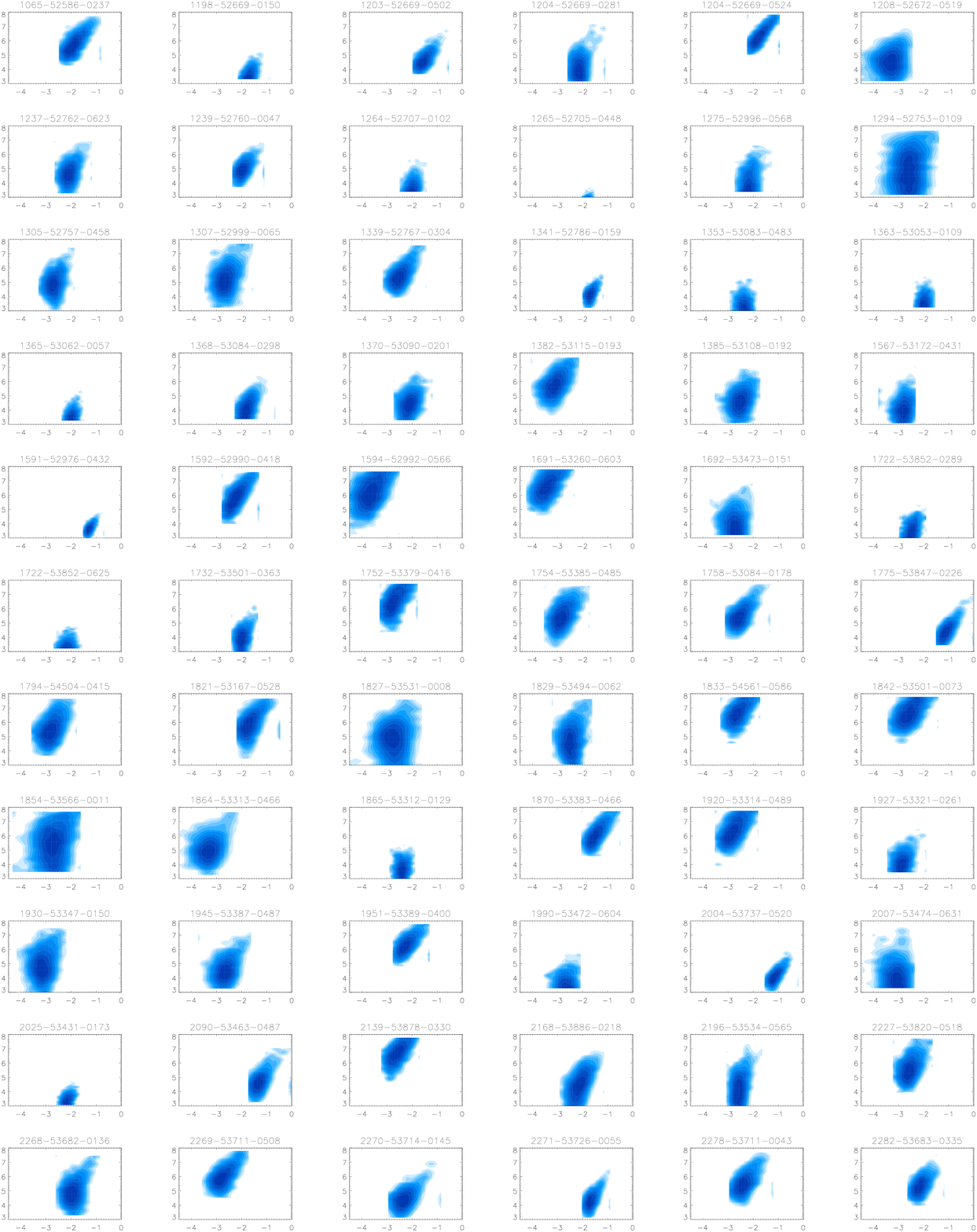}
\caption{--to be continued}
\end{figure*}

\setcounter{figure}{3}

\begin{figure*}
\centering\includegraphics[width = 18cm,height=10cm]{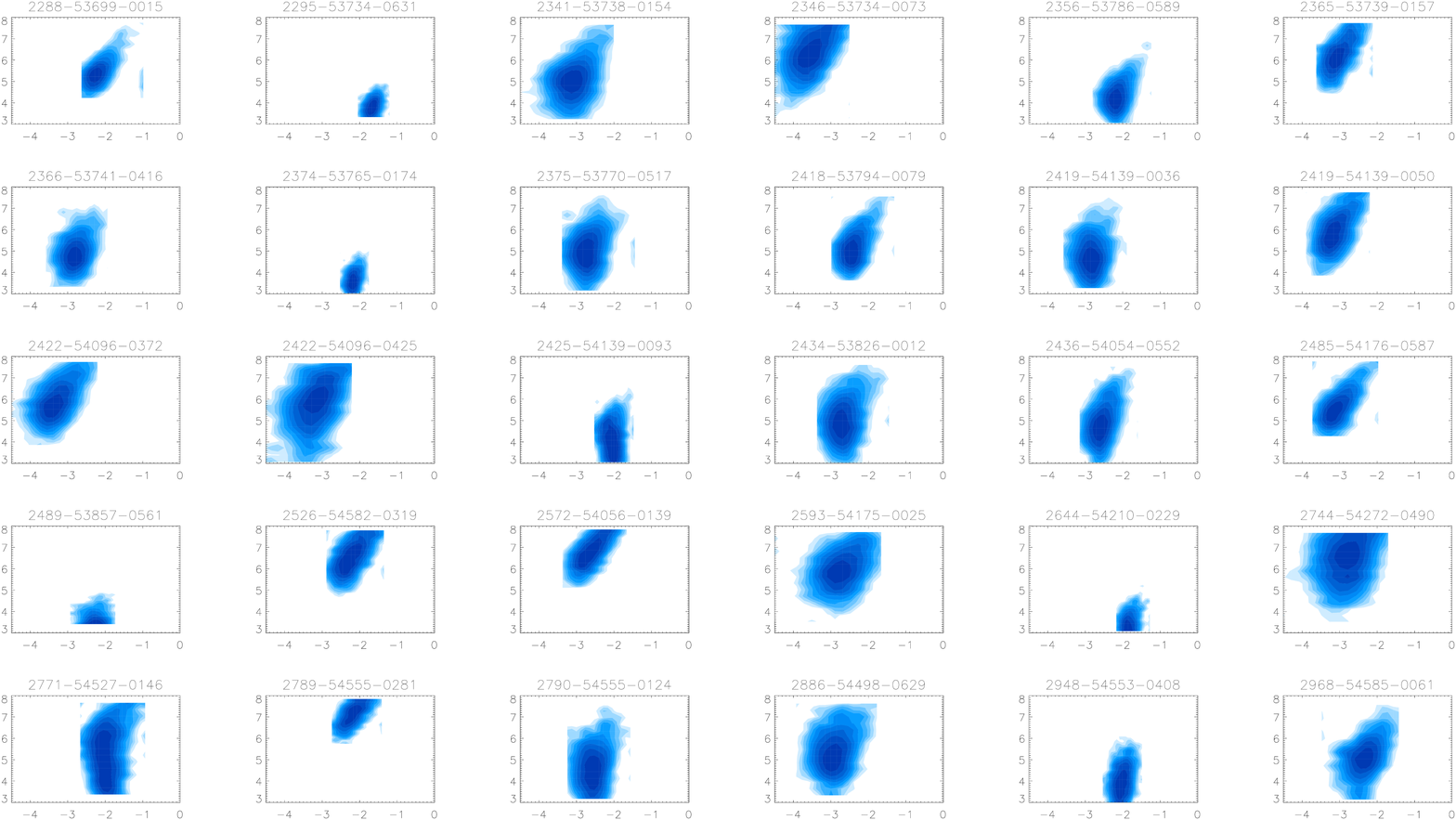}
\caption{--to be continued}
\end{figure*}

\begin{figure}
\centering\includegraphics[width = 8cm,height=5cm]{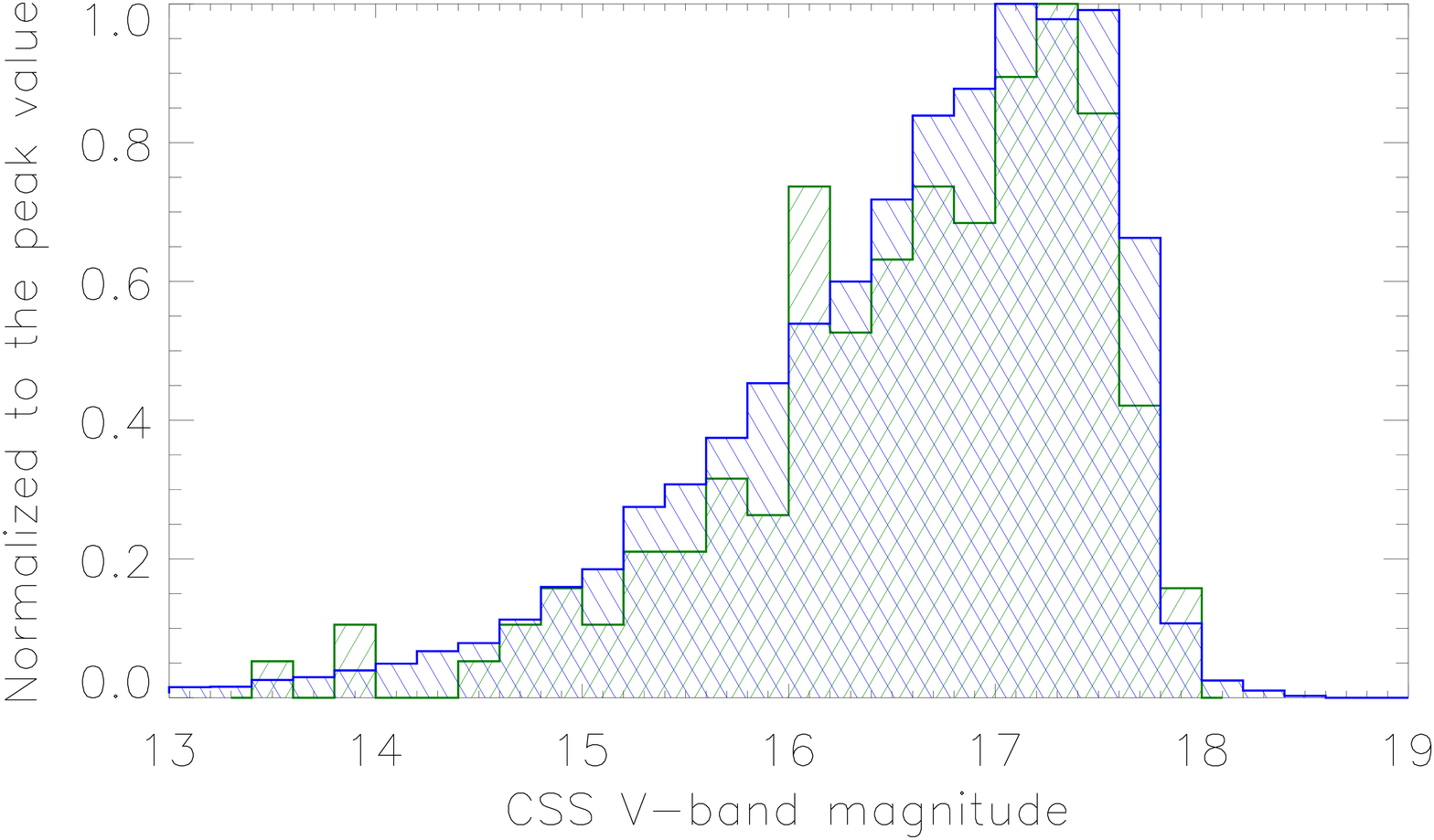}
\caption{Distributions of CSS V-band magnitude of the 156 Type-2 AGN (histogram filled by dark 
green lines) and with apparent long-term variabilities and the 12881 Type-2 AGN (histogram filled by blue lines).}
\label{db2}
\end{figure}

\begin{figure*}
\centering\includegraphics[width = 18cm,height=6cm]{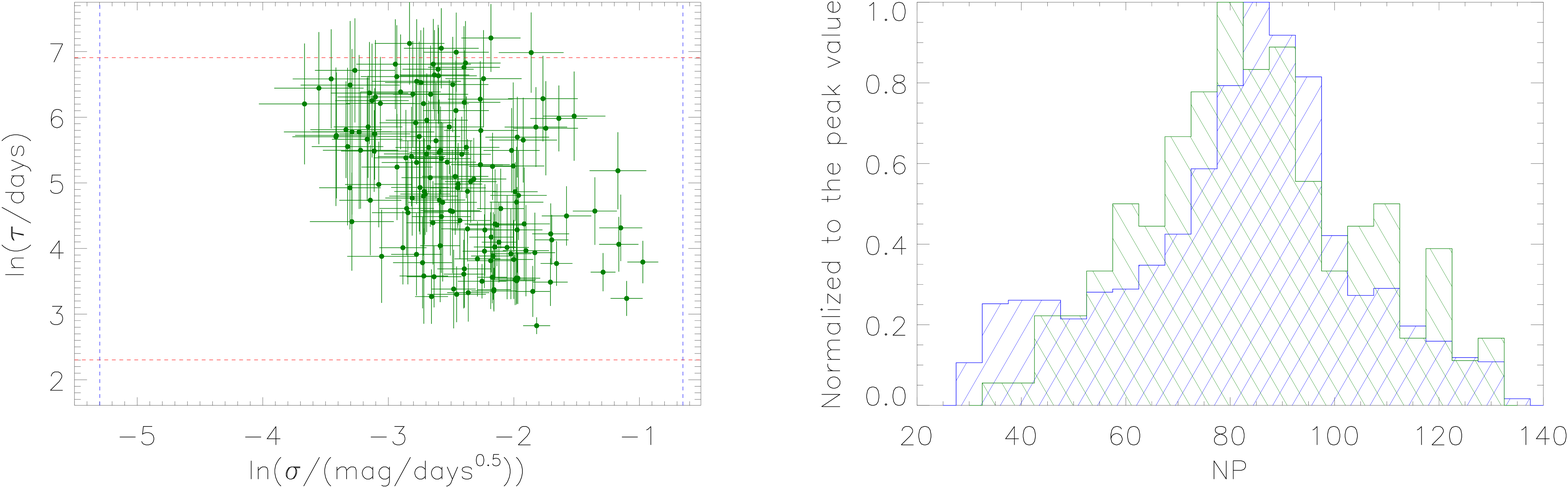}
\caption{Left panel shows properties of $\ln(\sigma/(mag/days^{0.5}))$ and $\ln(\tau/days)$ of the 156 
Type-2 AGN with apparent long-term variabilities. Vertical dashed blue lines and horizontal dashed red 
lines show the available ranges of $\ln(\sigma/(mag/days^{0.5}))$ and of $\ln(\tau/days)$ from the OGLE-III 
quasars in \citet{koz10}. Right panel shows distributions of $NP$ of the 156 Type-2 AGN (histogram filled 
by dark green lines) with apparent long-term variabilities and the other Type-2 AGN (histogram filled by 
blue lines).
}
\label{dis}
\end{figure*}

\begin{figure*}
\centering\includegraphics[width = 18cm,height=6cm]{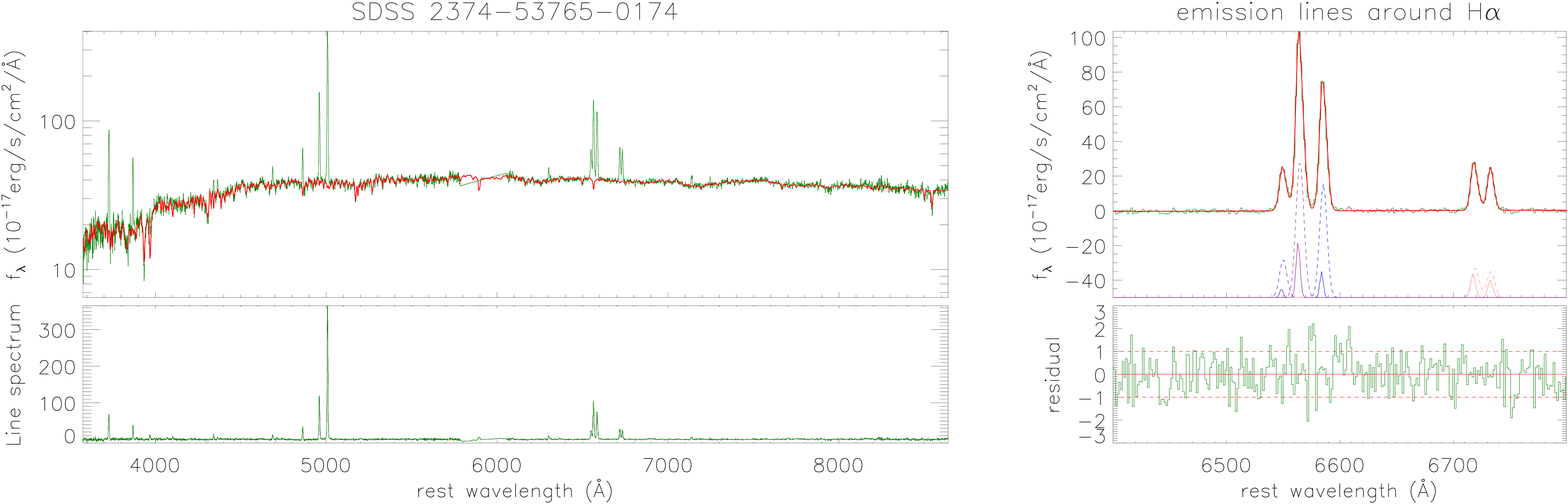}
\caption{Top left panel shows spectrum (in dark green) of SDSS 2374-53765-0174 and the corresponding SSP 
method determined host galaxy contributions (in red). Bottom left panel shows the line spectrum of SDSS 
2374-53765-0174 by the SDSS spectrum minus the host galaxy contributions. Top right panel shows the best 
descriptions (in red) to the emission lines (in dark green) around H$\alpha$ by multiple Gaussian functions, 
in the line spectrum shown in bottom left panel. In bottom region of top right panel, solid and dashed lines 
in blue show the determined first and second narrow [N~{\sc ii}] components, solid and dashed lines in pink 
show the determined first and second narrow [S~{\sc ii}] components, solid and dashed purple lines show 
the determined first and second narrow components in H$\alpha$. Bottom right panel shows the residuals 
calculated by the line spectrum minus the best fitting results and then divided by uncertainties of the 
spectrum. In the bottom right panel, solid and dashed red lines show $residual=0,\pm1$, respectively. 
}
\label{ssp}
\end{figure*}

\section{Spectroscopic Results of the 156 Type-2 AGN but with long-term variabilities}

	In order to confirm the 156 Type-2 AGN are well classified as Type-2 AGN by spectroscopic results, 
the spectra properties should be further discussed. Due to apparent host galaxy contributions to the SDSS 
spectra of the 156 Type-2 AGN, the starlight in the SDSS spectra should be firstly determined. Here, the 
commonly accepted SSP (simple stellar population) method has been applied. More detailed descriptions on 
the SSP method can be found in \citet{bc03, ka03, cm05, cm17}. And the SSP method has been applied in our 
previous papers \citet{zh14, zh16, ra17, zh19, zh21m, zh21a, zh21b, zh22}. Here, we do not show further 
discussions on the SSP method any more, but simple descriptions on SSP method are described as follows. 
The 39 simple stellar population templates in \citet{bc03} have been exploited, which can be used to 
well-describe the characteristics of almost all the SDSS galaxies as detailed discussions in \citet{bc03}. 
Meanwhile, there is an additional component, a fourth degree polynomial function describe component, which 
is applied to describe probable intrinsic AGN continuum emissions after considering the apparent variabilities 
or to modify continuum shapes when to describe the SDSS spectra by the templates. Meanwhile, when the SSP 
method is applied, the narrow emission lines listed in 
\url{http://classic.sdss.org/dr1/algorithms/speclinefits.html#linelist} are masked out by full width at 
zero intensity about 450${\rm km~s^{-1}}$, And the wavelength ranges from 4450 to 5600\AA~ and from 6250 
to 6750\AA~ are also masked out for the probably broad H$\beta$ and the broad H$\alpha$ emission lines.

\begin{figure*}
\centering\includegraphics[width = 8.5cm,height=3.2cm]{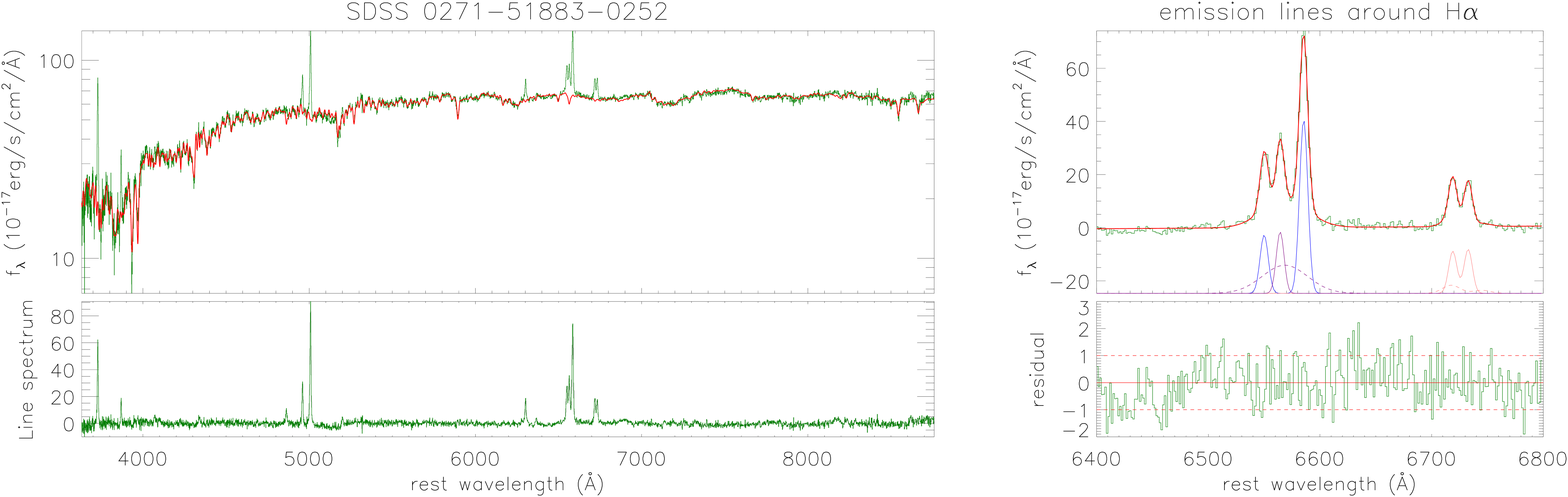}
\centering\includegraphics[width = 8.5cm,height=3.2cm]{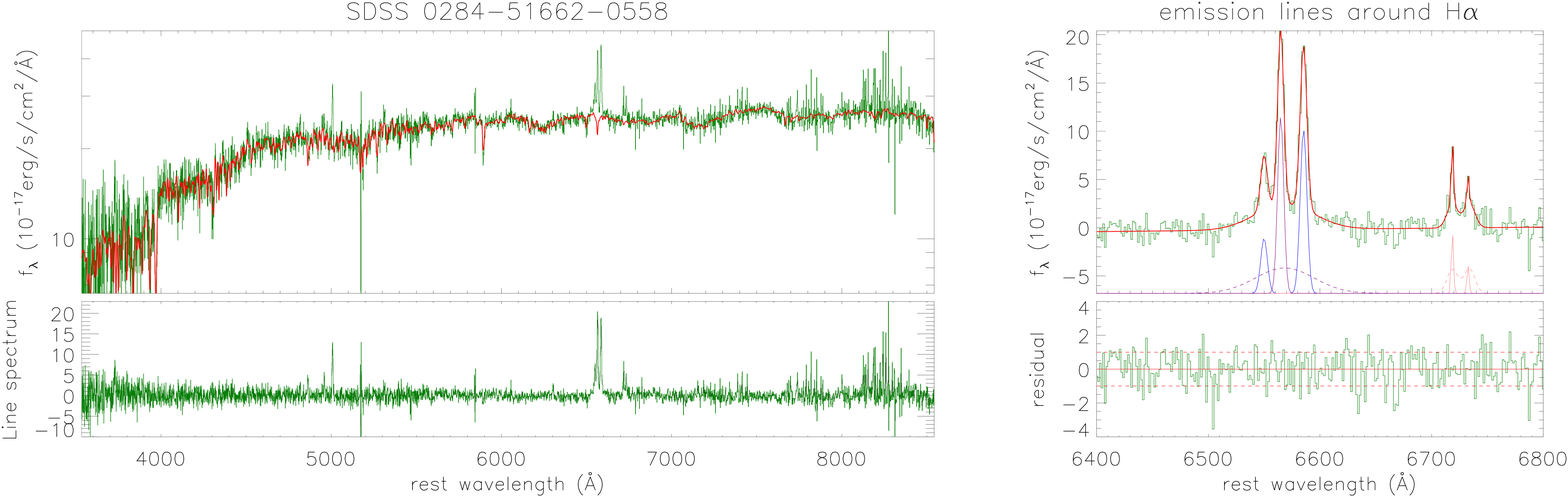}
\centering\includegraphics[width = 8.5cm,height=3.2cm]{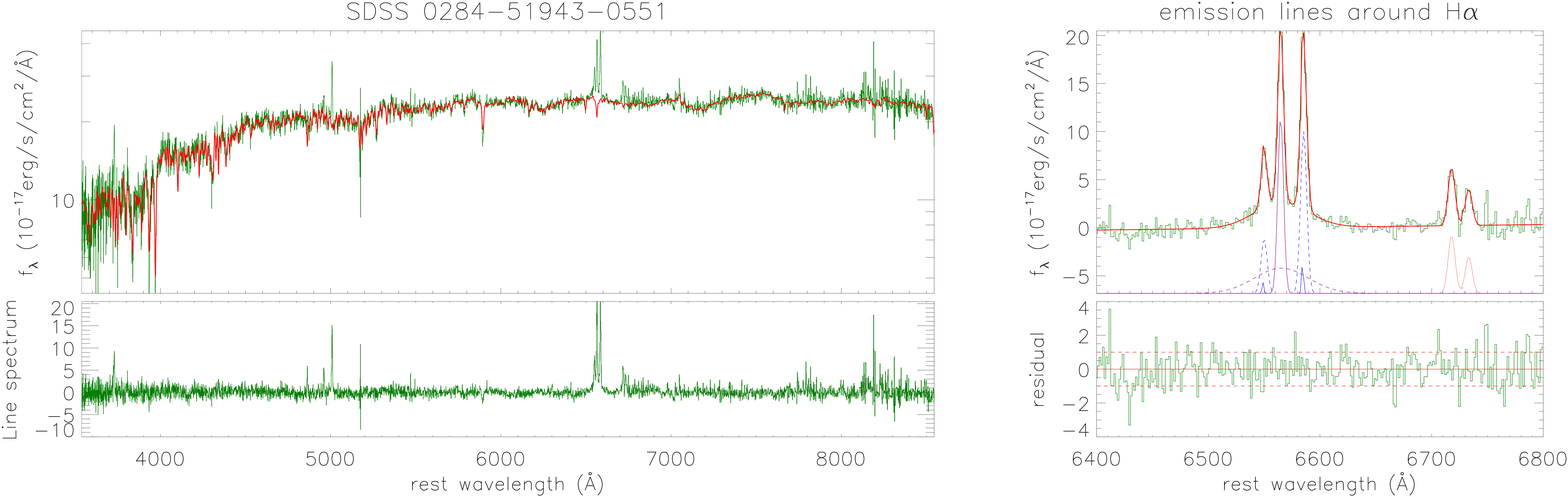}
\centering\includegraphics[width = 8.5cm,height=3.2cm]{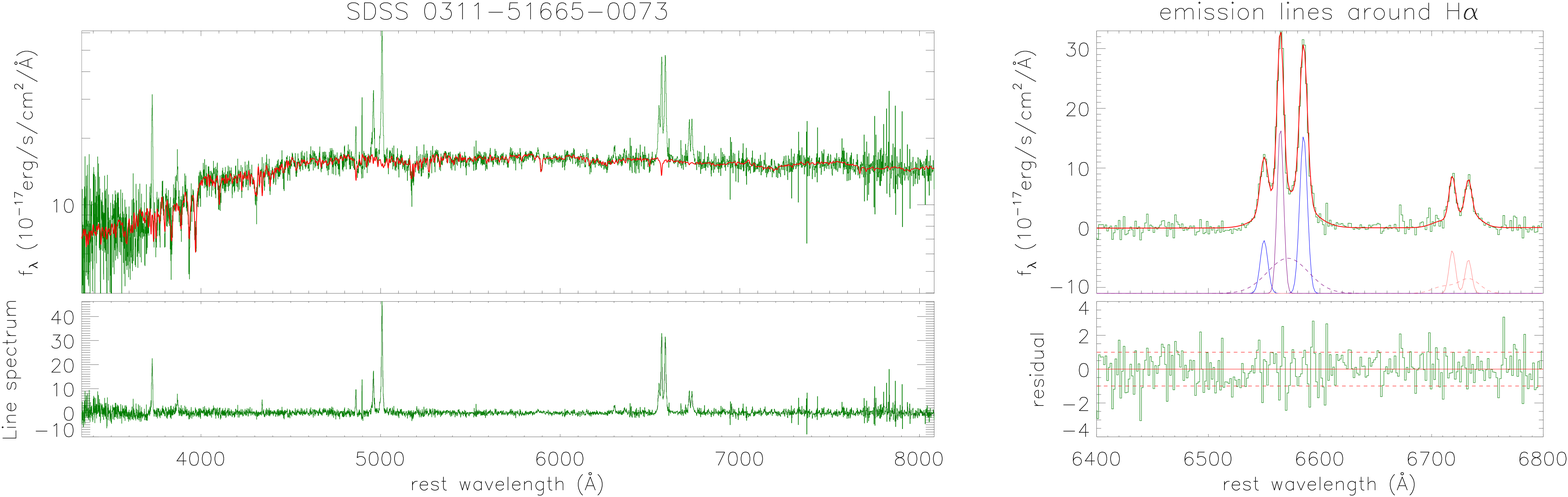}
\centering\includegraphics[width = 8.5cm,height=3.2cm]{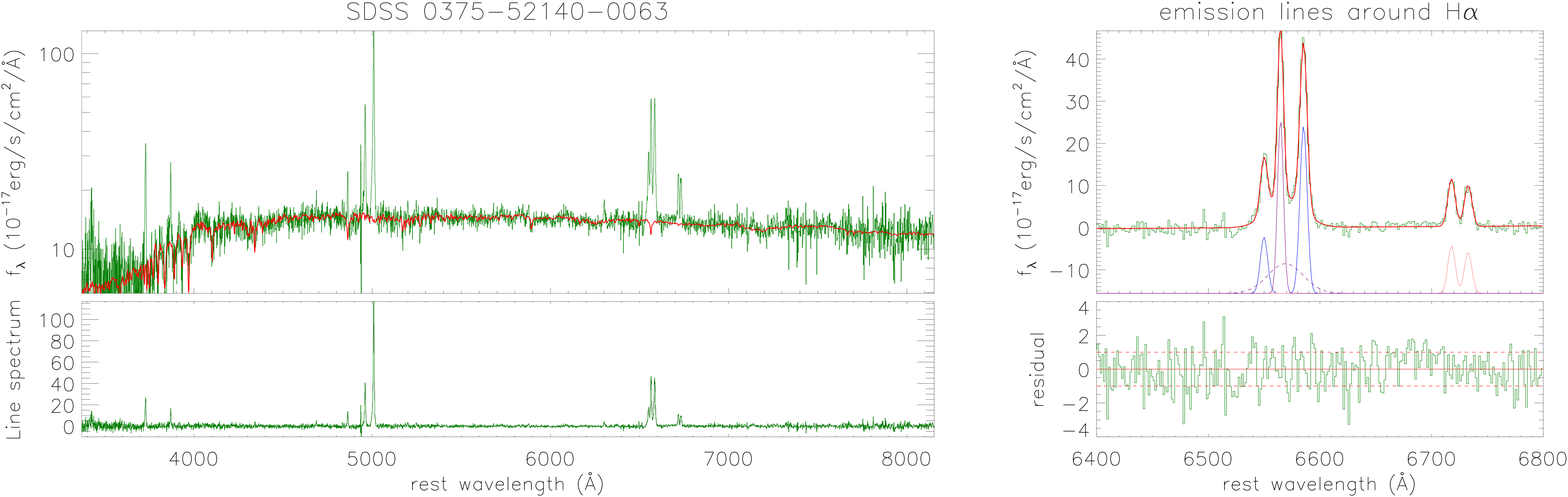}
\centering\includegraphics[width = 8.5cm,height=3.2cm]{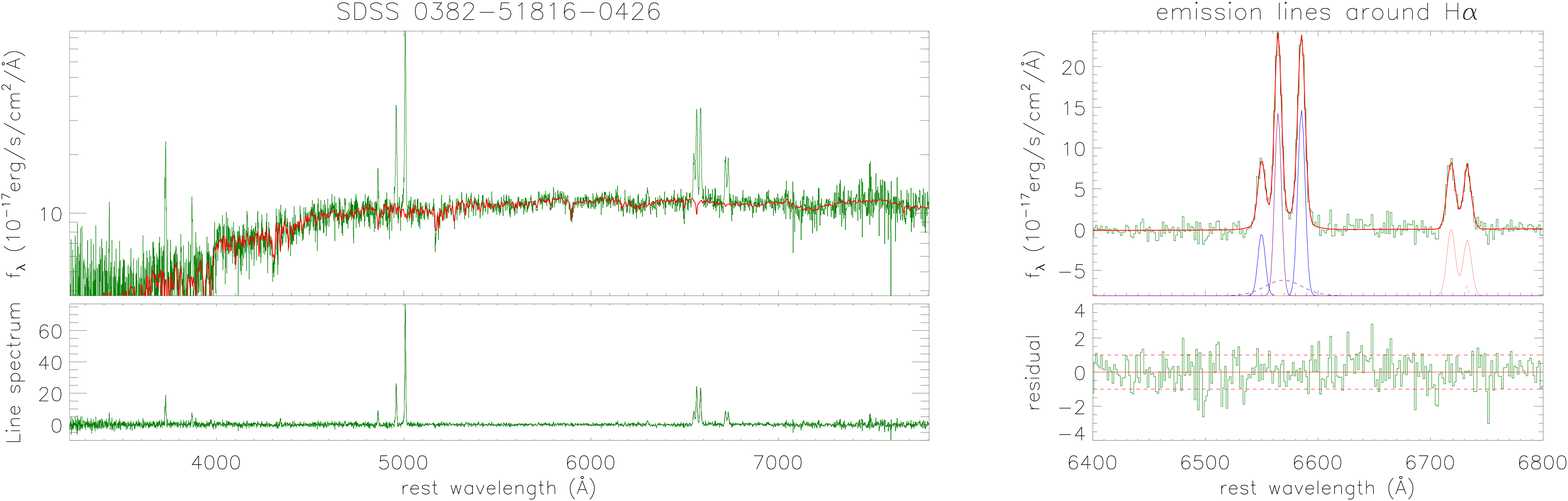}
\centering\includegraphics[width = 8.5cm,height=3.2cm]{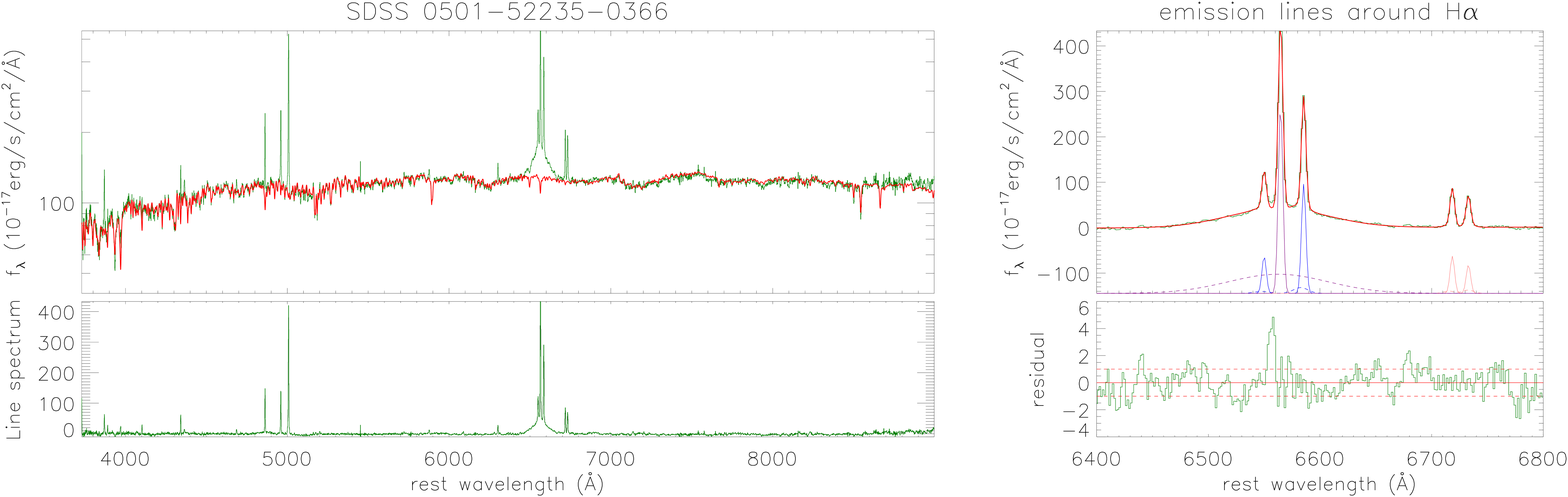}
\centering\includegraphics[width = 8.5cm,height=3.2cm]{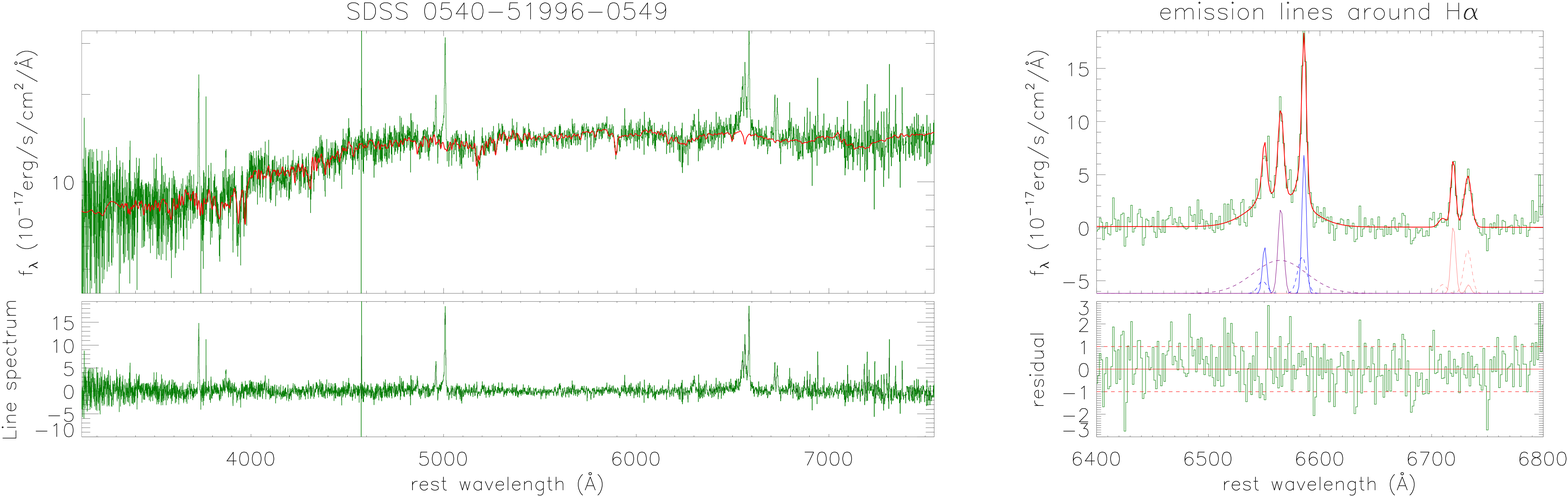}
\centering\includegraphics[width = 8.5cm,height=3.2cm]{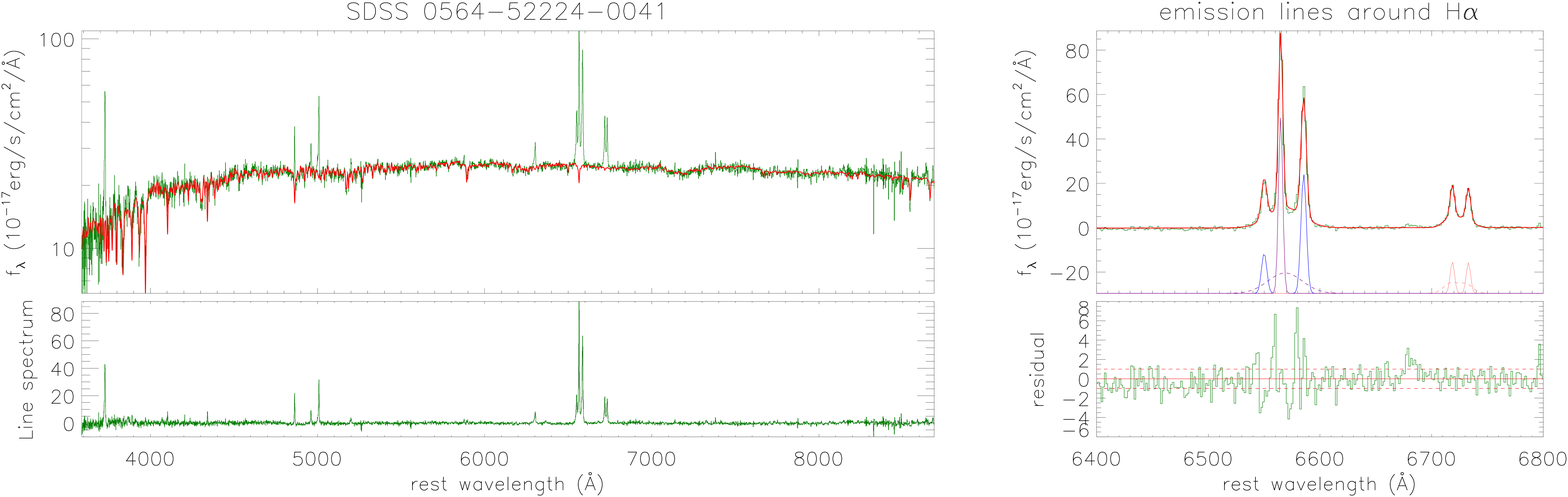}
\centering\includegraphics[width = 8.5cm,height=3.2cm]{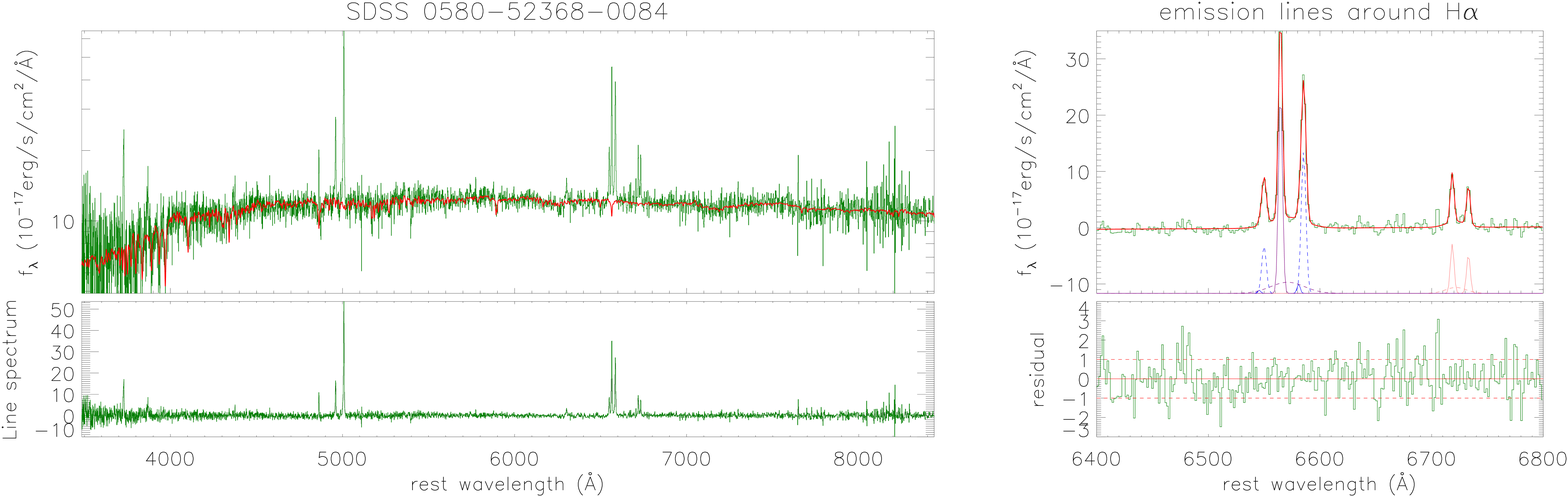}
\centering\includegraphics[width = 8.5cm,height=3.2cm]{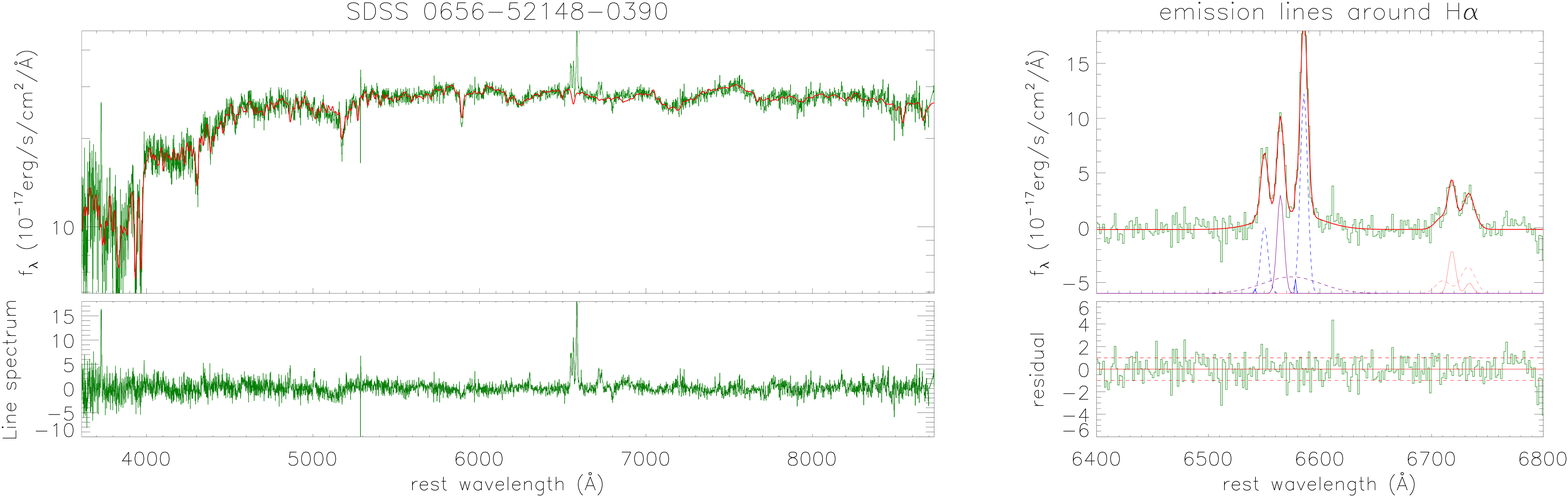}
\centering\includegraphics[width = 8.5cm,height=3.2cm]{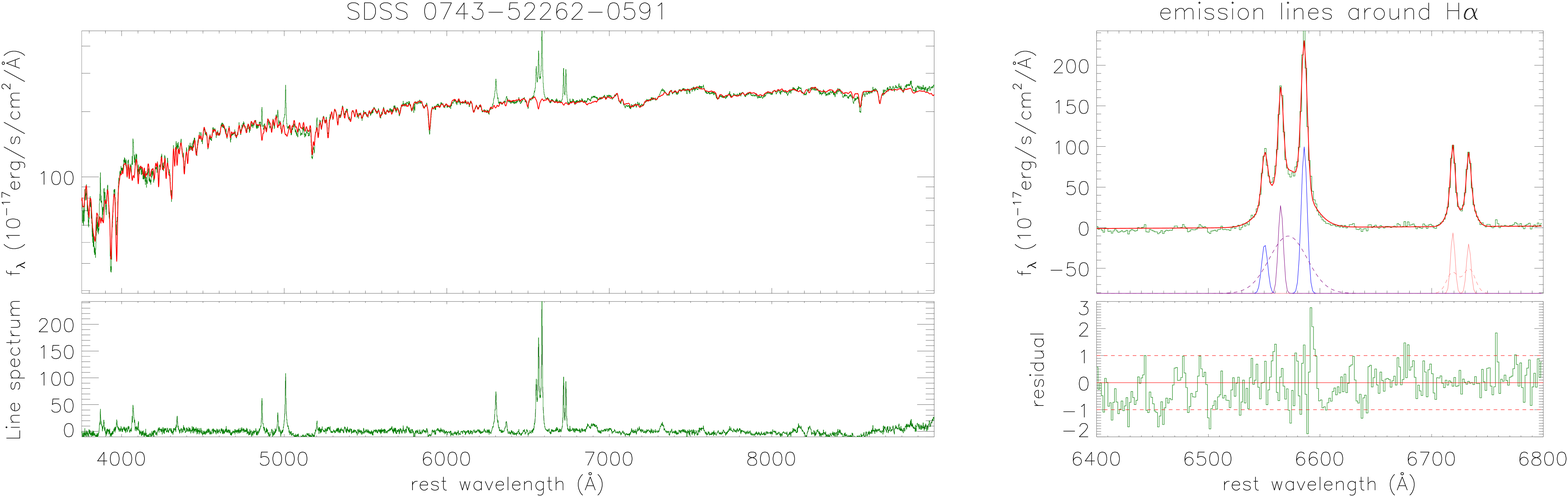}
\centering\includegraphics[width = 8.5cm,height=3.2cm]{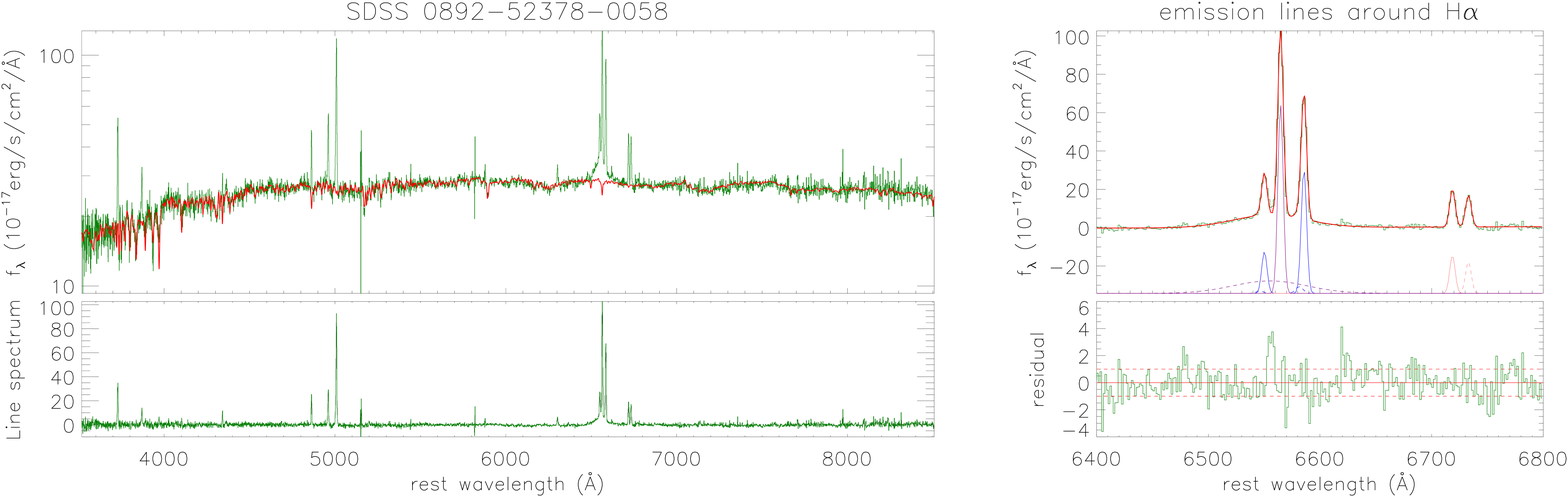}
\centering\includegraphics[width = 8.5cm,height=3.2cm]{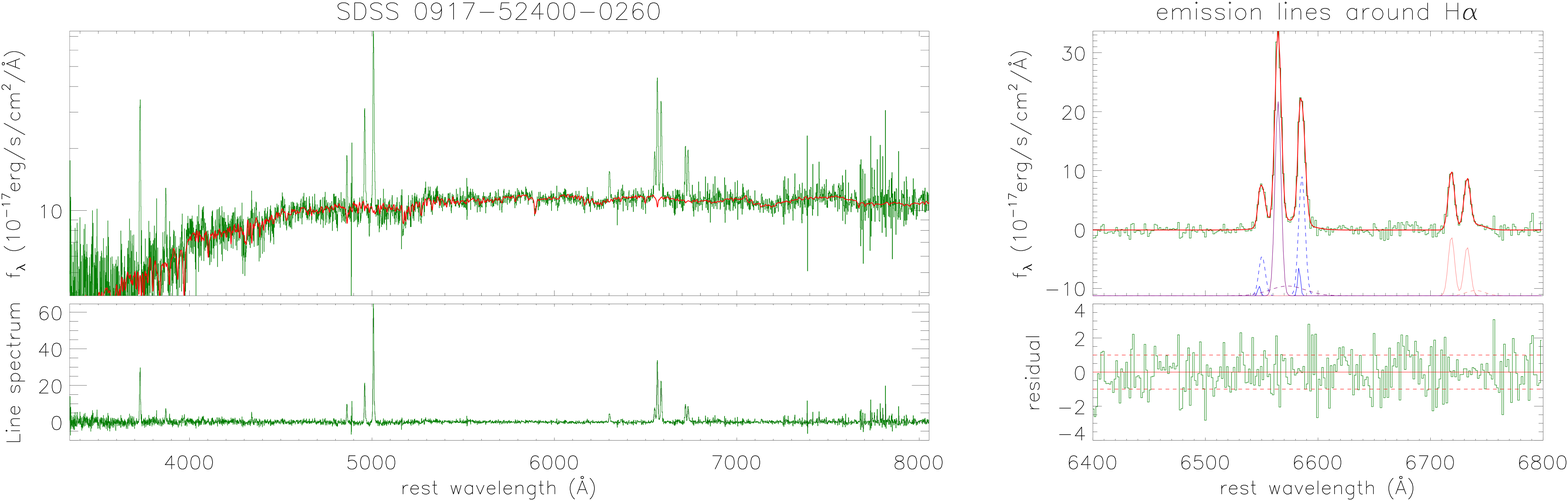}
\caption{Each two columns in each row show the similar results as those shown in Fig.~\ref{ssp}, but for the 
31 Type-2 AGN with both long-term variabilities and probable broad H$\alpha$ emission component described 
by dashed purple line. }
\label{line}
\end{figure*}

\setcounter{figure}{7}

\begin{figure*}
\centering\includegraphics[width = 8.5cm,height=3.2cm]{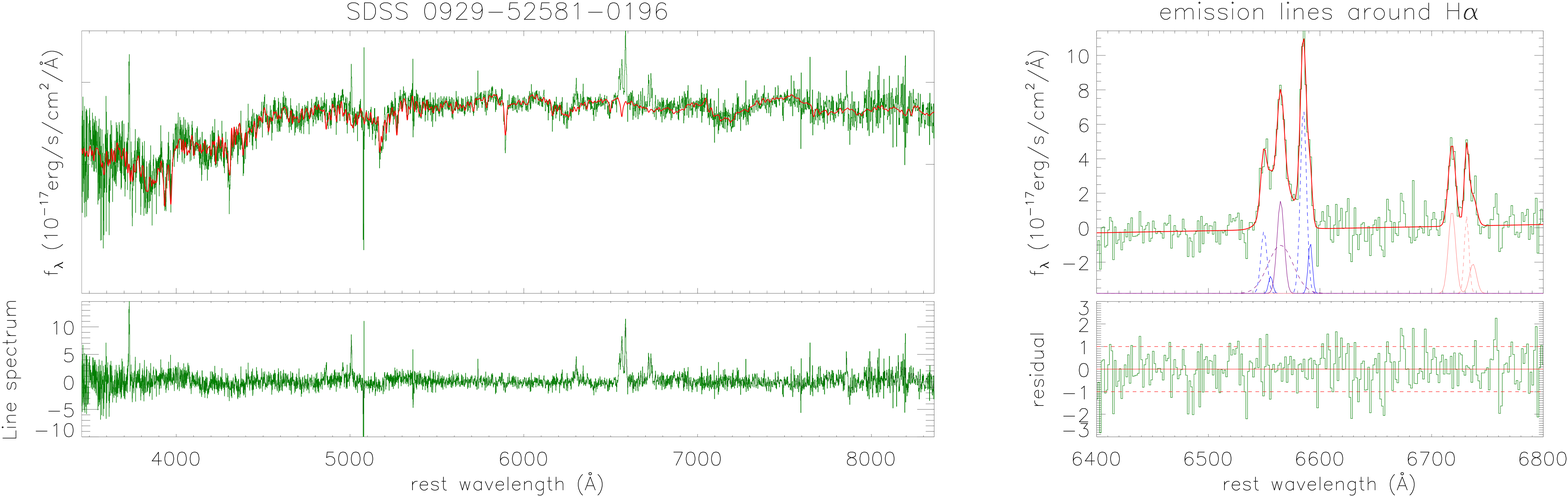}
\centering\includegraphics[width = 8.5cm,height=3.2cm]{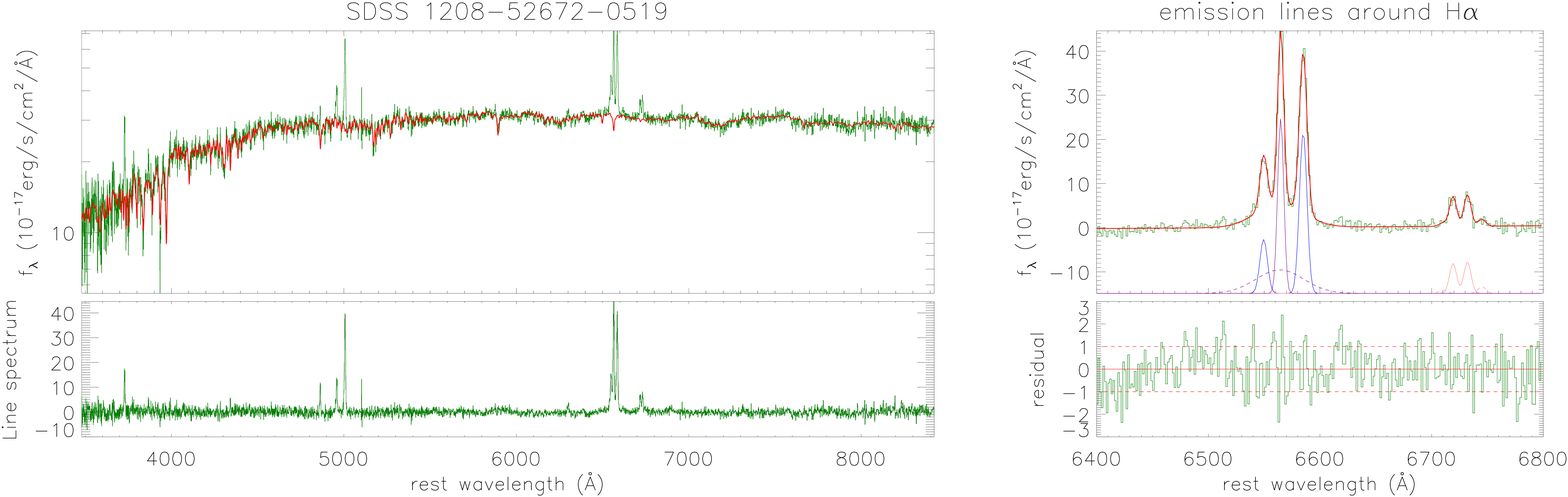}
\centering\includegraphics[width = 8.5cm,height=3.2cm]{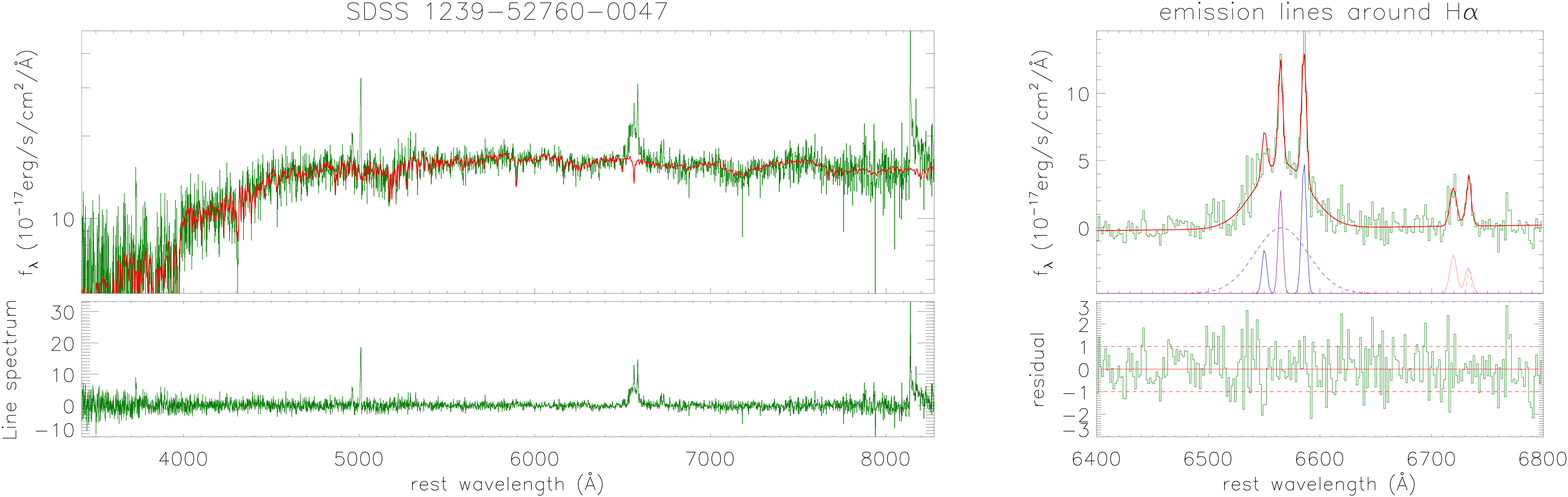}
\centering\includegraphics[width = 8.5cm,height=3.2cm]{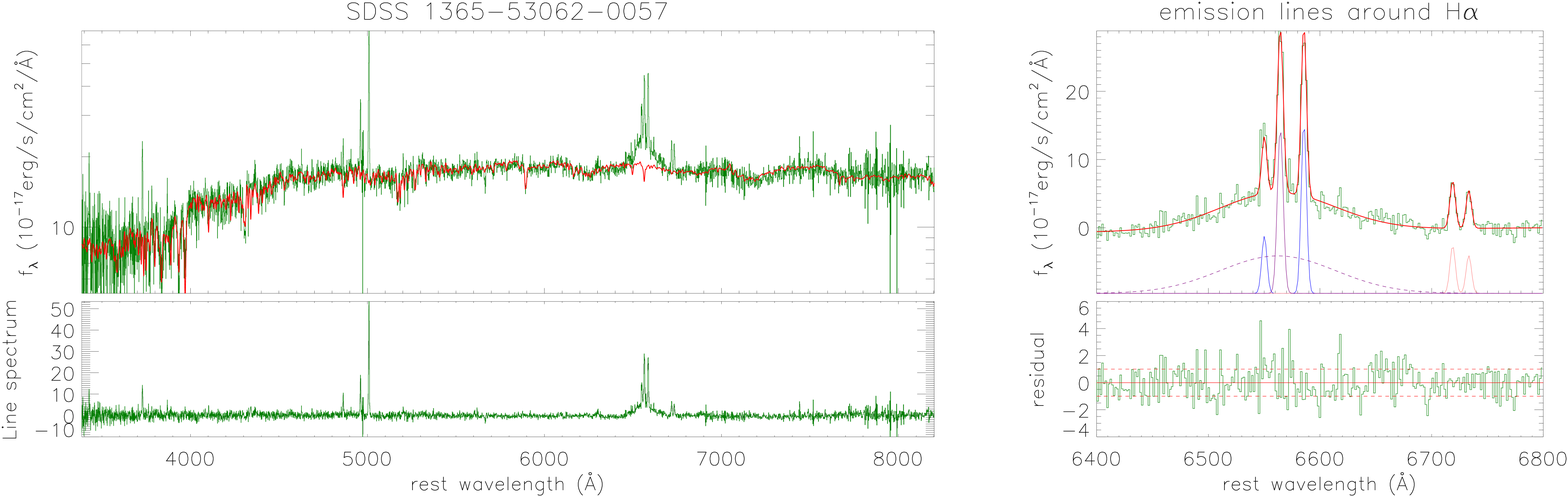}
\centering\includegraphics[width = 8.5cm,height=3.2cm]{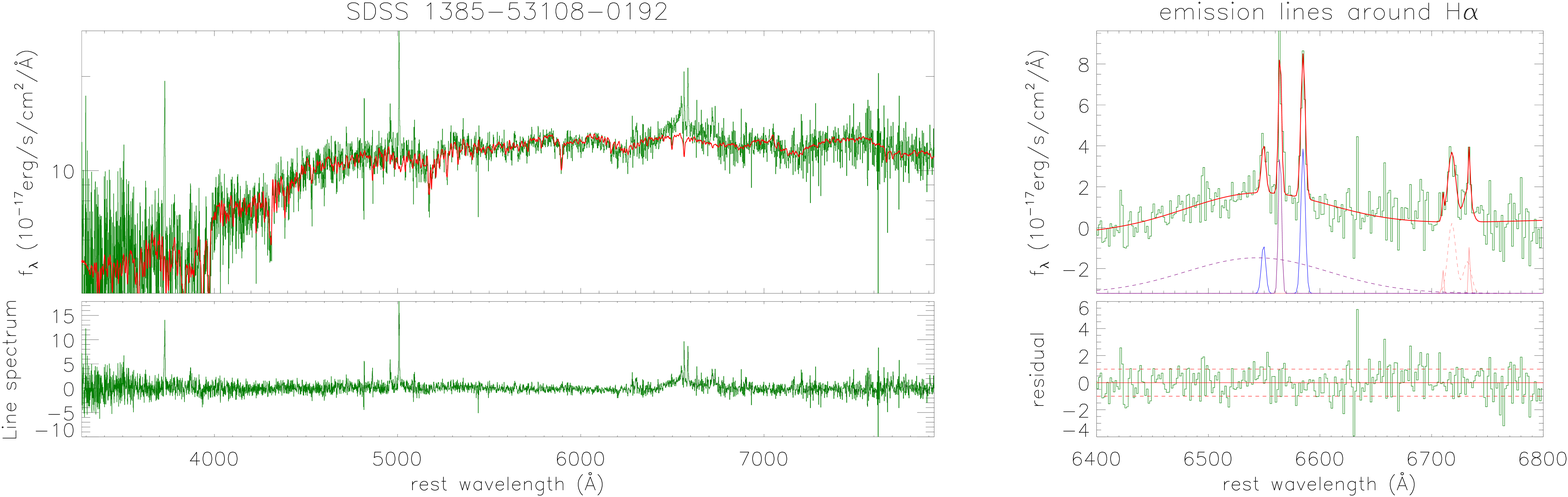}
\centering\includegraphics[width = 8.5cm,height=3.2cm]{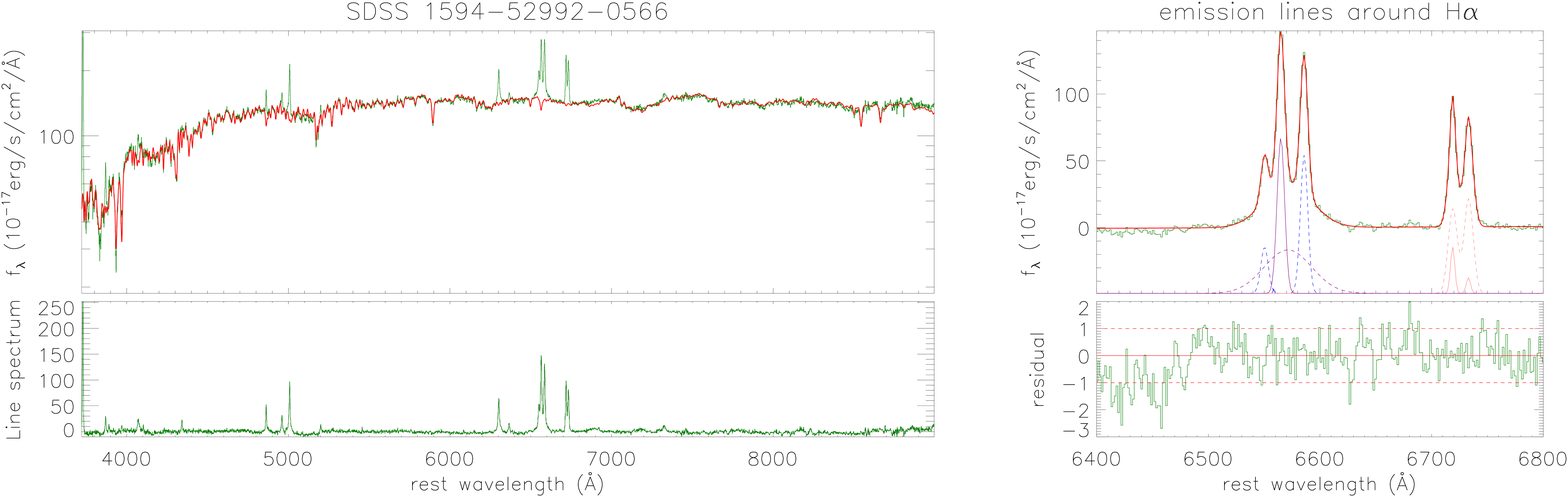}
\centering\includegraphics[width = 8.5cm,height=3.2cm]{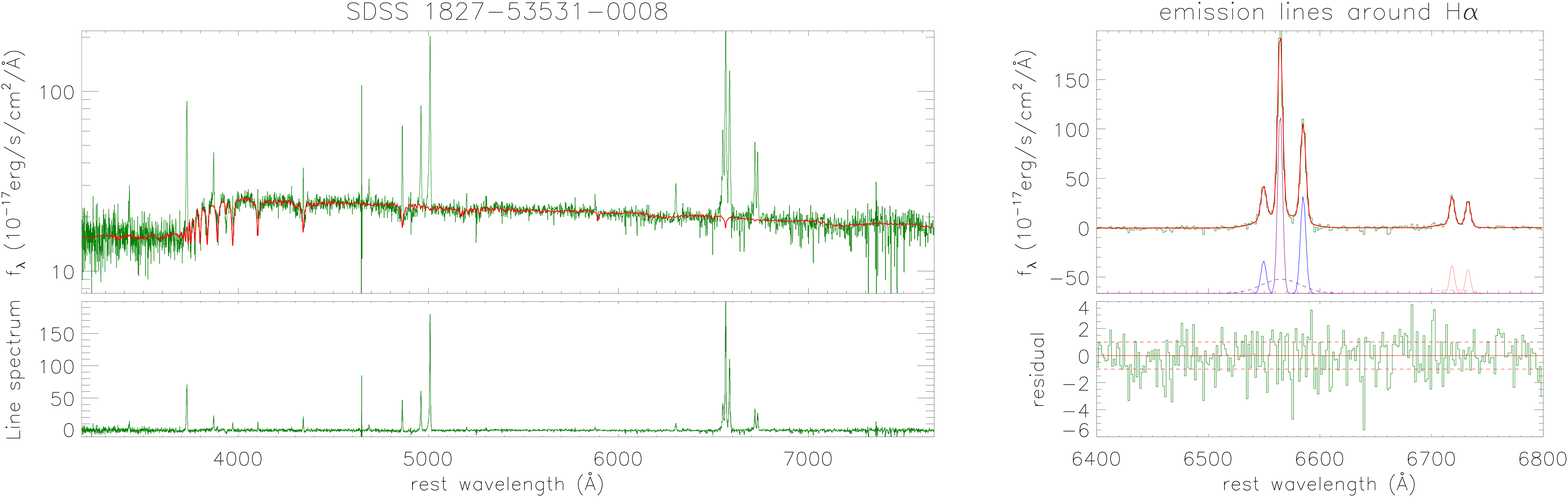}
\centering\includegraphics[width = 8.5cm,height=3.2cm]{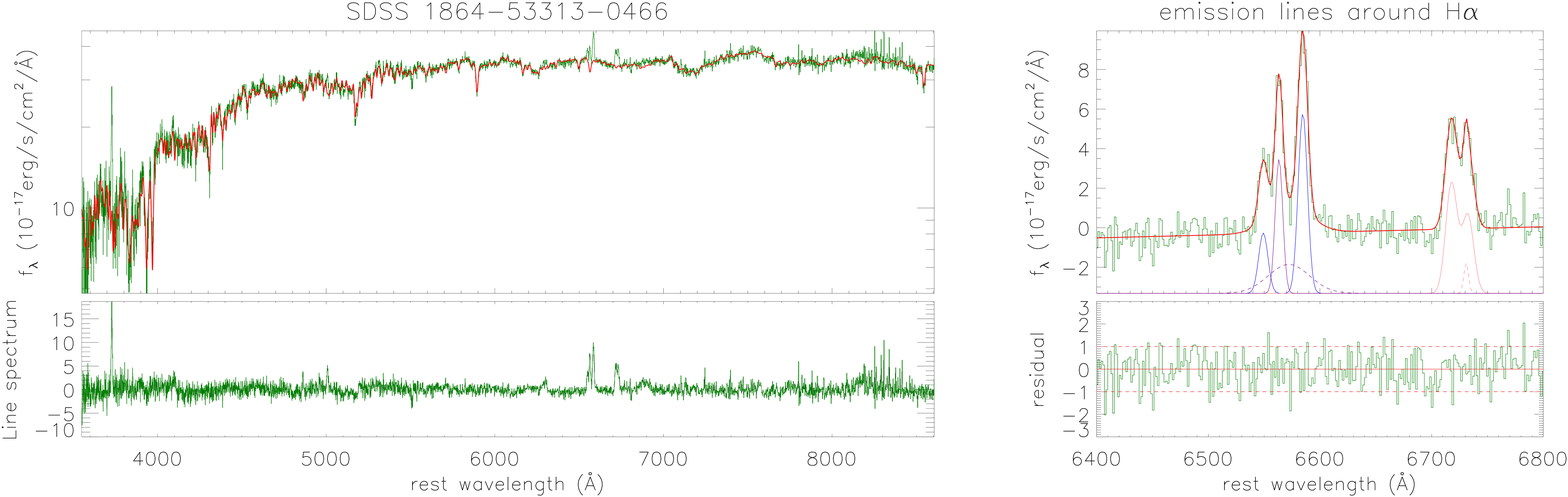}
\centering\includegraphics[width = 8.5cm,height=3.2cm]{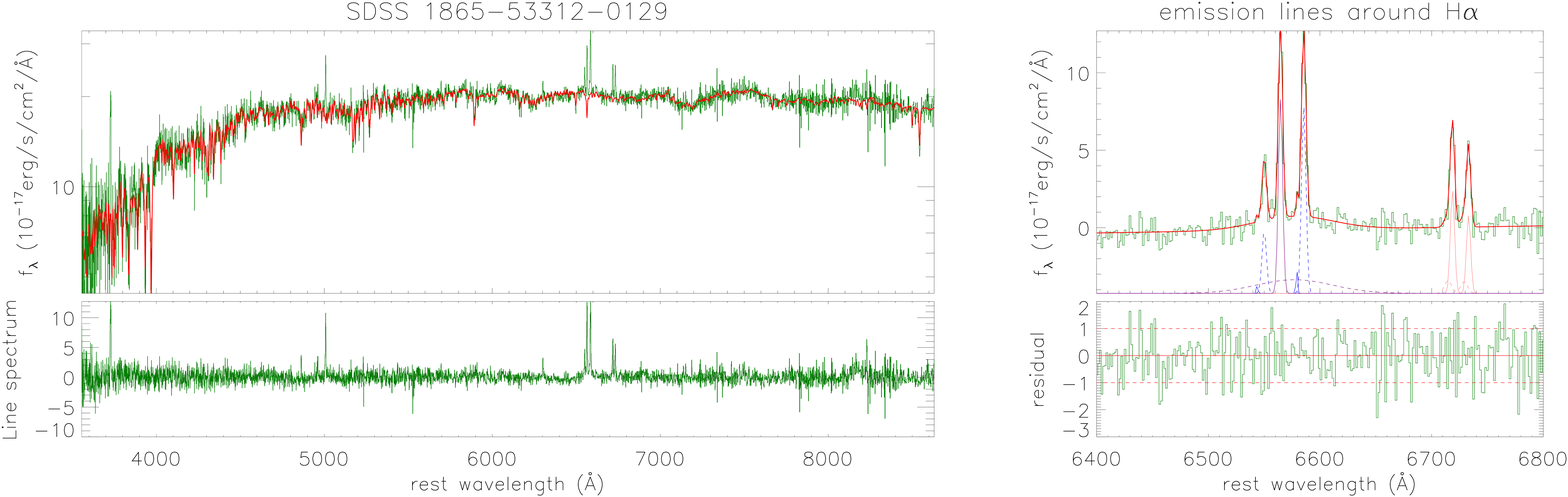}
\centering\includegraphics[width = 8.5cm,height=3.2cm]{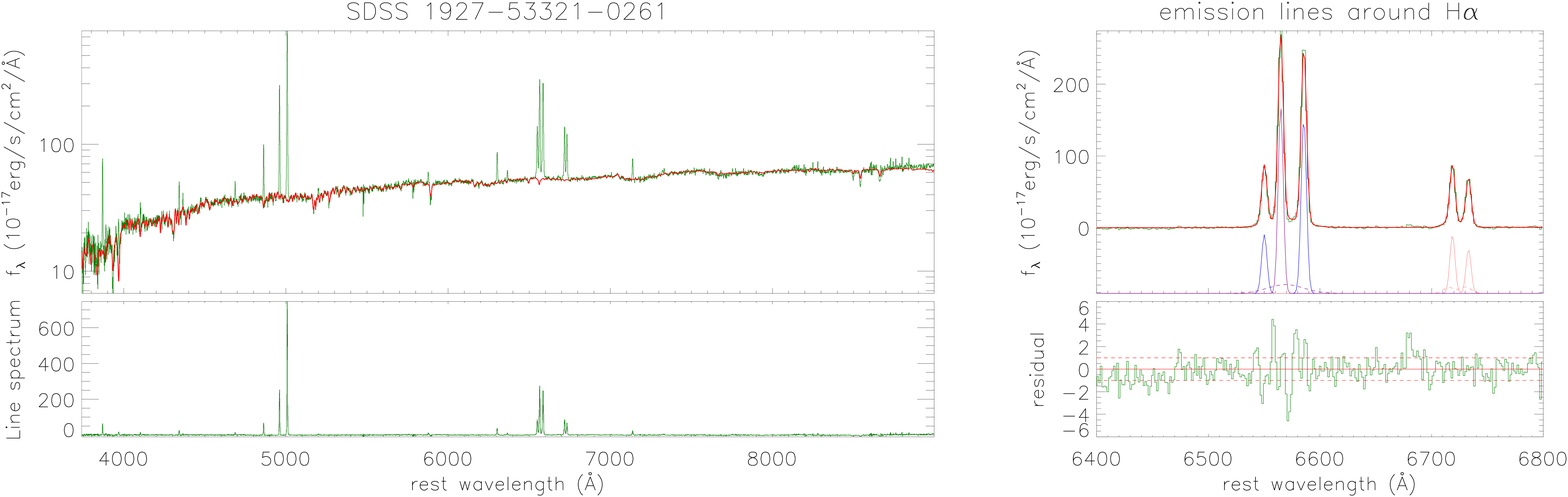}
\centering\includegraphics[width = 8.5cm,height=3.2cm]{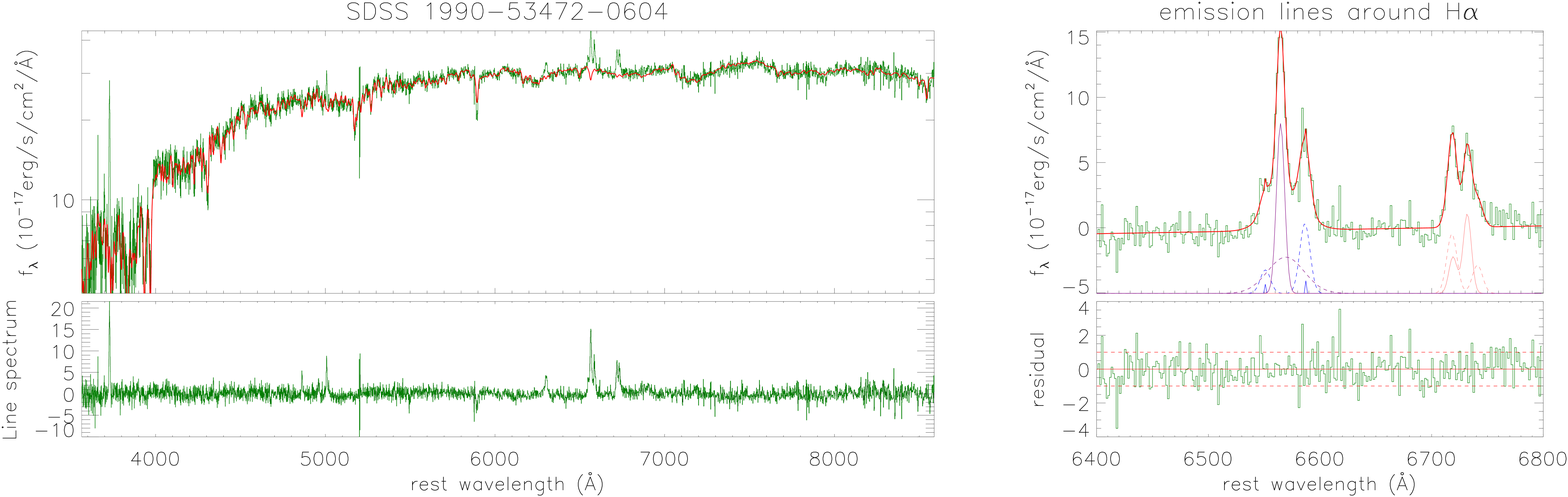}
\centering\includegraphics[width = 8.5cm,height=3.2cm]{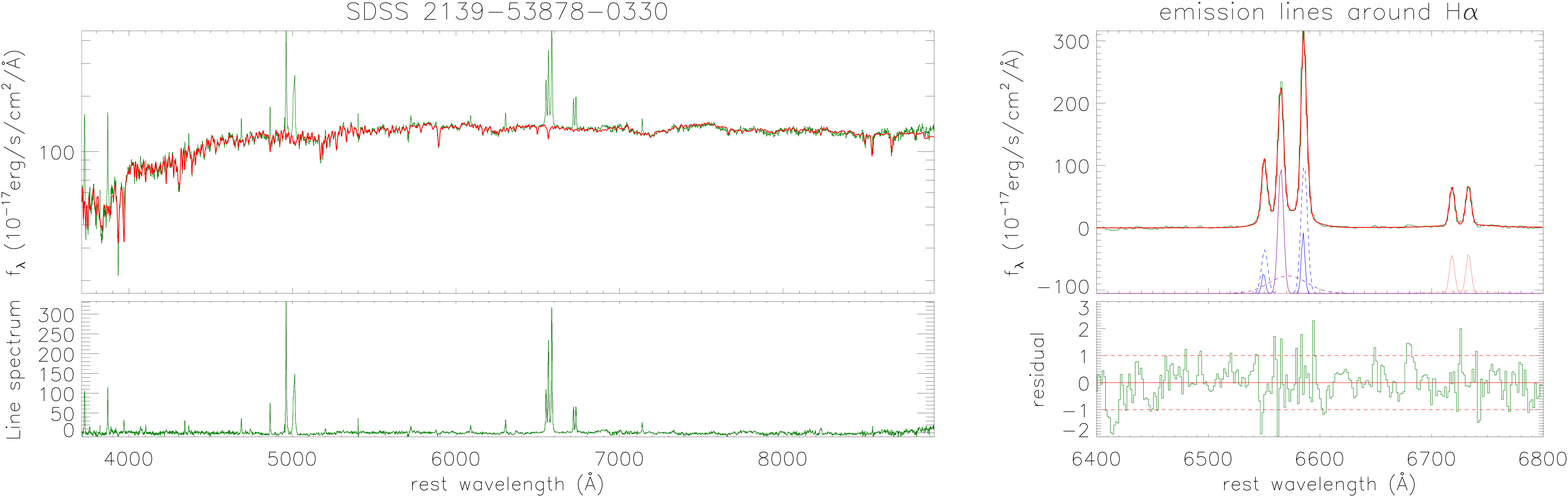}
\centering\includegraphics[width = 8.5cm,height=3.2cm]{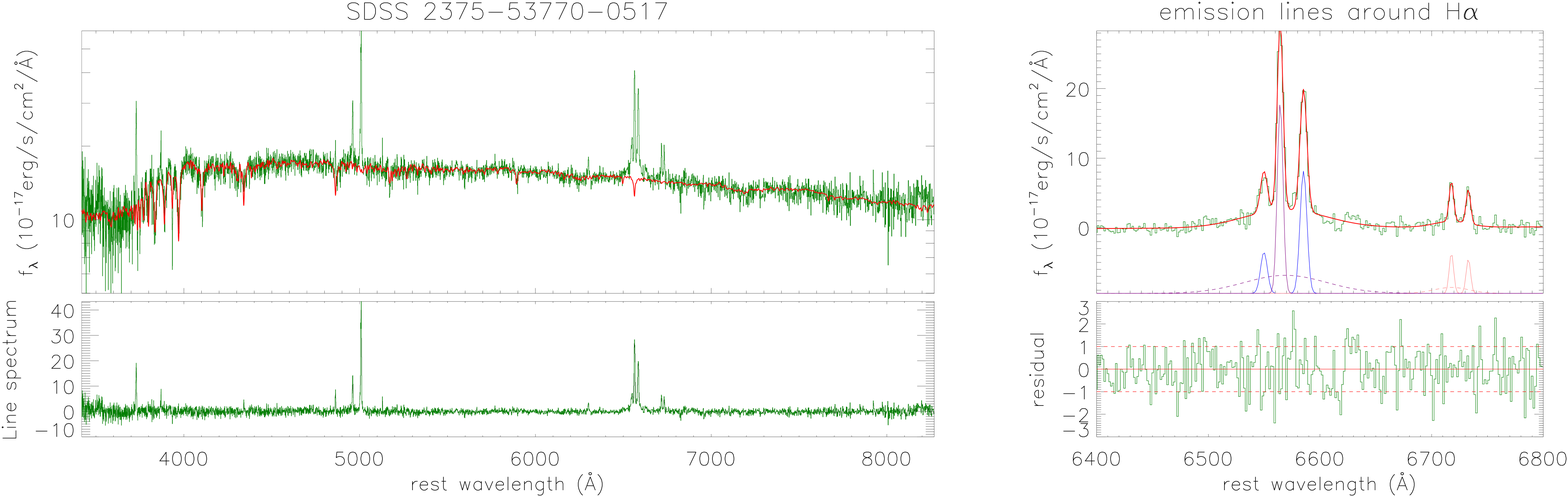}
\centering\includegraphics[width = 8.5cm,height=3.2cm]{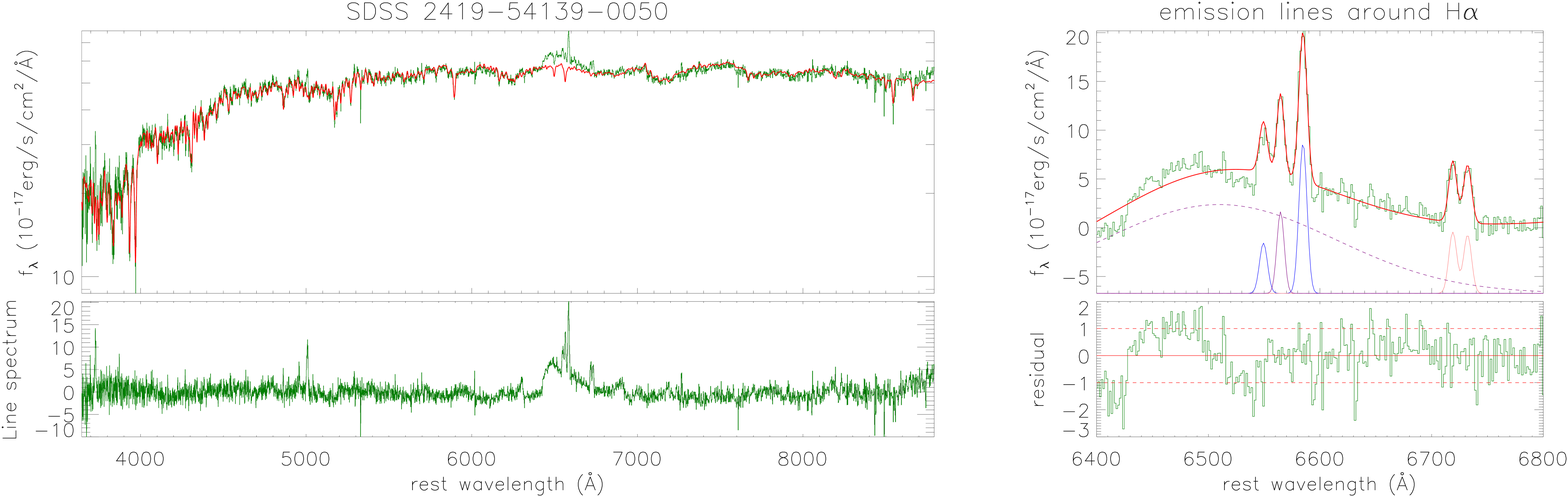}
\caption{--to be continued}
\end{figure*}

\setcounter{figure}{7}

\begin{figure*}
\centering\includegraphics[width = 8.5cm,height=3.2cm]{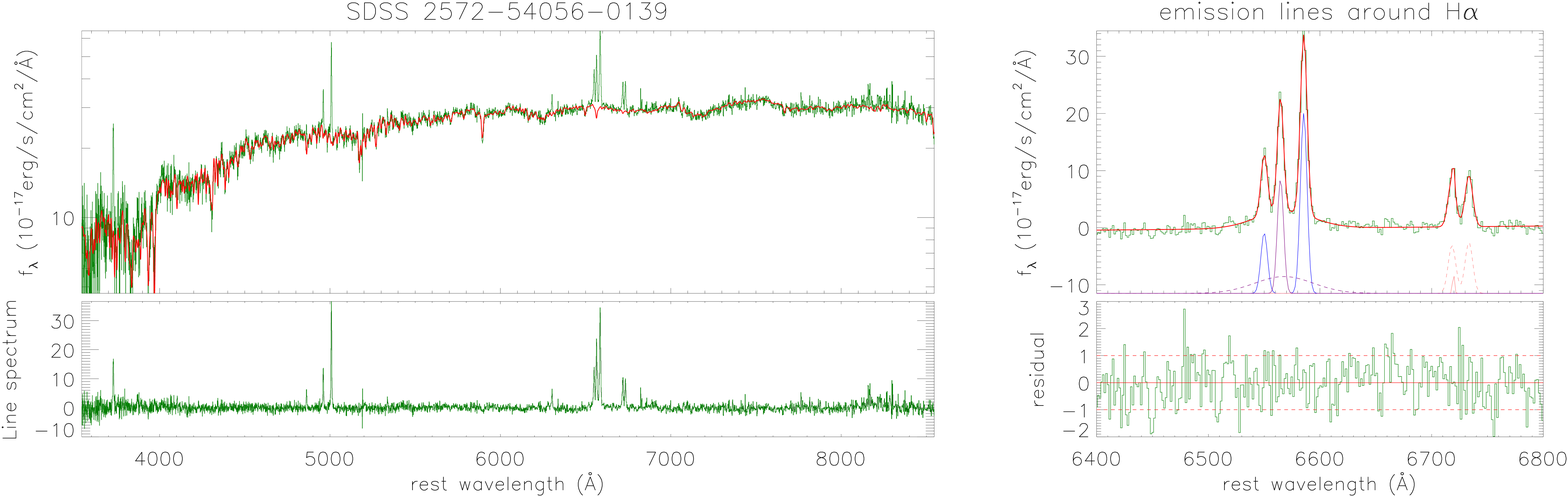}
\centering\includegraphics[width = 8.5cm,height=3.2cm]{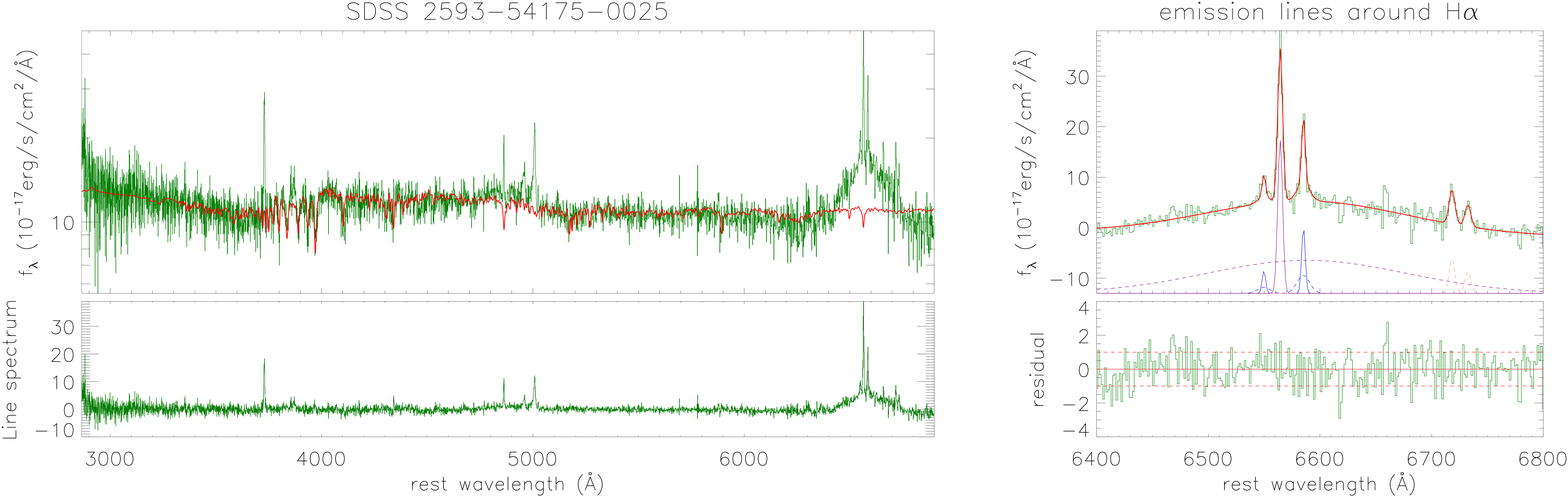}
\centering\includegraphics[width = 8.5cm,height=3.2cm]{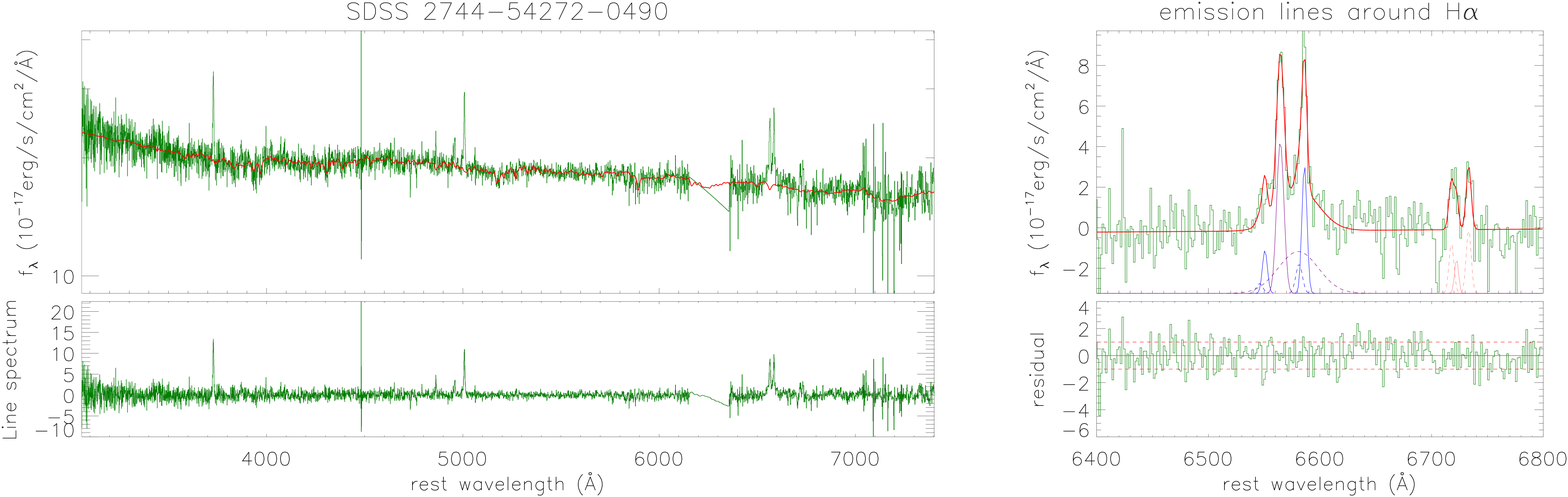}
\caption{--to be continued}
\end{figure*}

\begin{figure*}
\centering\includegraphics[width = 18cm,height=5cm]{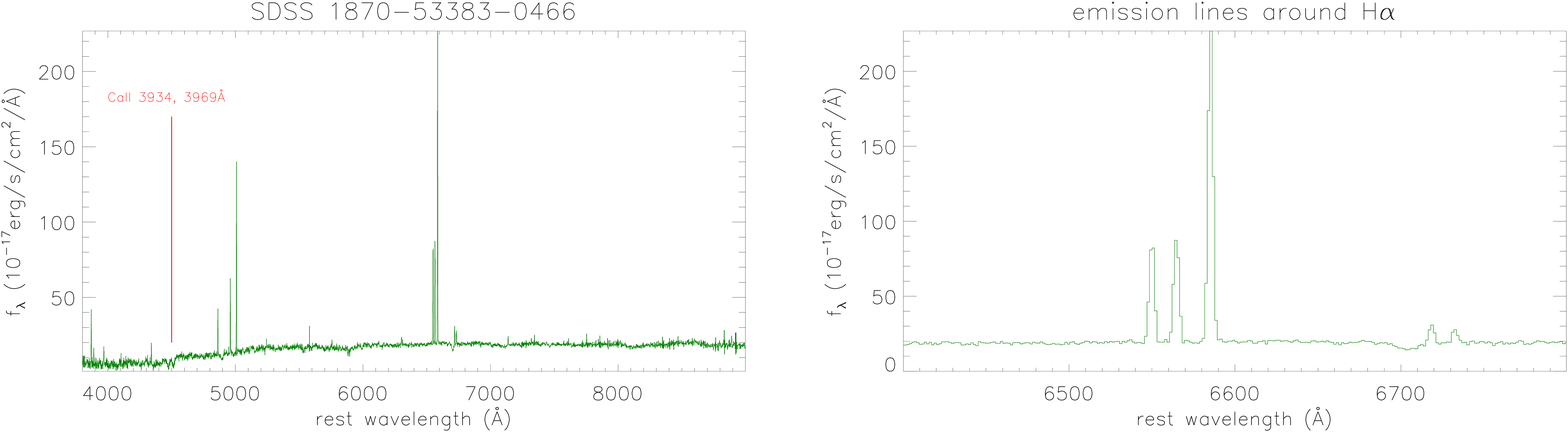}
\caption{Left panel shows spectrum (in dark green) of SDSS J075736.47+532557.1 (plate-mjd-fiberid = 
1870-53383-0466) with apparent Ca~{\sc ii}$\lambda3934, 3969$\AA~ absorption features marked by vertical 
red line. Right panel shows the spectrum around H$\alpha$ to ensure no broad component in H$\alpha$.
}
\label{s1870}
\end{figure*}

\begin{figure*}
\centering\includegraphics[width = 18cm,height=18cm]{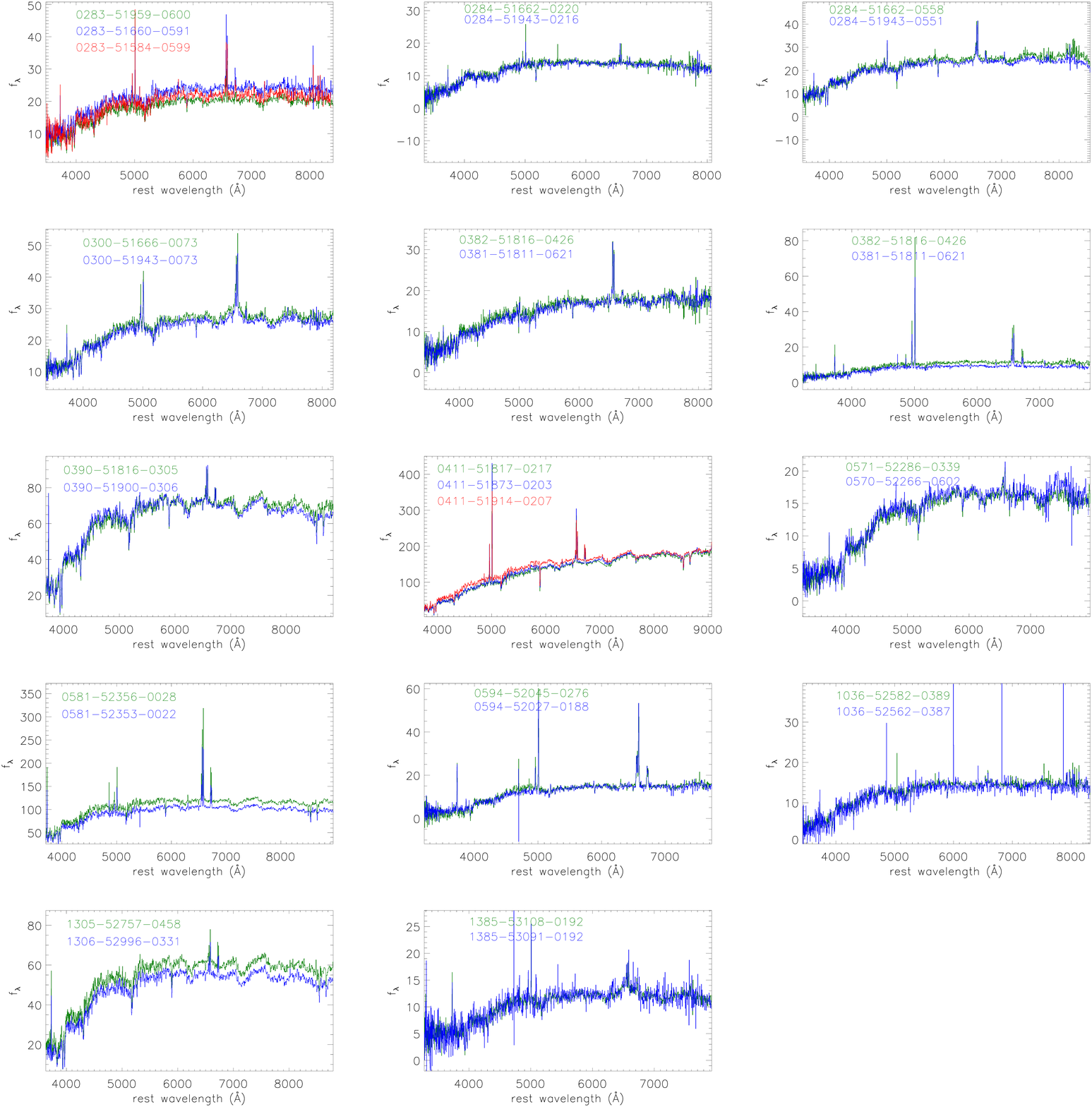}
\caption{Multi-epoch spectra in different colors of the 14 AGN. In each panel, plate-mjd-fiberid of each 
spectrum is marked in top left region by corresponding color.
}
\label{check}
\end{figure*}

\begin{figure}
\centering\includegraphics[width = 8cm,height=5cm]{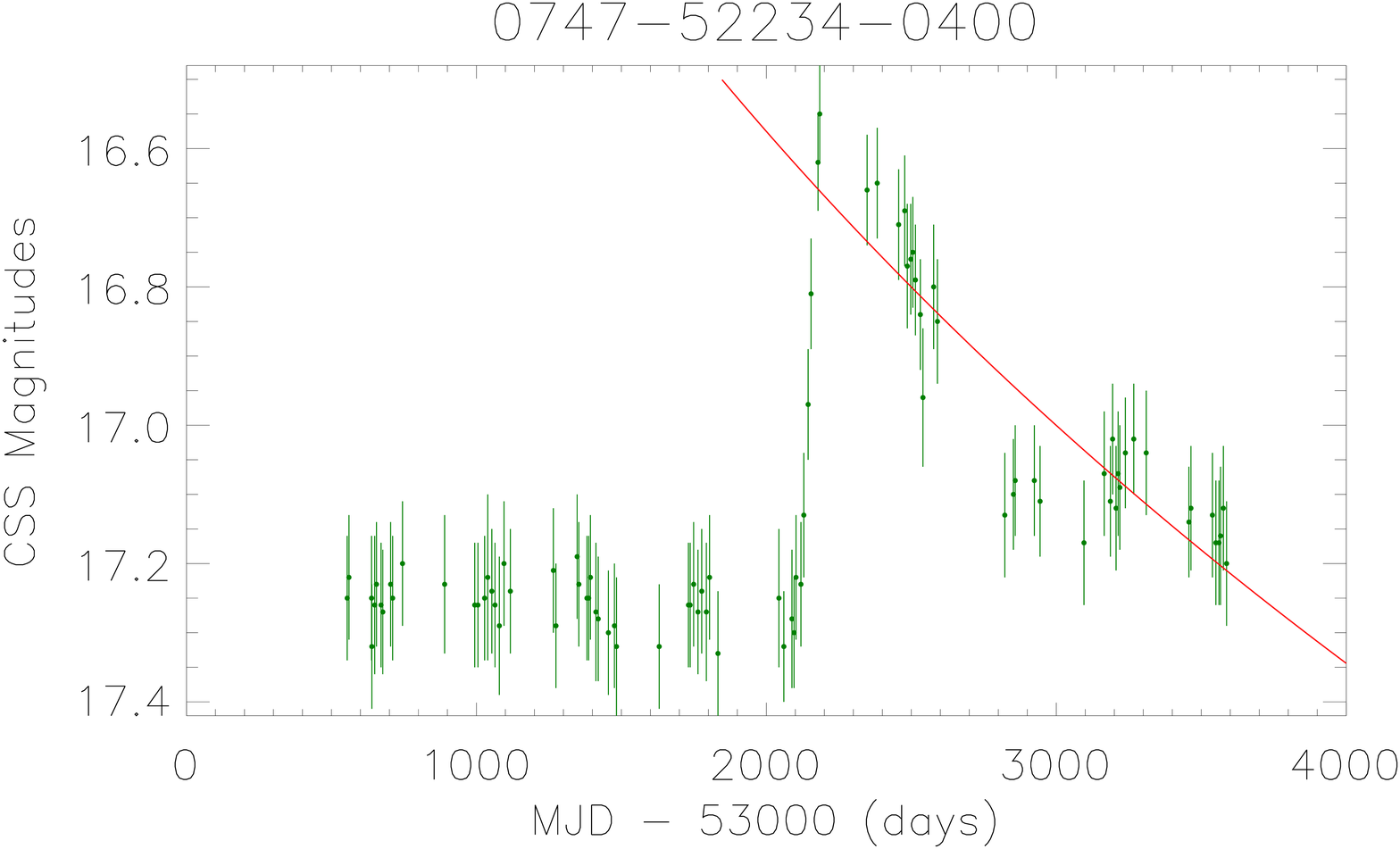}
\caption{TDE roughly expected variabilities described by $t^{-5/3}$ shown in solid red line in the long-term 
CSS light curve of SDSS 0747-52234-0400.
}
\label{tde}
\end{figure}

\begin{figure*}
\centering\includegraphics[width = 18cm,height=12cm]{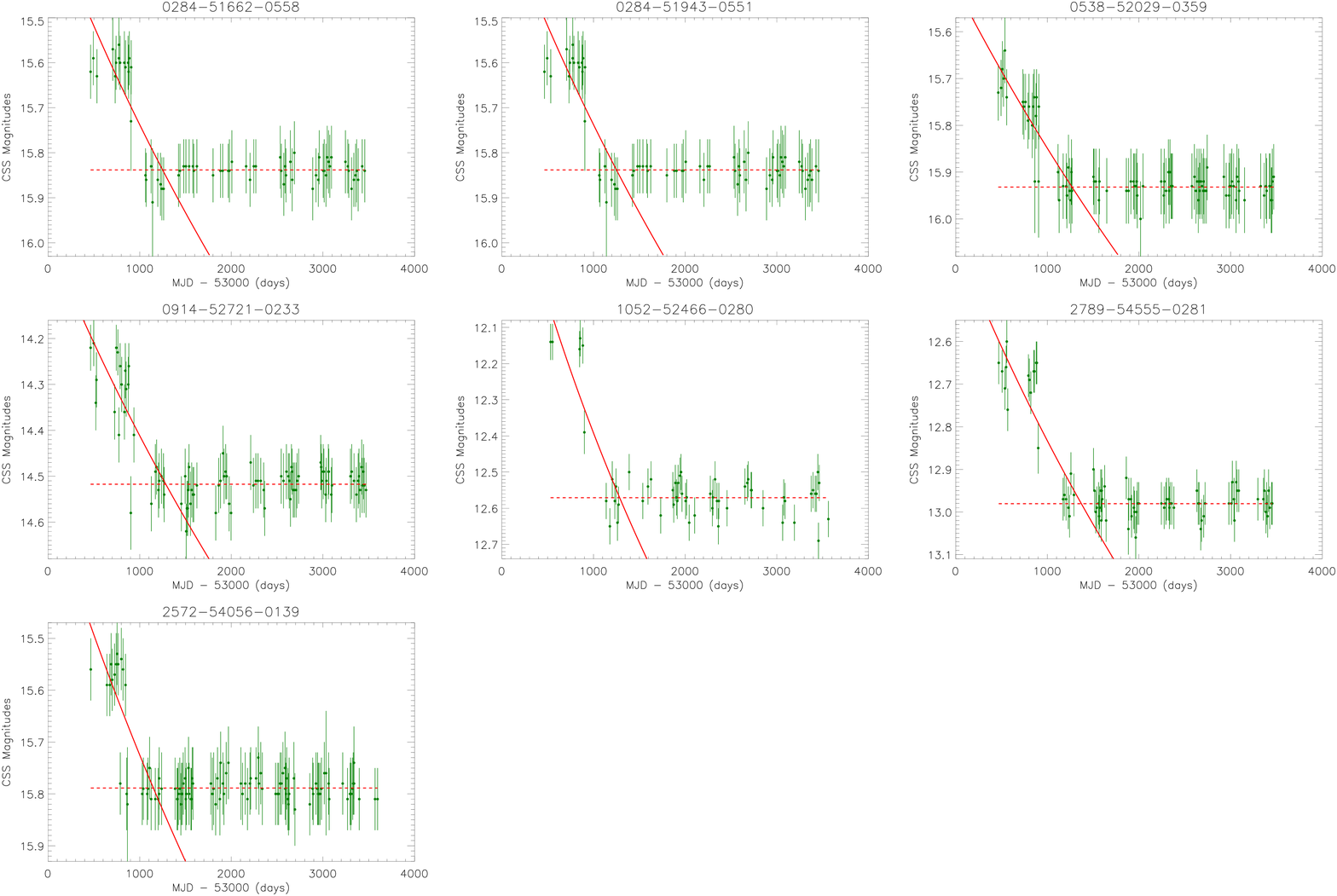}
\caption{TDE roughly expected variabilities described by $t^{-5/3}$ shown in solid red line followed by none 
variability component described by horizontal dashed red line in the long-term CSS light curve of the seven 
Type-2 AGN with apparent variabilities. 
}
\label{tde2}
\end{figure*}

	Moreover, when the SSP method is applied, there is only one criterion that the strengthen factor of 
each simple stellar population template is not smaller than zero. Then, through the Levenberg-Marquardt 
least-squares minimization technique (the MPFIT package), SDSS spectra with emission lines being masked out 
can be well described by the SSP method for the 155 Type-2 AGN with apparent variabilities, besides the SDSS 
1870-53383-0466 (plate-mjd-fiberid) with quite different redshifts estimated from emission lines and from 
absorption lines which will be individually discussed in the following section. Here, the SSP method determined 
host galaxy contributions are not shown in plots for all the 155 Type-2 AGN (SDSS 1870-53383-0466 not included), 
but left panels of Fig.~\ref{ssp} show an example on the SSP determined host galaxy contributions in the Type-2 
AGN SDSS 2374-53765-0174 due to its stronger narrow emission lines.

	After subtractions of starlight, emission lines around H$\alpha$ (rest wavelength range from 6400\AA~ 
to 6800\AA) are described by multiple Gaussian functions as follows, in order to check whether are there 
broad H$\alpha$ component in the optically selected 156 Type-2 AGN with apparent variabilities. In order to 
well describe the narrow emission lines, each two Gaussian functions, one component (first component) for 
common narrow component and the other component (second component which is assumed to be wider than the 
first component) for probable extended component, are applied to describe each narrow emission line, i.e, 
there are 10 Gaussian components applied to describe the [N~{\sc ii}]$\lambda6548,6583$\AA~ doublet, H$\alpha$ 
and [S~{\sc ii}]$\lambda6716,6731$\AA~ doublet. Properties of the second component in H$\alpha$ will provide 
clues to support whether are there broad components in H$\alpha$. Then, through the Levenberg-Marquardt 
least-squares minimization technique, the emission lines around H$\alpha$ can be well described. Meanwhile, 
when the Gaussian functions above are applied, the following three criteria are accepted. First, each Gaussian 
component has line intensity not smaller than zero. Second, the first (the second) components of the each 
doublet have the same redshift and the same line width. Third, the first (the second) components of the 
[N~{\sc ii}] doublet have the flux ratio to be fixed to the theoretical value 3. As an example, right panels 
of Fig.~\ref{ssp} show the best descriptions to the emission lines around H$\alpha$ in SDSS 2374-53765-0174, 
and the corresponding residuals calculated by line spectrum minus the best fitting results and then divided 
by uncertainties of the spectrum.

	Then, line parameters of [N~{\sc ii}] and H$\alpha$ of the 156 Type-2 AGN but with apparent 
variabilities can be well measured and listed in Table~2, after subtractions of host galaxy contributions. 
Then, based on the measured parameters, 31 out of the 156 Type-2 AGN have probable broad H$\alpha$, 
considering the following two criteria. On the one hand, the second moment of H$\alpha$ has its measured line 
parameters at least three times larger than corresponding uncertainties. On the other hand, the second 
component of H$\alpha$ has its measured second moment larger than 1000${\rm km/s}$, or at least three times 
larger than both the first moment of H$\alpha$ and the second moment of [N~{\sc ii}] doublet. There are 17 
Type-2 AGN with measured second moment of the second component of H$\alpha$ larger than 1000${\rm km/s}$, 
and 14 Type-2 AGN with the second component of H$\alpha$ at least three times larger than both the first 
moment of H$\alpha$ and the second moment of [N~{\sc ii}] doublet. Therefore, among the 156 Type-2 AGN, 
the 31 Type-2 AGN with probable broad component in H$\alpha$ should be preferred to be classified as 
Type-1.8/1.9 AGN, not as Type-2 AGN.

	Before proceeding further, two points are noted. On the one hand, as the shown results in 
Fig.~\ref{line}, due to complex broad emission lines in SDSS 2419-54139-0050, the best fitting results 
to the emission lines around H$\alpha$ are not so good. However, the fitting results can be clearly 
applied to support a broad emission component in H$\alpha$. Therefore, there are no further applications 
of more Gaussian components to describe the broad H$\alpha$ in SDSS 2419-54139-0050, because there are 
no further discussions on broad line properties in the manuscript. On the other hand, there are two 
Gaussian functions applied to describe each narrow emission line, in order to test whether are there 
broad emission line features. However, the 10 Gaussian functions should lead to larger uncertainties 
of each model parameters, leading the measured parameters (especially of the second component) 3 times 
smaller than the corresponding uncertainties. Therefore, in Table~2, there are no listed parameters in 
some collected Type-2 AGN, due to  measured line parameters smaller than corresponding uncertainties.

	Besides the 31 AGN of which spectra shown in Fig.~\ref{line}, the other Type-2 AGN have been 
carefully checked through the F-test technique, as what we have done in \citet{zh21d} by two different 
model functions. Besides the model function discussed above, the other model function has been 
considered by 10 Gaussian functions plus an additional broad Gaussian component. Then comparing with 
different estimated $\chi^2$ values by the two model functions, it can be confirmed that the broad 
Gaussian component is not preferred with confidence level higher than 5sigma. In other words, the other  
optically selected 125 Type-2 AGN (156-31) have no clues on broad emission lines but have apparent 
long-term optical variabilities.

\section{Discussions}

\subsection{SDSS J075736.47+532557.1}

	As described above, the AGN SDSS J075736.47+532557.1 is an unique object, due to its redshift 
determined to be zero through emission line features. However, as shown in Fig.~\ref{s1870}, there are 
apparent Ca~{\sc ii}$\lambda3934, 3969$\AA~ absorption features, leading the absorption line determined 
redshift to be about 0.13. The quite different redshifts from emission lines and absorption lines should 
indicate an interesting dual AGN system in SDSS J075736.47+532557.1, as more recent reviews in \citet{dv19}.

	Moreover, as the shown emission lines around H$\alpha$ in right panel of Fig.~\ref{s1870} in SDSS 
J075736.47+532557.1, there are no clues on probably broad emission component in H$\alpha$. Therefore, 
besides the quite different redshifts from absorption and emission lines, SDSS J075736.47+532557.1 is an 
optically selected Type-2 AGN but with apparent long-term optical variabilities. Further discussions on 
quite different redshifts from absorption and emission lines in SDSS J075736.47+532557.1 is beyond the scope 
of the manuscript, but will be give in our manuscript in preparation, and there are no further discussions 
on SDSS J075736.47+532557.1\ in the manuscript.

\subsection{Are the optically selected Type-2 AGN with apparent variabilities Changing-Look AGN?}

	As described in the Introduction, the optically selected Type-2 AGN but with apparent long-term 
variabilities could be candidates of changing-look AGN at quiet state. However, multi-epoch spectra are 
necessary to check whether an AGN was a changing-look AGN. Among the collected Type-2 AGN with apparent 
variabilities, there are 14 objects which have multi-epoch spectra, collected based on the SDSS pipeline 
provided parameter 'Nspecobs' which can be found in the second extension of Fits file of each SDSS spectrum. 
Then, the collected multi-epoch spectra of each Type-2 AGN are well checked and shown in Fig.~\ref{check}, 
and no signs can be found on appearance or disappearance of broad emission lines in different epochs. 
As the shown information of plate-mjd-fiberid in top-left corner in each panel of Fig.~\ref{check}, 
time differences among the epochs of the multiple spectra can be well determined as 375days, 281days, 
281days, 277days, 5days, 5days, 84days, 97days, 20days, 3days, 18days, 20days, 239days and 17days for the 
Type-2 AGN listed from left to right and from top to bottom. Therefore, at current stage, there are not 
enough evidence to support that the collected Type-2 AGN but with apparent long-term variabilities are 
changing-look AGN (probably due to smaller time differences among the multi-epochs?).

\subsection{Are the optically selected Type-2 AGN with apparent variabilities True Type-2 AGN?}

	As described in the Introduction, the optically selected Type-2 AGN but with apparent long-term 
optical variabilities could be candidates of true Type-2 AGN. The long-term optical variabilities can 
be well applied to confirm that the central regions have been directly observed, along with the loss 
of broad emission lines, the optically selected Type-2 AGN, especially the 125 objects with apparent 
variabilities but without broad emission lines, can be accepted as the good candidates of true Type-2 AGN.

	Actually, among the Type-2 AGN with apparent long-term optical variabilities, there is an unique 
TDE (Tidal disruption event) expected variabilities in the SDSS 0747-52234-0400. TDEs have been well 
studied in detail for more than four decades \citep{re88, lu97, gm06, ve11, ce12, gs12, gr13, gm14, 
ko15, ht16, st18, wy18, tc19, zh19, vg21, sg21, zh21n, zh22a}, with accreting fallback debris from stars 
tidally disrupted by central black holes (BHs) leading to apparent time-dependent variabilities. Here, 
systematic studying on TDE expected variabilities in SDSS 0747-52234-0400 with apparent long-term 
variabilities is beyond the scope of the manuscript. Therefore, there are no further discussions on 
SDSS 0747-52234-0400 with apparent long-term variabilities related to TDEs which will be discussed 
in detail in our manuscript in preparation, but Fig.~\ref{tde} shows the TDE expected variabilities 
roughly described by $t^{-5/3}$ in SDSS 0747-52234-0400. Besides SDSS 0747-52234-0400, there are 
several other Type-2 AGN with their long-term light curves having variability features including a smooth 
declining trend but with small decrease followed by none-variability component, such as in SDSS 
0284-51662-0558, SDSS 0284-51943-0551, SDSS 0538-52029-0359,  SDSS 0914-52721-0233, SDSS 1052-52466-0280, 
SDSS 2789-54555-0281, SDSS 2572-54056-0139, etc., of which light curves roughly described by $t^{-5/3}$ 
are shown in Fig.~\ref{tde2}. The probable TDEs provide strong and further evidence to support the 
objects as true Type-2 AGN, due to tidal disruption radii about tens of Schwarzschild radii quite 
close to central black hole. In our manuscript in preparation, detailed discussions on variabilities 
related to probable TDEs should be given for the small sample of optically selected Type-2 AGN but 
with apparent variabilities.

	Besides the long-term variabilities related to central AGN activities or to probable TDEs, 
another evidence should be reported to support the loss of broad emission lines. However, based on 
the collected SDSS spectra of the Type-2 AGN, there are only 31 AGN with probably broad emission 
lines, leading the 31 AGN to be classified as Type-1.8/1.9 AGN. For the other 125 Type-2 AGN with 
no broad emission lines in current collected SDSS spectra, it is confident to be accept as true 
Type-2 AGN. Certainly, high quality spectra should be necessary in the near future to confirm them 
as true Type-2 AGN without no broad emission lines.

\subsection{Comparing with previous work}

	As described in Introduction, \citet{ba14} have reported g-band variabilities in 17 out of 
173 Type-2 quasars covered in the Stripe 82 region, indicating about 5.8\% ((17-7)/173, with 7 objects 
having detected broad emission lines) of optical selected Type-2 AGN have apparent variabilities. 
However, in the manuscript, among the 12881 SDSS pipeline classified Type-2 AGN with CSS light curves, 
there are 156 objects with apparent long-term variabilities, indicating about 0.97\% ((156-31)/12881, 
with 31 objects having detected broad emission lines) optical selected low redshift Type-2 AGN have 
apparent variabilities, about six times smaller than the reported results in \citet{ba14}. Therefore, 
in the subsection, it is interesting to explain the different number ratios in \citet{ba14} and in 
the manuscript, for Type-2 AGN with and without long-term variabilities. The following two reasons 
are mainly considered.

	On the one hand, as well discussed in \citet{ba14}, the selected 7 of the 17 Type-2 quasars 
have detected broad emission lines in their re-observed Keck high quality spectra. Probably the other 
10 Type-2 quasars could have detected broad emission lines in high quality spectra, leading a number 
ratio quite smaller than 5.8\%.

	On the other hand, if Type-2 AGN with apparent long-term variabilities were evenly distributed 
in space, the quite different number ratios should give interesting clues on properties of central dust 
torus in AGN. As one of fundamental structures in Unified model of AGN, properties of dust torus have 
been well studied for more than four decades, especially on properties of opening angles (covering factor) 
of dust torus. \citet{at05} have proposed the receding torus model, based on the statistically significant 
correlation between the half opening angle of the torus and [O~{\sc iii}] emission-line luminosity. 
\citet{zo18} have reported that the half-opening angle of the torus declines with increasing accretion 
rate until the Eddington ratio reaches 0.5, above which the trend reverses. However, \citet{nl16, sr16} 
have found no evidence for a luminosity dependence of the torus covering factor in AGN. More recent 
review on dust torus can be found in \citet{ar17}.

	There are no definite conclusions on the receding torus model, however, the receding torus model 
can be well applied to explain the quite different number ratios on the Type-2 AGN with apparent 
variabilities. For our sample in lower redshift and in lower line luminosity than the Type-2 quasars 
in \citet{ba14}, smaller opening angles expected by the receding torus model are expected for the low 
redshift objects in our sample, leading to fewer Type-2 AGN with apparent variabilities which can be 
detected in our sample.

\section{Summaries and Conclusions}

   The main summaries and conclusions are as follows. 
\begin{itemize}   
\item Through the SDSS provided SQL search tool in DR16, 14354 SDSS pipeline classified low redshift 
	($z<0.3$) Type-2 AGN can be collected from SDSS main galaxies with spectral signal-to-noise 
	larger than 10, and located into the AGN region in the BPT diagram of O3HB versus N2HA.
\item Among the collected 14354 Type-2 AGN, long-term CSS V-band light curves of 12881 Type-2 AGN can 
	be collected from the CSS. 
\item The well-known DRW process is applied to describe the CSS light curves of the 12881 Type-2 AGN, 
	leading to the well determined process parameters of $\tau$ and $\sigma$. And based on the 
	measured $\ln(\sigma/(mag/days^{0.5}))$ larger than -4 and 3 times larger than corresponding 
	uncertainties, 156 Type-2 AGN are collected with apparent long-term variabilities.
\item For the 156 Type-2 AGN with apparent variabilities, the well-applied SSP method is accepted to 
	determine host galaxy contributions in SDSS spectra. After subtractions of starlight, emission 
	lines around H$\alpha$ can be well measured by multiple Gaussian functions, leading to 31 objects 
	with detected broad H$\alpha$. There are no signs (confidence level higher than 5sigma by the 
	F-test technique) for broad emission lines in the other 125 Type-2 AGN with long-term variabilities.
\item For the 156 optically selected Type-2 AGN with apparent variabilities, there are 14 objects having 
	multi-epoch SDSS spectra. After checking the multi-epoch SDSS spectra, there are no clues to 
	appearance or disappearance of broad lines. The results indicate that the collected Type-2 AGN 
	classified as the changing-look AGN  should be not preferred at current stage.
\item For the 125 (156-31) optically selected Type-2 AGN without broad lines but with apparent variabilities, 
	they could be well accepted as candidates of True Type-2 AGN (AGN without hidden central broad 
	emission line regions). 
\item For the 156 optically selected Type-2 AGN with apparent variabilities, there are a small sample 
	of objects with their long-term variabilities having features roughly described by TDE expected 
	$t^{-5/3}$, indicating probable central TDEs which can be applied as further evidence to support 
	true Type-2 AGN.
\item The smaller number ratio of Type-2 AGN with variabilities to normal Type-2 AGN than in high 
	redshift and high luminosity Type-2 quasars reported in \citet{ba14} could be applied to probably 
	support the receding torus model in AGN, if the reported Type-2 quasars with variabilities in 
	\citet{ba14} were true Type-2 quasar without central hidden BLRs.  
\end{itemize}

\begin{table*}
\caption{Basic information of the 156 Type-2 AGN with apparent long-term variabilities}
\begin{center}
\begin{tabular}{cccccccccc}
\hline\hline
pmf & $z$ & name & m1 & m2 & $\log(L_{O3})$ & $\log(O3HB)$ & $\log(N2HA)$ & $\ln(\sigma)$ & $\ln(\tau)$ \\
\hline
0271-51883-0252 & 0.049 & J101439.5-004951.2 & 13.85 & 14.71 & 40.55$\pm$0.01 & 0.767$\pm$0.034 & 
	0.2379$\pm$0.018 & -3.15$\pm$0.31 & 6.36$\pm$0.77 \\
0283-51959-0600 & 0.096 & J114815.9-000328.4 & 16.18 & 17.03 & 40.55$\pm$0.01 & 0.675$\pm$0.058 & 
	-0.072$\pm$0.029 & -2.68$\pm$0.26 & 5.53$\pm$0.71 \\
0284-51662-0220 & 0.138 & J115135.0-005655.9 & 16.51 & 18.89 & 40.51$\pm$0.02 & 0.369$\pm$0.084 & 
	-0.237$\pm$0.055 & -2.84$\pm$0.28 & 4.54$\pm$1.09 \\
0284-51662-0558 & 0.078 & J115601.1+001237.9 & 15.78 & 23.86 & 40.03$\pm$0.02 & 0.398$\pm$0.081 & 
	-0.028$\pm$0.024 & -2.39$\pm$0.28 & 6.76$\pm$0.62 \\
0284-51943-0551 & 0.078 & J115601.1+001237.9 & 15.78 & 23.86 & 40.06$\pm$0.01 & 0.445$\pm$0.061 & 
	-0.023$\pm$0.019 & -2.38$\pm$0.27 & 6.82$\pm$0.63 \\
0288-52000-0152 & 0.154 & J121958.1-003530.6 & 17.39 & 24.31 & 40.57$\pm$0.02 & 0.460$\pm$0.094 & 
	-0.223$\pm$0.045 & -0.97$\pm$0.12 & 3.79$\pm$0.32 \\
0300-51666-0073 & 0.124 & J135244.6-001526.1 & 16.56 & 17.67 & 40.83$\pm$0.01 & 0.534$\pm$0.048 & 
	0.0593$\pm$0.022 & -3.30$\pm$0.39 & 6.49$\pm$0.97 \\
0305-51613-0044 & 0.086 & J142934.0-000237.8 & 16.03 & 16.86 & 40.41$\pm$0.01 & 0.480$\pm$0.049 & 
	-0.019$\pm$0.020 & -2.43$\pm$0.24 & 4.42$\pm$0.58 \\
0305-51613-0146 & 0.150 & J142557.7-002218.9 & 17.50 & 18.76 & 40.18$\pm$0.05 & 0.182$\pm$0.138 & 
	-0.201$\pm$0.040 & -1.82$\pm$0.24 & 5.85$\pm$0.61 \\
0311-51665-0073 & 0.139 & J151235.5-000007.3 & 17.11 & 18.01 & 41.28$\pm$0.01 & 0.740$\pm$0.030 & 
	0.0425$\pm$0.014 & -2.23$\pm$0.26 & 4.28$\pm$0.69 \\
0315-51663-0549 & 0.072 & J154121.3+003336.1 & 16.99 & 18.25 & 39.13$\pm$0.12 & 0.455$\pm$0.851 & 
	-0.018$\pm$0.119 & -2.66$\pm$0.29 & 6.34$\pm$0.74 \\
0375-52140-0063 & 0.130 & J222707.1-001836.4 & 16.38 & 19.08 & 41.62$\pm$0.01 & 0.956$\pm$0.024 & 
	0.0233$\pm$0.012 & -1.10$\pm$0.12 & 3.23$\pm$0.26 \\
0379-51789-0271 & 0.117 & J225053.0-000041.2 & 17.25 & 18.49 & 40.24$\pm$0.04 & 0.203$\pm$0.121 & 
	-0.137$\pm$0.035 & -1.96$\pm$0.17 & 3.54$\pm$0.39 \\
0379-51789-0401 & 0.071 & J225222.3+010700.0 & 15.30 & 17.15 & 39.94$\pm$0.04 & 0.116$\pm$0.110 & 
	-0.174$\pm$0.041 & -3.22$\pm$0.29 & 5.49$\pm$0.88 \\
0382-51816-0426 & 0.178 & J231501.7+001613.9 & 17.23 & 18.04 & 41.65$\pm$0.01 & 0.988$\pm$0.035 & 
	0.0104$\pm$0.018 & -2.45$\pm$0.23 & 3.30$\pm$0.42 \\
0390-51816-0305 & 0.039 & J001558.2-001812.6 & 14.00 & 14.91 & 39.48$\pm$0.03 & 0.198$\pm$0.084 & 
	-0.111$\pm$0.038 & -3.32$\pm$0.28 & 5.55$\pm$1.20 \\
0390-51900-0306 & 0.039 & J001558.2-001812.6 & 14.00 & 14.91 & 39.47$\pm$0.02 & 0.152$\pm$0.048 & 
	-0.061$\pm$0.020 & -3.28$\pm$0.31 & 5.77$\pm$1.26 \\
0411-51817-0217 & 0.013 & J030349.1-010613.4 & 12.53 & 13.69 & 39.98$\pm$0.01 & 0.967$\pm$0.020 & 
	-0.302$\pm$0.016 & -3.10$\pm$0.30 & 6.30$\pm$0.86 \\
0411-51873-0203 & 0.013 & J030349.1-010613.4 & 12.53 & 13.69 & 39.98$\pm$0.01 & 0.955$\pm$0.020 & 
	-0.302$\pm$0.016 & -3.13$\pm$0.29 & 6.25$\pm$0.86 \\
0424-51893-0634 & 0.045 & J012839.8+144553.8 & 14.76 & 15.67 & 40.70$\pm$0.01 & 0.827$\pm$0.016 & 
	0.0378$\pm$0.009 & -2.48$\pm$0.28 & 6.49$\pm$0.72 \\
0447-51877-0287 & 0.096 & J084012.3+512127.0 & 16.07 & 16.93 & 39.42$\pm$0.11 & 0.009$\pm$0.262 & 
	-0.008$\pm$0.061 & -2.37$\pm$0.22 & 4.87$\pm$0.79 \\
0451-51908-0092 & 0.049 & J092043.1+553804.1 & 16.15 & 16.85 & 39.29$\pm$0.04 & 0.499$\pm$0.234 & 
	0.0628$\pm$0.077 & -1.98$\pm$0.21 & 4.70$\pm$0.84 \\
0465-51910-0378 & 0.129 & J040403.3-052345.4 & 16.89 & 18.06 & 39.92$\pm$0.07 & 0.118$\pm$0.212 & 
	0.0711$\pm$0.058 & -2.29$\pm$0.18 & 3.84$\pm$0.57 \\
0500-51994-0390 & 0.098 & J095456.3+022403.3 & 17.35 & 18.33 & 39.86$\pm$0.05 & 0.024$\pm$0.127 & 
	0.0045$\pm$0.051 & -2.77$\pm$0.27 & 5.31$\pm$0.63 \\
0501-52235-0366 & 0.023 & J100207.0+030327.6 & 13.94 & 14.79 & 40.34$\pm$0.01 & 0.456$\pm$0.010 & 
	-0.182$\pm$0.008 & -3.26$\pm$0.31 & 6.71$\pm$0.79 \\
0504-52316-0492 & 0.071 & J102813.5+024657.5 & 16.41 & 17.27 & 39.92$\pm$0.02 & 0.288$\pm$0.075 & 
	-0.200$\pm$0.030 & -2.62$\pm$0.28 & 5.64$\pm$0.88 \\
0533-51994-0187 & 0.127 & J141445.2+015000.3 & 16.74 & 17.87 & 41.27$\pm$0.01 & 0.513$\pm$0.020 & 
	-0.308$\pm$0.012 & -2.26$\pm$0.23 & 5.27$\pm$0.68 \\
0538-52029-0359 & 0.069 & J145019.0+015205.2 & 15.89 & 16.79 & 41.20$\pm$0.01 & 0.780$\pm$0.014 & 
	-0.220$\pm$0.011 & -2.57$\pm$0.29 & 7.05$\pm$0.61 \\
0540-51996-0549 & 0.219 & J151215.7+020316.9 & 17.24 & 18.48 & 41.41$\pm$0.01 & 0.564$\pm$0.075 & 
	0.0898$\pm$0.038 & -1.17$\pm$0.22 & 5.18$\pm$0.58 \\
0543-52017-0342 & 0.117 & J075057.0+380035.3 & 15.96 & 17.10 & 40.23$\pm$0.03 & 0.083$\pm$0.088 & 
	-0.056$\pm$0.034 & -2.65$\pm$0.13 & 3.26$\pm$0.41 \\
0564-52224-0019 & 0.068 & J085009.2+021606.2 & 17.70 & 18.25 & 39.14$\pm$0.06 & 0.086$\pm$0.169 & 
	0.0054$\pm$0.057 & -1.92$\pm$0.20 & 4.37$\pm$0.86 \\
0564-52224-0041 & 0.060 & J084654.8+021457.7 & 16.91 & 17.51 & 40.04$\pm$0.01 & 0.182$\pm$0.020 & 
	-0.070$\pm$0.007 & -2.23$\pm$0.17 & 3.95$\pm$0.56 \\
0571-52286-0339 & 0.159 & J094739.4+044128.4 & 17.13 & 18.63 & 40.14$\pm$0.07 & 0.172$\pm$0.201 & 
	0.3545$\pm$0.194 & -2.46$\pm$0.26 & 5.09$\pm$0.88 \\
0576-52325-0639 & 0.155 & J103448.5+042419.7 & 15.63 & 16.48 & 40.13$\pm$0.10 & -0.11$\pm$0.201 & 
	0.3116$\pm$0.066 & -2.51$\pm$0.27 & 5.85$\pm$0.71 \\
0580-52368-0084 & 0.091 & J110027.1+034047.4 & 16.89 & 17.61 & 40.66$\pm$0.01 & 0.740$\pm$0.036 & 
	-0.079$\pm$0.018 & -2.12$\pm$0.13 & 4.09$\pm$0.61 \\
0581-52356-0028 & 0.028 & J111044.8+043039.0 & 14.01 & 14.86 & 39.97$\pm$0.01 & 0.181$\pm$0.015 & 
	-0.015$\pm$0.008 & -2.93$\pm$0.28 & 6.61$\pm$0.78 \\
0589-52055-0263 & 0.043 & J150011.9+033819.4 & 15.81 & 16.67 & 39.13$\pm$0.06 & -0.11$\pm$0.107 & 
	0.0419$\pm$0.024 & -2.90$\pm$0.32 & 6.38$\pm$0.87 \\
0594-52045-0276 & 0.187 & J154317.4+023552.2 & 17.26 & 18.51 & 41.62$\pm$0.01 & 0.951$\pm$0.061 & 
	0.3197$\pm$0.019 & -1.90$\pm$0.18 & 3.96$\pm$0.69 \\
0630-52050-0241 & 0.121 & J164716.6+384327.6 & 16.57 & 17.45 & 41.57$\pm$0.01 & 0.948$\pm$0.013 & 
	-0.103$\pm$0.008 & -2.37$\pm$0.26 & 5.54$\pm$0.77 \\
0656-52148-0390 & 0.055 &  J004150.4-091811.3 & 13.15 & 19.11 & 39.21$\pm$0.05 & -0.07$\pm$0.120 & 
	0.2844$\pm$0.033 & -3.42$\pm$0.34 & 5.70$\pm$1.05 \\
0696-52209-0234 & 0.017 & J012116.5-003240.1 & 12.64 & 13.68 & 38.95$\pm$0.02 & 0.121$\pm$0.053 & 
	0.2841$\pm$0.018 & -3.17$\pm$0.28 & 5.66$\pm$0.94 \\
\hline
\end{tabular}
\end{center}
\tablecomments{
The first column shows the pmf information of plate-mjd-fiberid of each Type-2 AGN, the second column 
shows the redshift information of each Type-2 AGN, the third column shows the SDSS coordinate 
based name (Jhhmmss.s$\pm$ddmmss.s) of each Type-2 AGN, the fourth column shows the mean photometric 
magnitude of the CSS light curve of each Type-2 AGN, the fifth column shows the SDSS provided photometric 
g-band Petrosian magnitude of each Type-2 AGN, the sixth column shows the [O~{\sc iii}] line luminosity 
$\log(L_{O3}/(erg/s))$ of each Type-2 AGN, the seventh column and the eighth column show the $\log(O3HB)$ 
and $\log(N2HA)$ of each Type-2 AGN, the last two columns show the determined $\ln(\sigma/(mag/days^{0.5}))$ 
and $\ln(\tau/days)$ of each Type-2 AGN. 
}
\end{table*}

\setcounter{table}{0}

\begin{table*}
\caption{--to be continued}
\begin{center}
\begin{tabular}{cccccccccc}
\hline\hline
pmf & $z$ & name & m1 & m2 & $\log(L_{O3})$ & $\log(O3HB)$ & $\log(N2HA)$ & $\ln(\sigma)$ & $\ln(\tau)$ \\
\hline
0699-52202-0163 & 0.083 & J014638.4-002527.3 & 15.96 & 16.99 & 40.92$\pm$0.01 & 0.688$\pm$0.026 & 
	-0.135$\pm$0.016 & -2.71$\pm$0.19 & 3.58$\pm$0.72 \\
0720-52206-0191 & 0.085 & J222219.6-085607.1 & 14.84 & 16.01 & 39.67$\pm$0.07 & 0.154$\pm$0.241 & 
	0.1478$\pm$0.069 & -2.46$\pm$0.26 & 6.10$\pm$0.76 \\
0743-52262-0591 & 0.016 & J231221.4+143633.7 & 12.67 & 14.11 & 39.66$\pm$0.01 & 0.198$\pm$0.025 & 
	0.0590$\pm$0.015 & -2.59$\pm$0.23 & 5.46$\pm$0.80 \\
0747-52234-0400 & 0.107 & J233454.0+145712.8 & 17.13 & 18.18 & 40.90$\pm$0.01 & 0.423$\pm$0.025 & 
	0.0085$\pm$0.009 & -1.64$\pm$0.22 & 5.98$\pm$0.50 \\
0758-52253-0514 & 0.129 & J080728.1+391135.3 & 16.87 & 18.12 & 40.51$\pm$0.02 & 0.737$\pm$0.136 & 
	0.0955$\pm$0.051 & -2.44$\pm$0.22 & 4.92$\pm$0.61 \\
0817-52381-0412 & 0.176 & J163124.2+401735.7 & 17.05 & 18.23 & 40.46$\pm$0.07 & 0.313$\pm$0.196 & 
	0.3812$\pm$0.119 & -1.97$\pm$0.25 & 5.69$\pm$0.61 \\
0892-52378-0058 & 0.082 & J081013.0+345136.8 & 16.69 & 17.47 & 40.93$\pm$0.01 & 0.568$\pm$0.018 & 
	-0.149$\pm$0.009 & -2.94$\pm$0.31 & 6.80$\pm$0.68 \\
0914-52721-0233 & 0.075 & J135251.6-023630.5 & 14.48 & 15.52 & 39.98$\pm$0.03 & 0.214$\pm$0.091 & 
	0.1092$\pm$0.039 & -2.45$\pm$0.25 & 6.99$\pm$0.60 \\
0917-52400-0260 & 0.141 & J141726.0-024948.4 & 17.44 & 18.45 & 41.34$\pm$0.01 & 0.808$\pm$0.023 & 
	-0.144$\pm$0.011 & -2.71$\pm$0.36 & 6.20$\pm$1.15 \\
0929-52581-0196 & 0.098 & J075846.9+270515.5 & 16.34 & 18.09 & 39.96$\pm$0.03 & -0.14$\pm$0.071 & 
	-0.170$\pm$0.041 & -2.01$\pm$0.22 & 5.49$\pm$0.64 \\
0931-52619-0094 & 0.097 & J081917.5+301935.7 & 16.62 & 17.54 & 40.63$\pm$0.01 & 0.789$\pm$0.066 & 
	-0.000$\pm$0.024 & -2.83$\pm$0.27 & 7.12$\pm$0.62 \\
0938-52708-0404 & 0.073 & J091829.0+405803.5 & 16.83 & 17.09 & 41.03$\pm$0.01 & 0.986$\pm$0.025 & 
	0.0416$\pm$0.013 & -2.15$\pm$0.16 & 4.02$\pm$0.50 \\
0979-52427-0095 & 0.178 & J172000.4+261855.0 & 16.79 & 18.02 & 41.44$\pm$0.01 & 0.411$\pm$0.022 & 
	-0.068$\pm$0.010 & -2.34$\pm$0.20 & 5.01$\pm$0.62 \\
0983-52443-0594 & 0.135 & J205720.3+001207.3 & 16.88 & 17.98 & 40.79$\pm$0.02 & 0.356$\pm$0.057 & 
	-0.050$\pm$0.024 & -1.97$\pm$0.20 & 4.28$\pm$0.65 \\
0984-52442-0348 & 0.157 & J205510.0+003533.3 & 16.63 & 17.67 & 40.88$\pm$0.01 & 0.451$\pm$0.064 & 
	-0.016$\pm$0.026 & -2.39$\pm$0.28 & 6.22$\pm$0.79 \\
1001-52670-0389 & 0.026 & J104729.5+071503.8 & 13.77 & 14.25 & 39.01$\pm$0.04 & -0.11$\pm$0.086 & 
	0.0537$\pm$0.026 & -3.41$\pm$0.32 & 5.72$\pm$0.95 \\
1036-52582-0135 & 0.058 & J223044.8-010030.5 & 17.93 & 18.94 & 39.79$\pm$0.01 & 0.595$\pm$0.039 & 
	-0.395$\pm$0.028 & -2.39$\pm$0.21 & 3.68$\pm$0.59 \\
1036-52582-0389 & 0.107 & J222651.8+003023.0 & 17.54 & 18.98 & 39.44$\pm$0.08 & 0.154$\pm$0.230 & 
	-0.032$\pm$0.062 & -2.75$\pm$0.32 & 5.70$\pm$1.44 \\
1052-52466-0280 & 0.018 & J152945.0+425507.1 & 12.53 & 13.54 & 39.18$\pm$0.02 & 0.275$\pm$0.058 & 
	0.2313$\pm$0.023 & -1.86$\pm$0.25 & 6.98$\pm$0.61 \\
1065-52586-0237 & 0.065 & J031634.4-002419.1 & 16.42 & 17.72 & 39.83$\pm$0.02 & 0.331$\pm$0.080 & 
	-0.131$\pm$0.026 & -1.92$\pm$0.26 & 5.65$\pm$0.64 \\
1198-52669-0150 & 0.095 & J085716.7+391545.8 & 17.46 & 17.52 & 40.19$\pm$0.02 & 0.484$\pm$0.070 & 
	-0.025$\pm$0.025 & -1.70$\pm$0.13 & 3.48$\pm$0.36 \\
1203-52669-0502 & 0.128 & J074713.0+221721.6 & 16.79 & 17.47 & 40.09$\pm$0.05 & 0.119$\pm$0.141 & 
	0.0067$\pm$0.031 & -1.58$\pm$0.19 & 4.49$\pm$0.45 \\
1204-52669-0281 & 0.181 & J075043.8+225353.3 & 17.78 & 18.89 & 41.32$\pm$0.01 & 1.012$\pm$0.076 & 
	0.1156$\pm$0.036 & -2.16$\pm$0.17 & 3.88$\pm$0.66 \\
1204-52669-0524 & 0.016 & J075820.4+250857.1 & 12.35 & 13.49 & 38.83$\pm$0.03 & 0.219$\pm$0.083 & 
	0.1854$\pm$0.031 & -1.76$\pm$0.27 & 6.28$\pm$0.65 \\
1208-52672-0519 & 0.093 & J083318.0+310238.7 & 15.68 & 16.64 & 40.80$\pm$0.01 & 0.592$\pm$0.038 & 
	0.0255$\pm$0.016 & -3.29$\pm$0.33 & 4.40$\pm$0.74 \\
1237-52762-0623 & 0.103 & J101805.5+084225.7 & 17.40 & 17.74 & 40.89$\pm$0.01 & 0.434$\pm$0.027 & 
	-0.148$\pm$0.013 & -2.10$\pm$0.18 & 4.60$\pm$0.60 \\
1239-52760-0047 & 0.113 & J103213.9+080048.5 & 17.81 & 18.36 & 40.52$\pm$0.02 & 0.620$\pm$0.147 & 
	-0.185$\pm$0.045 & -1.99$\pm$0.17 & 4.86$\pm$0.44 \\
1264-52707-0102 & 0.091 & J075502.8+214953.5 & 16.46 & 18.36 & 41.11$\pm$0.01 & 0.829$\pm$0.025 & 
	-0.402$\pm$0.021 & -1.99$\pm$0.16 & 3.52$\pm$0.38 \\
1265-52705-0448 & 0.142 & J080738.6+244116.8 & 16.88 & 18.36 & 40.06$\pm$0.06 & 0.233$\pm$0.176 & 
	0.1669$\pm$0.059 & -1.81$\pm$0.10 & 2.82$\pm$0.12 \\
1275-52996-0568 & 0.150 & J093740.2+374048.2 & 17.04 & 18.05 & 40.45$\pm$0.03 & 0.298$\pm$0.108 & 
	-0.146$\pm$0.030 & -2.18$\pm$0.17 & 3.81$\pm$0.73 \\
1294-52753-0109 & 0.134 & J153758.5+373450.8 & 17.56 & 18.28 & 40.74$\pm$0.01 & 0.510$\pm$0.047 & 
	-0.158$\pm$0.025 & -2.56$\pm$0.27 & 4.70$\pm$1.20 \\
1305-52757-0458 & 0.047 & J094732.8+104508.5 & 15.76 & 17.04 & 39.47$\pm$0.03 & 0.057$\pm$0.088 & 
	-0.097$\pm$0.048 & -2.70$\pm$0.19 & 4.82$\pm$0.62 \\
1307-52999-0065 & 0.149 & J100422.9+095143.6 & 17.28 & 18.56 & 41.22$\pm$0.01 & 0.844$\pm$0.048 & 
	-0.157$\pm$0.025 & -2.66$\pm$0.22 & 5.07$\pm$0.74 \\
1339-52767-0304 & 0.096 & J163330.6+340921.1 & 17.05 & 17.86 & 39.73$\pm$0.06 & 0.332$\pm$0.285 & 
	0.0291$\pm$0.064 & -2.53$\pm$0.22 & 5.31$\pm$0.57 \\
1341-52786-0159 & 0.136 & J164359.9+321009.8 & 17.27 & 18.46 & 40.97$\pm$0.01 & 0.792$\pm$0.079 & 
	0.1142$\pm$0.034 & -1.69$\pm$0.13 & 4.13$\pm$0.38 \\
1353-53083-0483 & 0.065 & J151437.4+362938.8 & 16.67 & 17.47 & 39.81$\pm$0.02 & 0.176$\pm$0.066 & 
	-0.272$\pm$0.029 & -2.36$\pm$0.16 & 3.32$\pm$0.44 \\
1363-53053-0109 & 0.114 & J110539.4+420834.0 & 17.70 & 18.54 & 40.51$\pm$0.01 & 0.250$\pm$0.039 & 
	-0.149$\pm$0.018 & -1.98$\pm$0.15 & 3.54$\pm$0.40 \\
1365-53062-0057 & 0.121 & J112532.9+435354.1 & 17.06 & 18.16 & 41.01$\pm$0.01 & 0.723$\pm$0.042 & 
	-0.057$\pm$0.022 & -1.97$\pm$0.14 & 3.50$\pm$0.34 \\
1368-53084-0298 & 0.299 & J114538.5+442021.9 & 17.89 & 18.66 & 41.51$\pm$0.01 & 0.365$\pm$0.076 & 
	0.0173$\pm$0.087 & -1.83$\pm$0.16 & 3.93$\pm$0.60 \\
1370-53090-0201 & 0.066 & J121101.5+443205.2 & 17.57 & 18.07 & 40.44$\pm$0.01 & 0.852$\pm$0.034 & 
	-0.097$\pm$0.019 & -2.13$\pm$0.22 & 4.35$\pm$0.62 \\
1382-53115-0193 & 0.014 & J143722.1+363404.2 & 12.52 & 13.28 & 40.44$\pm$0.01 & 0.992$\pm$0.014 & 
	0.1352$\pm$0.010 & -3.10$\pm$0.28 & 5.74$\pm$0.69 \\
1385-53108-0192 & 0.158 & J150745.0+341122.0 & 17.44 & 18.71 & 40.73$\pm$0.02 & 0.657$\pm$0.101 & 
	0.0805$\pm$0.110 & -2.50$\pm$0.23 & 4.57$\pm$0.68 \\
1567-53172-0431 & 0.053 & J165642.0+184605.3 & 16.16 & 17.21 & 39.22$\pm$0.06 & 0.528$\pm$0.364 & 
	0.2051$\pm$0.085 & -2.77$\pm$0.20 & 3.91$\pm$0.62 \\
1591-52976-0432 & 0.048 & J090901.0+321658.9 & 14.92 & 15.56 & 39.58$\pm$0.02 & 0.329$\pm$0.062 & 
	-0.175$\pm$0.034 & -1.28$\pm$0.09 & 3.63$\pm$0.29 \\
\hline
\end{tabular}
\end{center}
\end{table*}

\setcounter{table}{0}

\begin{table*}
\caption{--to be continued}
\begin{center}
\begin{tabular}{cccccccccc}
\hline\hline
	pmf & $z$ & name & m1 & m2 & $\log(L_{O3})$ & $\log(O3HB)$ & $\log(N2HA)$ & $\ln(\sigma)$ & $\ln(\tau)$ \\
\hline
1592-52990-0418 & 0.020 & J091941.4+334417.3 & 13.44 & 14.79 & 39.58$\pm$0.01 & 0.677$\pm$0.021 & 
	-0.133$\pm$0.013 & -2.26$\pm$0.29 & 5.79$\pm$0.77 \\
1594-52992-0566 & 0.022 & J094319.1+361452.1 & 12.92 & 13.74 & 39.82$\pm$0.01 & 0.236$\pm$0.020 & 
	-0.072$\pm$0.011 & -3.66$\pm$0.36 & 6.20$\pm$0.92 \\
1691-53260-0603 & 0.035 & J165630.5+275839.0 & 13.37 & 14.74 & 39.65$\pm$0.02 & 0.218$\pm$0.076 & 
	0.0918$\pm$0.040 & -3.45$\pm$0.30 & 6.58$\pm$0.75 \\
1692-53473-0151 & 0.052 & J165721.8+253350.0 & 16.73 & 17.84 & 40.38$\pm$0.01 & 0.707$\pm$0.023 & 
	-0.116$\pm$0.009 & -2.72$\pm$0.23 & 3.78$\pm$0.69 \\
1722-53852-0289 & 0.132 & J152614.9+095307.6 & 15.88 & 17.57 & 40.48$\pm$0.02 & 0.429$\pm$0.097 & 
	0.2224$\pm$0.034 & -2.47$\pm$0.17 & 3.38$\pm$0.59 \\
1722-53852-0625 & 0.087 & J153550.4+104420.2 & 15.31 & 16.74 & 40.56$\pm$0.01 & 0.430$\pm$0.022 & 
	-0.034$\pm$0.011 & -2.16$\pm$0.17 & 3.35$\pm$0.28 \\
1732-53501-0363 & 0.107 & J161907.2+081701.3 & 16.82 & 18.21 & 39.28$\pm$0.18 & -0.11$\pm$0.431 & 
	0.1642$\pm$0.220 & -2.00$\pm$0.15 & 3.83$\pm$0.60 \\
1752-53379-0416 & 0.051 & J111353.7+145900.0 & 15.32 & 16.44 & 39.45$\pm$0.04 & 0.038$\pm$0.092 & 
	0.0302$\pm$0.045 & -2.74$\pm$0.28 & 6.52$\pm$0.79 \\
1754-53385-0485 & 0.080 & J113122.6+151513.1 & 17.28 & 18.04 & 40.81$\pm$0.01 & 0.491$\pm$0.016 & 
	-0.150$\pm$0.009 & -2.86$\pm$0.25 & 5.38$\pm$0.73 \\
1758-53084-0178 & 0.111 & J082832.1+075923.7 & 17.45 & 18.22 & 40.50$\pm$0.01 & 0.348$\pm$0.044 & 
	-0.124$\pm$0.018 & -2.57$\pm$0.23 & 5.37$\pm$0.55 \\
1775-53847-0226 & 0.077 & J133127.7+134005.8 & 15.89 & 16.20 & 39.82$\pm$0.05 & 0.083$\pm$0.123 & 
	-0.128$\pm$0.033 & -1.14$\pm$0.17 & 4.31$\pm$0.50 \\
1794-54504-0415 & 0.094 & J130116.2+080310.6 & 17.16 & 18.13 & 39.88$\pm$0.03 & 0.327$\pm$0.127 & 
	-0.187$\pm$0.064 & -2.81$\pm$0.26 & 5.40$\pm$0.75 \\
1821-53167-0528 & 0.134 & J154601.0+063121.7 & 17.41 & 18.55 & 40.58$\pm$0.01 & 0.153$\pm$0.039 & 
	-0.205$\pm$0.015 & -1.74$\pm$0.25 & 5.83$\pm$0.71 \\
1827-53531-0008 & 0.200 & J143238.6+051921.8 & 17.57 & 18.26 & 42.02$\pm$0.01 & 0.577$\pm$0.014 & 
	-0.232$\pm$0.009 & -2.72$\pm$0.27 & 4.79$\pm$0.90 \\
1829-53494-0062 & 0.082 & J144947.2+053205.3 & 16.92 & 17.60 & 40.23$\pm$0.01 & 0.551$\pm$0.079 & 
	-0.100$\pm$0.037 & -2.48$\pm$0.23 & 4.56$\pm$0.94 \\
1833-54561-0586 & 0.102 & J151845.7+061356.1 & 15.19 & 16.45 & 39.83$\pm$0.09 & -0.53$\pm$0.144 & 
	-0.065$\pm$0.089 & -2.63$\pm$0.27 & 6.64$\pm$0.68 \\
1842-53501-0073 & 0.053 & J144552.2+304854.0 & 16.07 & 16.89 & 39.45$\pm$0.03 & 0.053$\pm$0.088 & 
	0.0886$\pm$0.044 & -2.60$\pm$0.30 & 6.63$\pm$0.73 \\
1854-53566-0011 & 0.103 & J163018.5+193017.0 & 17.36 & 18.40 & 40.51$\pm$0.01 & 0.742$\pm$0.121 & 
	0.0201$\pm$0.051 & -2.69$\pm$0.33 & 5.44$\pm$1.19 \\
1864-53313-0466 & 0.070 & J072520.5+413900.8 & 15.78 & 16.63 & 39.75$\pm$0.03 & 0.097$\pm$0.093 & 
	0.1225$\pm$0.041 & -3.30$\pm$0.24 & 4.92$\pm$0.63 \\
1865-53312-0129 & 0.066 & J072932.9+424145.7 & 16.57 & 17.12 & 39.65$\pm$0.02 & 0.304$\pm$0.077 & 
	-0.038$\pm$0.035 & -2.39$\pm$0.15 & 3.60$\pm$0.52 \\
1870-53383-0466 & 0.000 & J075736.4+532557.0 & 16.43 & 18.98 & 0.000$\pm$0.01 & 0.560$\pm$0.023 & 
	0.4368$\pm$0.011 & -1.52$\pm$0.24 & 6.01$\pm$0.67 \\
1920-53314-0489 & 0.044 & J075238.9+181917.7 & 15.21 & 15.85 & 40.72$\pm$0.01 & 0.562$\pm$0.011 & 
	-0.112$\pm$0.007 & -2.80$\pm$0.31 & 6.35$\pm$0.80 \\
1927-53321-0261 & 0.018 & J081937.8+210651.4 & 13.77 & 14.43 & 40.33$\pm$0.01 & 1.011$\pm$0.010 & 
	-0.032$\pm$0.006 & -2.88$\pm$0.18 & 4.01$\pm$0.55 \\
1930-53347-0150 & 0.070 & J083949.2+250458.4 & 16.71 & 17.96 & 39.14$\pm$0.08 & 0.336$\pm$0.432 & 
	0.0637$\pm$0.104 & -3.14$\pm$0.24 & 4.73$\pm$0.83 \\
1945-53387-0487 & 0.133 & J094713.9+334535.4 & 17.57 & 18.61 & 40.79$\pm$0.01 & 0.590$\pm$0.045 & 
	-0.278$\pm$0.025 & -2.64$\pm$0.21 & 4.39$\pm$0.63 \\
1951-53389-0400 & 0.118 & J100323.4+352503.8 & 17.57 & 18.67 & 40.89$\pm$0.01 & 0.812$\pm$0.048 & 
	-0.234$\pm$0.033 & -2.26$\pm$0.25 & 6.27$\pm$0.57 \\
1990-53472-0604 & 0.072 & J110742.3+310627.5 & 16.23 & 17.31 & 39.84$\pm$0.03 & 0.195$\pm$0.086 & 
	-0.295$\pm$0.038 & -2.63$\pm$0.21 & 3.56$\pm$0.46 \\
2004-53737-0520 & 0.192 & J121329.4+322332.1 & 17.32 & 18.26 & 41.36$\pm$0.01 & 0.508$\pm$0.026 & 
	-0.115$\pm$0.015 & -1.16$\pm$0.15 & 4.06$\pm$0.41 \\
2007-53474-0631 & 0.145 & J105801.5+383912.0 & 16.29 & 17.55 & 40.48$\pm$0.02 & 0.231$\pm$0.070 & 
	0.2100$\pm$0.024 & -3.05$\pm$0.27 & 3.88$\pm$0.70 \\
2025-53431-0173 & 0.082 & J104424.1+334642.8 & 16.52 & 17.03 & 39.98$\pm$0.03 & 0.132$\pm$0.088 & 
	-0.050$\pm$0.033 & -2.15$\pm$0.13 & 3.37$\pm$0.33 \\
2090-53463-0487 & 0.134 & J104758.9+362518.0 & 17.49 & 17.94 & 40.76$\pm$0.01 & 0.612$\pm$0.074 & 
	0.1504$\pm$0.024 & -1.35$\pm$0.17 & 4.57$\pm$0.51 \\
2139-53878-0330 & 0.032 & J143727.8+254556.0 & 13.91 & 14.66 & 41.34$\pm$0.01 & 1.381$\pm$0.026 & 
	0.1180$\pm$0.009 & -2.60$\pm$0.28 & 6.73$\pm$0.67 \\
2168-53886-0218 & 0.129 & J154528.1+191934.1 & 17.45 & 18.58 & 40.34$\pm$0.03 & 0.309$\pm$0.112 & 
	0.0628$\pm$0.041 & -2.14$\pm$0.24 & 4.37$\pm$0.83 \\
2196-53534-0565 & 0.031 & J160523.3+174525.8 & 14.27 & 15.10 & 39.08$\pm$0.03 & -0.02$\pm$0.063 & 
	-0.091$\pm$0.019 & -2.58$\pm$0.15 & 4.04$\pm$0.85 \\
2227-53820-0518 & 0.028 & J120638.9+281028.1 & 11.74 & 25.11 & 38.75$\pm$0.06 & -0.19$\pm$0.117 & 
	0.0126$\pm$0.026 & -2.58$\pm$0.22 & 5.49$\pm$0.67 \\
2268-53682-0136 & 0.108 & J080750.5+130723.7 & 17.57 & 18.50 & 39.56$\pm$0.09 & 0.163$\pm$0.269 & 
	-0.283$\pm$0.307 & -1.96$\pm$0.22 & 4.80$\pm$0.67 \\
2269-53711-0508 & 0.179 & J081437.9+172208.3 & 17.35 & 18.65 & 40.56$\pm$0.03 & 0.602$\pm$0.184 & 
	0.0931$\pm$0.078 & -2.78$\pm$0.28 & 5.91$\pm$0.62 \\
2270-53714-0145 & 0.131 & J081635.7+143538.7 & 17.20 & 18.35 & 39.94$\pm$0.06 & 0.088$\pm$0.152 & 
	-0.092$\pm$0.036 & -2.36$\pm$0.20 & 4.29$\pm$0.49 \\
2271-53726-0055 & 0.153 & J082448.2+164044.7 & 17.55 & 18.33 & 41.11$\pm$0.01 & 0.733$\pm$0.047 & 
	-0.078$\pm$0.017 & -1.70$\pm$0.14 & 4.22$\pm$0.46 \\
2278-53711-0043 & 0.073 & J084230.7+160618.6 & 17.51 & 18.18 & 40.93$\pm$0.01 & 0.694$\pm$0.016 & 
	-0.198$\pm$0.010 & -2.41$\pm$0.23 & 5.43$\pm$0.57 \\
2282-53683-0335 & 0.082 & J085149.1+221209.9 & 17.49 & 18.15 & 39.83$\pm$0.03 & 0.415$\pm$0.119 & 
	0.0267$\pm$0.028 & -2.17$\pm$0.18 & 5.25$\pm$0.51 \\
2288-53699-0015 & 0.029 & J092045.6+193327.5 & 16.72 & 17.64 & 39.45$\pm$0.01 & 0.506$\pm$0.047 & 
	0.1366$\pm$0.013 & -2.26$\pm$0.17 & 5.27$\pm$0.42 \\
2295-53734-0631 & 0.152 & J095307.4+223307.7 & 17.08 & 18.49 & 40.06$\pm$0.07 & -0.01$\pm$0.162 & 
	-0.117$\pm$0.043 & -1.66$\pm$0.12 & 3.77$\pm$0.34 \\
2341-53738-0154 & 0.034 & J095308.5+241756.8 & 15.84 & 16.43 & 38.97$\pm$0.03 & 0.150$\pm$0.072 & 
	0.0812$\pm$0.021 & -3.07$\pm$0.26 & 4.97$\pm$0.65 \\
2346-53734-0073 & 0.037 & J102153.8+234017.4 & 14.21 & 15.14 & 39.54$\pm$0.02 & 0.428$\pm$0.077 & 
	0.0913$\pm$0.027 & -3.55$\pm$0.35 & 6.44$\pm$0.85 \\
2356-53786-0589 & 0.168 & J104158.6+274237.5 & 17.35 & 18.75 & 40.90$\pm$0.01 & 0.611$\pm$0.068 & 
	-0.004$\pm$0.030 & -2.18$\pm$0.17 & 4.17$\pm$0.53 \\
\hline
\end{tabular}
\end{center}
\end{table*}

\setcounter{table}{0}

\begin{table*}
\caption{--to be continued}
\begin{center}
\begin{tabular}{cccccccccc}
\hline\hline
pmf & $z$ & name & m1 & m2 & $\log(L_{O3})$ & $\log(O3HB)$ & $\log(N2HA)$ & $\ln(\sigma)$ & $\ln(\tau)$ \\
\hline
2365-53739-0157 & 0.004 & J101738.5+214117.6 & 11.55 & 13.93 & 38.92$\pm$0.01 & 0.566$\pm$0.012 & 
	-0.102$\pm$0.008 & -3.06$\pm$0.27 & 6.21$\pm$0.70 \\
2366-53741-0416 & 0.134 & J102313.2+225438.8 & 17.16 & 18.21 & 40.08$\pm$0.05 & -0.08$\pm$0.116 & 
	-0.053$\pm$0.044 & -2.81$\pm$0.21 & 4.76$\pm$0.57 \\
2374-53765-0174 & 0.064 & J101936.8+193313.5 & 16.24 & 17.09 & 41.27$\pm$0.01 & 1.025$\pm$0.016 & 
	-0.152$\pm$0.010 & -2.17$\pm$0.11 & 3.56$\pm$0.39 \\
2375-53770-0517 & 0.114 & J102654.3+201914.4 & 17.48 & 18.36 & 40.90$\pm$0.01 & 0.783$\pm$0.051 & 
	-0.127$\pm$0.022 & -2.74$\pm$0.22 & 4.92$\pm$0.71 \\
2418-53794-0079 & 0.131 & J080611.5+110113.3 & 17.16 & 18.31 & 40.72$\pm$0.01 & 0.398$\pm$0.034 & 
	-0.225$\pm$0.017 & -2.44$\pm$0.19 & 4.98$\pm$0.62 \\
2419-54139-0036 & 0.052 & J080819.4+090050.0 & 14.05 & 15.12 & 39.79$\pm$0.02 & 0.130$\pm$0.071 & 
	0.0146$\pm$0.031 & -2.85$\pm$0.21 & 4.60$\pm$0.71 \\
2419-54139-0050 & 0.046 & J080431.3+083920.1 & 15.72 & 16.79 & 39.56$\pm$0.02 & 0.566$\pm$0.133 & 
	0.2181$\pm$0.055 & -3.16$\pm$0.27 & 5.85$\pm$0.77 \\
2422-54096-0372 & 0.066 & J082013.0+134108.3 & 15.63 & 16.33 & 39.57$\pm$0.03 & 0.249$\pm$0.091 & 
	-0.202$\pm$0.030 & -3.33$\pm$0.28 & 5.81$\pm$0.73 \\
2422-54096-0425 & 0.173 & J081909.8+132817.0 & 16.43 & 18.03 & 40.31$\pm$0.07 & 0.363$\pm$0.308 & 
	0.0102$\pm$0.135 & -3.23$\pm$0.59 & 5.77$\pm$1.17 \\
2425-54139-0093 & 0.140 & J083021.7+122533.5 & 17.99 & 18.86 & 40.35$\pm$0.02 & 0.270$\pm$0.056 & 
	-0.289$\pm$0.027 & -2.05$\pm$0.17 & 4.01$\pm$0.78 \\
2434-53826-0012 & 0.125 & J090950.5+133248.4 & 17.01 & 17.74 & 39.92$\pm$0.08 & -0.02$\pm$0.182 & 
	-0.076$\pm$0.040 & -2.71$\pm$0.23 & 4.86$\pm$0.80 \\
2436-54054-0552 & 0.113 & J091231.4+150559.8 & 16.42 & 17.33 & 40.08$\pm$0.04 & 0.087$\pm$0.091 & 
	0.1267$\pm$0.034 & -2.59$\pm$0.20 & 4.73$\pm$0.75 \\
2485-54176-0587 & 0.047 & J110156.4+193648.1 & 15.74 & 16.60 & 39.00$\pm$0.06 & 0.249$\pm$0.213 & 
	-0.035$\pm$0.085 & -3.11$\pm$0.26 & 5.48$\pm$0.62 \\
2489-53857-0561 & 0.166 & J111045.0+250801.9 & 17.29 & 18.54 & 41.12$\pm$0.01 & 0.799$\pm$0.065 & 
	-0.113$\pm$0.037 & -2.25$\pm$0.19 & 3.50$\pm$0.26 \\
2526-54582-0319 & 0.123 & J160427.4+092933.9 & 17.00 & 18.25 & 41.08$\pm$0.01 & 0.636$\pm$0.032 & 
	-0.133$\pm$0.011 & -2.24$\pm$0.30 & 6.58$\pm$0.73 \\
2572-54056-0139 & 0.075 & J083202.7+093759.1 & 15.76 & 17.82 & 40.44$\pm$0.01 & 0.727$\pm$0.055 & 
	0.1424$\pm$0.021 & -2.64$\pm$0.29 & 6.80$\pm$0.68 \\
2593-54175-0025 & 0.335 & J104043.7+174705.7 & 18.41 & 19.26 & 41.50$\pm$0.01 & 0.389$\pm$0.043 & 
	-0.127$\pm$0.044 & -2.69$\pm$0.32 & 5.95$\pm$0.71 \\
2644-54210-0229 & 0.023 & J120807.0+220604.1 & 16.64 & 16.68 & 38.90$\pm$0.02 & 0.251$\pm$0.061 & 
	-0.312$\pm$0.030 & -1.84$\pm$0.13 & 3.34$\pm$0.38 \\
2744-54272-0490 & 0.244 & J140244.5+155956.6 & 17.30 & 18.18 & 41.17$\pm$0.01 & 0.020$\pm$0.041 & 
	-0.263$\pm$0.050 & -2.77$\pm$0.35 & 6.54$\pm$0.94 \\
2771-54527-0146 & 0.126 & J140728.1+194946.1 & 16.97 & 18.09 & 40.98$\pm$0.01 & 0.443$\pm$0.026 & 
	-0.281$\pm$0.015 & -2.00$\pm$0.24 & 5.25$\pm$1.27 \\
2789-54555-0281 & 0.030 & J143916.1+200237.3 & 12.93 & 13.96 & 39.73$\pm$0.02 & 0.114$\pm$0.059 & 
	0.1780$\pm$0.029 & -2.18$\pm$0.23 & 7.20$\pm$0.51 \\
2790-54555-0124 & 0.044 & J145120.9+184028.9 & 16.60 & 16.72 & 39.05$\pm$0.04 & 0.235$\pm$0.148 & 
	-0.103$\pm$0.051 & -2.57$\pm$0.23 & 4.48$\pm$0.85 \\
2886-54498-0629 & 0.072 & J105742.9+095255.7 & 16.24 & 17.66 & 39.83$\pm$0.02 & 0.319$\pm$0.077 & 
	-0.318$\pm$0.030 & -2.93$\pm$0.28 & 5.24$\pm$0.80 \\
2948-54553-0408 & 0.121 & J163447.6+284323.3 & 17.57 & 17.83 & 40.28$\pm$0.02 & 0.443$\pm$0.099 & 
	0.1082$\pm$0.040 & -2.02$\pm$0.16 & 3.92$\pm$0.69 \\
2968-54585-0061 & 0.083 & J161601.6+175653.8 & 16.44 & 17.17 & 41.13$\pm$0.01 & 0.726$\pm$0.020 & 
	-0.045$\pm$0.009 & -2.31$\pm$0.25 & 5.05$\pm$0.62 \\
\hline
\end{tabular}
\end{center}
\end{table*}

\begin{table*}
\begin{center}
\caption{Line parameters of the 156 Type-2 AGN with apparent long-term variabilities}
\begin{tabular}{ccccccccccc}
\hline\hline
pmf & \multicolumn{2}{c}{first [N~{\sc ii}]} & \multicolumn{2}{c}{second [N~{\sc ii}]} & \multicolumn{2}{c}{first H$\alpha$} & 
	\multicolumn{2}{c}{second H$\alpha$} & $\chi^2$ & broad? \\
	& $\sigma$ & flux & $\sigma$ & flux & $\sigma$ & flux & $\sigma$ & flux &  & \\	
\hline
0271-51883-0252 & 3.72$\pm$0.09 & 609$\pm$18 & $\dots$ & $\dots$ & 3.50$\pm$0.26 & 203$\pm$20 & 20.2$\pm$1.14 & 537$\pm$45 & 0.80 &  yes  \\
0283-51959-0600 & 2.65$\pm$0.10 & 112$\pm$3 & $\dots$ & $\dots$ & 2.81$\pm$0.10 & 121$\pm$4 & $\dots$ & $\dots$ & 0.78 &  no  \\
0284-51662-0220 & 2.46$\pm$0.24 & 39$\pm$4 & $\dots$ & $\dots$ & 2.04$\pm$0.16 & 52$\pm$3 & $\dots$ & $\dots$ & 0.90 &  no  \\
0284-51662-0558 & 3.02$\pm$0.13 & 128$\pm$6 & $\dots$ & $\dots$ & 2.75$\pm$0.13 & 126$\pm$6 & 26.2$\pm$2.68 & 173$\pm$18 & 0.98 &  yes  \\
0284-51943-0551 & 2.79$\pm$0.13 & 117$\pm$6 & $\dots$ & $\dots$ & 2.58$\pm$0.10 & 118$\pm$5 & 24.1$\pm$1.96 & 158$\pm$14 & 0.94 &  yes  \\
0288-52000-0152 & 2.87$\pm$0.24 & 71$\pm$5 & $\dots$ & $\dots$ & 1.98$\pm$0.31 & 63$\pm$18 & 2.07$\pm$0.44 & 48$\pm$18 & 1.02 &  no  \\
0300-51666-0073 & 4.40$\pm$0.11 & 280$\pm$6 & $\dots$ & $\dots$ & 4.18$\pm$0.13 & 242$\pm$6 & $\dots$ & $\dots$ & 1.78 &  no  \\
0305-51613-0044 & 3.98$\pm$0.13 & 204$\pm$7 & $\dots$ & $\dots$ & 4.30$\pm$0.30 & 188$\pm$15 & $\dots$ & $\dots$ & 0.74 &  no  \\
0305-51613-0146 & 2.99$\pm$0.34 & 54$\pm$7 & $\dots$ & $\dots$ & 2.64$\pm$0.10 & 89$\pm$3 & $\dots$ & $\dots$ & 1.21 &  no  \\
0311-51665-0073 & 3.44$\pm$0.09 & 228$\pm$7 & $\dots$ & $\dots$ & 2.86$\pm$0.09 & 200$\pm$7 & 18.3$\pm$0.81 & 270$\pm$17 & 1.16 &  yes  \\
0315-51663-0549 & 2.88$\pm$0.84 & 17$\pm$5 & $\dots$ & $\dots$ & 1.38$\pm$0.31 & 10$\pm$2 & $\dots$ & $\dots$ & 0.89 &  no  \\
0375-52140-0063 & 3.24$\pm$0.07 & 324$\pm$8 & $\dots$ & $\dots$ & 2.64$\pm$0.07 & 275$\pm$8 & 15.8$\pm$0.79 & 279$\pm$18 & 1.45 &  yes  \\
0379-51789-0271 & 4.52$\pm$0.25 & 140$\pm$7 & $\dots$ & $\dots$ & 4.30$\pm$0.19 & 192$\pm$7 & $\dots$ & $\dots$ & 2.04 &  no  \\
0379-51789-0401 & 2.68$\pm$0.17 & 153$\pm$9 & $\dots$ & $\dots$ & 2.27$\pm$0.10 & 216$\pm$8 & $\dots$ & $\dots$ & 1.03 &  no  \\
0382-51816-0426 & 3.26$\pm$0.10 & 186$\pm$7 & $\dots$ & $\dots$ & 3.00$\pm$0.11 & 169$\pm$8 & 16.2$\pm$2.81 & 77$\pm$19 & 0.87 &  yes  \\
0390-51816-0305 & 3.17$\pm$0.18 & 177$\pm$30 & $\dots$ & $\dots$ & 3.25$\pm$0.25 & 208$\pm$15 & $\dots$ & $\dots$ & 0.84 &  no  \\
0390-51900-0306 & 3.83$\pm$0.27 & 162$\pm$17 & $\dots$ & $\dots$ & 1.72$\pm$0.24 & 85$\pm$26 & 3.52$\pm$0.41 & 127$\pm$30 & 0.79 &  no  \\
0411-51817-0217 & 3.27$\pm$0.12 & 760$\pm$29 & $\dots$ & $\dots$ & 3.22$\pm$0.28 & 1308$\pm$150 & $\dots$ & $\dots$ & 0.61 &  no  \\
0411-51873-0203 & 3.30$\pm$0.18 & 728$\pm$56 & $\dots$ & $\dots$ & 3.12$\pm$0.09 & 1351$\pm$41 & $\dots$ & $\dots$ & 0.47 &  no  \\        0424-51893-0634 & 2.66$\pm$0.07 & 619$\pm$35 & 6.54$\pm$0.91 & 189$\pm$37 & 2.41$\pm$0.10 & 494$\pm$46 & 5.84$\pm$0.96 & 225$\pm$38 & 1.36 &  no  \\
0447-51877-0287 & 2.32$\pm$0.20 & 56$\pm$4 & $\dots$ & $\dots$ & 1.78$\pm$0.24 & 47$\pm$8 & $\dots$ & $\dots$ & 0.74 &  no  \\
0451-51908-0092 & 2.86$\pm$0.25 & 63$\pm$5 & $\dots$ & $\dots$ & 3.39$\pm$0.49 & 50$\pm$10 & $\dots$ & $\dots$ & 0.63 &  no  \\
0465-51910-0378 & 3.83$\pm$0.32 & 55$\pm$4 & $\dots$ & $\dots$ & 3.63$\pm$0.71 & 47$\pm$4 & $\dots$ & $\dots$ & 1.10 &  no  \\
\hline
\end{tabular}
\end{center}
\tablecomments{
The first column shows the pmf information of plate-mjd-fiberid of each Type-2 AGN, the second column and 
the third column show the second moment (in unit of \AA) and line flux (in unit of $10^{-17}{\rm erg/s/cm^2}$) 
of the first component of [N~{\sc ii}]$\lambda6583$\AA~ of each Type-2 AGN, the fourth column and the fifth column 
show the second moment and line flux of the second component of [N~{\sc ii}]$\lambda6583$\AA~ of each Type-2 AGN, 
the sixth column, the seventh column, the eighth column and the ninth column show the second moment and line flux 
of the first and the second component of H$\alpha$ of each Type-2 AGN, the tenth column shows the determined 
$\chi^2$ (calculated by the summed squared residuals divided by degree of freedom) related to the best fitting 
results to the emission lines, the last column shows whether are there broad components in H$\alpha$: yes means 
there is probable broad H$\alpha$. \\
Symbol of $\dots$ means no reliable parameter which is three times smaller than its corresponding uncertainty.\\
Symbol of $**$ means the unique spectroscopic features in SDSS 1870-53383-0466 with different redshifts from emission 
lines and from absorption features.
}
\end{table*}

\setcounter{table}{1}

\begin{table*}
\begin{center}
\caption{--to be continued.}
\begin{tabular}{ccccccccccc}
\hline\hline
pmf & \multicolumn{2}{c}{first [N~{\sc ii}]} & \multicolumn{2}{c}{second [N~{\sc ii}]} & \multicolumn{2}{c}{first H$\alpha$} &
	\multicolumn{2}{c}{second H$\alpha$} & $\chi^2$ & broad?\\
	& $\sigma$ & flux & $\sigma$ & flux & $\sigma$ & flux & $\sigma$ & flux &  & \\
\hline
0500-51994-0390 & 5.07$\pm$0.35 & 76$\pm$4 & $\dots$ & $\dots$ & 1.97$\pm$0.34 & 26$\pm$6 & 2.91$\pm$0.54 & 37$\pm$7 & 0.80 &  no  \\
0501-52235-0366 & 1.98$\pm$0.04 & 1197$\pm$33 & 6.48$\pm$1.15 & 206$\pm$43 & 2.06$\pm$0.02 & 2134$\pm$22 & 44.6$\pm$0.74 & 4705$\pm$76 & 1.30 &  yes  \\
0504-52316-0492 & 2.52$\pm$0.57 & 80$\pm$26 & $\dots$ & $\dots$ & 3.03$\pm$0.12 & 158$\pm$7 & $\dots$ & $\dots$ & 0.68 &  no  \\
0533-51994-0187 & 2.89$\pm$0.41 & 197$\pm$64 & $\dots$ & $\dots$ & 4.03$\pm$0.04 & 616$\pm$7 & $\dots$ & $\dots$ & 1.56 &  no  \\
0538-52029-0359 & 2.70$\pm$0.04 & 564$\pm$8 & $\dots$ & $\dots$ & 2.82$\pm$0.03 & 970$\pm$11 & $\dots$ & $\dots$ & 1.90 &  no  \\
0540-51996-0549 & 1.92$\pm$0.31 & 63$\pm$24 & $\dots$ & $\dots$ & 2.50$\pm$0.29 & 50$\pm$7 & 23.8$\pm$2.37 & 185$\pm$25 & 1.00 &  yes  \\
0543-52017-0342 & 4.83$\pm$0.23 & 157$\pm$6 & $\dots$ & $\dots$ & 4.24$\pm$0.21 & 154$\pm$7 & $\dots$ & $\dots$ & 1.12 &  no  \\
0564-52224-0019 & 2.58$\pm$0.68 & 30$\pm$2 & $\dots$ & $\dots$ & 3.32$\pm$0.26 & 38$\pm$2 & $\dots$ & $\dots$ & 1.12 &  no  \\
0564-52224-0041 & 2.41$\pm$0.03 & 325$\pm$4 & $\dots$ & $\dots$ & 1.90$\pm$0.02 & 382$\pm$4 & 15.3$\pm$0.41 & 353$\pm$10 & 2.23 &  yes  \\
0571-52286-0339 & 2.87$\pm$0.89 & 34$\pm$9 & $\dots$ & $\dots$ & 5.53$\pm$1.93 & 27$\pm$8 & $\dots$ & $\dots$ & 1.02 &  no  \\
0576-52325-0639 & 4.20$\pm$0.29 & 157$\pm$10 & $\dots$ & $\dots$ & 3.00$\pm$0.30 & 66$\pm$5 & $\dots$ & $\dots$ & 1.16 &  no  \\
0580-52368-0084 & 2.10$\pm$0.12 & 132$\pm$8 & $\dots$ & $\dots$ & 1.77$\pm$0.04 & 160$\pm$4 & 15.1$\pm$1.97 & 74$\pm$11 & 0.95 &  yes  \\
0581-52356-0028 & 2.09$\pm$0.04 & 1081$\pm$32 & 6.82$\pm$0.55 & 449$\pm$29 & 1.80$\pm$0.04 & 1145$\pm$47 & 4.60$\pm$0.51 & 306$\pm$44 & 0.63 &  no  \\
0589-52055-0263 & 2.56$\pm$0.07 & 231$\pm$6 & $\dots$ & $\dots$ & 2.46$\pm$0.08 & 192$\pm$5 & $\dots$ & $\dots$ & 0.72 &  no  \\
0594-52045-0276 & 3.80$\pm$0.11 & 378$\pm$15 & $\dots$ & $\dots$ & 3.89$\pm$0.21 & 171$\pm$13 & $\dots$ & $\dots$ & 1.07 &  no  \\
0630-52050-0241 & 2.06$\pm$0.06 & 265$\pm$16 & 5.39$\pm$0.41 & 174$\pm$15 & 1.99$\pm$0.07 & 297$\pm$23 & 4.87$\pm$0.30 & 260$\pm$21 & 1.58 &  no  \\
0656-52148-0390 & 3.28$\pm$0.12 & 149$\pm$6 & $\dots$ & $\dots$ & 3.09$\pm$0.26 & 69$\pm$7 & 25.5$\pm$4.48 & 96$\pm$19 & 1.15 &  yes  \\
0696-52209-0234 & 2.99$\pm$0.05 & 861$\pm$14 & $\dots$ & $\dots$ & 2.64$\pm$0.09 & 388$\pm$12 & $\dots$ & $\dots$ & 0.76 &  no  \\
0699-52202-0163 & 3.81$\pm$0.09 & 260$\pm$5 & $\dots$ & $\dots$ & 3.85$\pm$0.07 & 343$\pm$6 & $\dots$ & $\dots$ & 0.74 &  no  \\
0720-52206-0191 & 2.67$\pm$0.38 & 91$\pm$22 & $\dots$ & $\dots$ & 3.11$\pm$1.08 & 46$\pm$12 & $\dots$ & $\dots$ & 0.57 &  no  \\
0743-52262-0591 & 2.66$\pm$0.08 & 1204$\pm$37 & $\dots$ & $\dots$ & 2.25$\pm$0.12 & 613$\pm$36 & 16.9$\pm$0.42 & 2998$\pm$94 & 0.87 &  yes  \\
0747-52234-0400 & 2.77$\pm$0.07 & 417$\pm$16 & 9.70$\pm$0.46 & 439$\pm$16 & 2.76$\pm$0.13 & 513$\pm$66 & $\dots$ & $\dots$ & 1.06 &  no  \\
0758-52253-0514 & 4.24$\pm$0.25 & 79$\pm$4 & $\dots$ & $\dots$ & 3.31$\pm$0.27 & 50$\pm$3 & $\dots$ & $\dots$ & 1.13 &  no  \\
0817-52381-0412 & 6.08$\pm$0.49 & 93$\pm$6 & $\dots$ & $\dots$ & 5.68$\pm$0.82 & 49$\pm$6 & $\dots$ & $\dots$ & 1.02 &  no  \\
0892-52378-0058 & 2.39$\pm$0.15 & 389$\pm$125 & $\dots$ & $\dots$ & 2.40$\pm$0.02 & 595$\pm$7 & 33.1$\pm$1.15 & 538$\pm$21 & 1.39 &  yes  \\
0914-52721-0233 & 2.84$\pm$0.13 & 183$\pm$7 & $\dots$ & $\dots$ & 2.94$\pm$0.17 & 143$\pm$7 & $\dots$ & $\dots$ & 0.58 &  no  \\
0917-52400-0260 & 1.22$\pm$0.30 & 14$\pm$7 & 2.87$\pm$0.10 & 145$\pm$9 & 2.72$\pm$0.04 & 225$\pm$4 & 15.5$\pm$1.95 & 63$\pm$10 & 1.39 &  yes  \\
0929-52581-0196 & 2.74$\pm$0.40 & 73$\pm$12 & $\dots$ & $\dots$ & 3.18$\pm$0.62 & 42$\pm$13 & 11.2$\pm$3.38 & 78$\pm$15 & 0.75 &  yes  \\
0931-52619-0094 & 3.75$\pm$0.12 & 189$\pm$5 & $\dots$ & $\dots$ & 3.32$\pm$0.11 & 177$\pm$5 & $\dots$ & $\dots$ & 0.86 &  no  \\
0938-52708-0404 & 4.13$\pm$0.07 & 435$\pm$9 & $\dots$ & $\dots$ & 4.12$\pm$0.07 & 407$\pm$6 & $\dots$ & $\dots$ & 0.92 &  no  \\
0979-52427-0095 & 4.09$\pm$0.05 & 461$\pm$6 & $\dots$ & $\dots$ & 2.96$\pm$0.03 & 521$\pm$5 & $\dots$ & $\dots$ & 2.52 &  no  \\
0983-52443-0594 & 4.50$\pm$0.15 & 186$\pm$5 & $\dots$ & $\dots$ & 4.11$\pm$0.16 & 182$\pm$11 & 1.40$\pm$0.58 & $\dots$ & 1.18 &  no  \\
0984-52442-0348 & 4.11$\pm$0.12 & 212$\pm$5 & $\dots$ & $\dots$ & 1.95$\pm$0.22 & 76$\pm$13 & 2.29$\pm$0.27 & 124$\pm$15 & 1.07 &  no  \\
1001-52670-0389 & 3.33$\pm$0.10 & 473$\pm$14 & $\dots$ & $\dots$ & 3.00$\pm$0.11 & 382$\pm$13 & $\dots$ & $\dots$ & 0.42 &  no  \\
1036-52582-0135 & 1.67$\pm$0.08 & 30$\pm$1 & $\dots$ & $\dots$ & 1.63$\pm$0.04 & 71$\pm$1 & $\dots$ & $\dots$ & 1.05 &  no  \\
1036-52582-0389 & 3.15$\pm$0.48 & 32$\pm$4 & $\dots$ & $\dots$ & 3.37$\pm$0.43 & 27$\pm$3 & $\dots$ & $\dots$ & 1.32 &  no  \\
1052-52466-0280 & 2.77$\pm$0.07 & 850$\pm$22 & $\dots$ & $\dots$ & 2.78$\pm$0.10 & 530$\pm$19 & $\dots$ & $\dots$ & 0.57 &  no  \\
1065-52586-0237 & 4.17$\pm$0.16 & 145$\pm$5 & $\dots$ & $\dots$ & 4.49$\pm$0.16 & 176$\pm$5 & $\dots$ & $\dots$ & 0.91 &  no  \\
1198-52669-0150 & 2.33$\pm$0.55 & 90$\pm$3 & $\dots$ & $\dots$ & 2.07$\pm$0.07 & 99$\pm$3 & $\dots$ & $\dots$ & 0.94 &  no  \\
1203-52669-0502 & 3.98$\pm$0.20 & 131$\pm$6 & $\dots$ & $\dots$ & 6.02$\pm$1.32 & 69$\pm$32 & $\dots$ & $\dots$ & 0.98 &  no  \\
1204-52669-0281 & 2.55$\pm$0.11 & 105$\pm$5 & $\dots$ & $\dots$ & 2.50$\pm$0.19 & 77$\pm$7 & $\dots$ & $\dots$ & 0.96 &  no  \\
1204-52669-0524 & 2.93$\pm$0.10 & 544$\pm$17 & $\dots$ & $\dots$ & 2.24$\pm$0.12 & 287$\pm$14 & $\dots$ & $\dots$ & 0.50 &  no  \\
1208-52672-0519 & 3.39$\pm$0.10 & 309$\pm$10 & $\dots$ & $\dots$ & 2.75$\pm$0.09 & 273$\pm$10 & 21.3$\pm$1.48 & 283$\pm$25 & 0.85 &  yes  \\
1237-52762-0623 & 3.61$\pm$0.19 & 354$\pm$46 & $\dots$ & $\dots$ & 3.01$\pm$0.17 & 377$\pm$51 & $\dots$ & $\dots$ & 1.26 &  no  \\
1239-52760-0047 & 2.20$\pm$0.20 & 53$\pm$5 & $\dots$ & $\dots$ & 2.00$\pm$0.27 & 38$\pm$5 & 23.9$\pm$1.41 & 293$\pm$18 & 0.88 &  yes  \\
\hline
\end{tabular}
\end{center}
\end{table*}

\setcounter{table}{1}

\begin{table*}
\begin{center}
\caption{Line parameters of the 156 Type-2 AGN with apparent long-term variabilities}
\begin{tabular}{ccccccccccc}
\hline\hline
pmf & \multicolumn{2}{c}{first [N~{\sc ii}]} & \multicolumn{2}{c}{second [N~{\sc ii}]} & \multicolumn{2}{c}{first H$\alpha$} &
	\multicolumn{2}{c}{second H$\alpha$} & $\chi^2$ & broad?\\
	& $\sigma$ & flux & $\sigma$ & flux & $\sigma$ & flux & $\sigma$ & flux & & \\
\hline
1264-52707-0102 & 2.73$\pm$0.10 & 143$\pm$5 & $\dots$ & $\dots$ & 2.63$\pm$0.05 & 327$\pm$6 & $\dots$ & $\dots$ & 1.03 &  no  \\
1265-52705-0448 & 2.76$\pm$0.18 & 71$\pm$4 & $\dots$ & $\dots$ & 2.93$\pm$0.33 & 44$\pm$6 & $\dots$ & $\dots$ & 0.98 &  no  \\
1275-52996-0568 & 4.80$\pm$0.21 & 156$\pm$6 & $\dots$ & $\dots$ & 2.16$\pm$0.38 & 43$\pm$12 & 3.51$\pm$0.30 & 151$\pm$13 & 0.90 &  no  \\
1294-52753-0109 & 1.87$\pm$0.22 & 68$\pm$18 & $\dots$ & $\dots$ & 1.90$\pm$0.06 & 131$\pm$4 & $\dots$ & $\dots$ & 0.88 &  no  \\
1305-52757-0458 & 3.65$\pm$0.25 & 144$\pm$9 & $\dots$ & $\dots$ & 2.72$\pm$0.17 & 143$\pm$8 & $\dots$ & $\dots$ & 0.50 &  no  \\
1307-52999-0065 & 3.67$\pm$0.21 & 117$\pm$18 & $\dots$ & $\dots$ & 2.95$\pm$0.35 & 138$\pm$27 & $\dots$ & $\dots$ & 0.94 &  no  \\
1339-52767-0304 & 3.46$\pm$0.25 & 61$\pm$4 & $\dots$ & $\dots$ & 3.98$\pm$0.88 & 43$\pm$10 & $\dots$ & $\dots$ & 1.50 &  no  \\
1341-52786-0159 & 2.97$\pm$0.27 & 89$\pm$13 & 9.16$\pm$1.32 & 87$\pm$15 & 4.23$\pm$0.26 & 102$\pm$6 & $\dots$ & $\dots$ & 0.83 &  no  \\
1353-53083-0483 & 3.94$\pm$0.19 & 110$\pm$5 & $\dots$ & $\dots$ & 2.01$\pm$0.27 & 71$\pm$24 & 2.95$\pm$0.43 & 117$\pm$25 & 0.90 &  no  \\
1363-53053-0109 & 1.89$\pm$0.07 & 178$\pm$8 & 8.47$\pm$1.93 & 60$\pm$9 & 1.51$\pm$0.08 & 225$\pm$32 & 2.98$\pm$0.60 & 67$\pm$32 & 0.95 &  no  \\
1365-53062-0057 & 2.36$\pm$0.08 & 148$\pm$5 & $\dots$ & $\dots$ & 2.62$\pm$0.10 & 159$\pm$6 & 51.7$\pm$2.01 & 711$\pm$28 & 1.33 &  yes  \\
1368-53084-0298 & 3.89$\pm$0.41 & 83$\pm$11 & $\dots$ & $\dots$ & 9.04$\pm$2.31 & 75$\pm$15 & $\dots$ & $\dots$ & 1.55 &  no  \\
1370-53090-0201 & 1.85$\pm$0.56 & 113$\pm$3 & $\dots$ & $\dots$ & 2.08$\pm$0.55 & 141$\pm$13 & $\dots$ & $\dots$ & 0.80 &  no  \\
1382-53115-0193 & 4.05$\pm$0.04 & 4012$\pm$44 & $\dots$ & $\dots$ & 4.23$\pm$0.06 & 2905$\pm$42 & $\dots$ & $\dots$ & 0.56 &  no  \\
1385-53108-0192 & 1.96$\pm$0.23 & 34$\pm$3 & $\dots$ & $\dots$ & 1.40$\pm$0.31 & 25$\pm$4 & 64.2$\pm$6.22 & 279$\pm$34 & 1.27 &  yes  \\
1567-53172-0431 & 1.72$\pm$0.36 & 48$\pm$4 & $\dots$ & $\dots$ & 1.73$\pm$0.35 & 18$\pm$3 & $\dots$ & $\dots$ & 0.80 &  no  \\
1591-52976-0432 & 2.57$\pm$0.16 & 101$\pm$7 & $\dots$ & $\dots$ & 2.53$\pm$0.10 & 145$\pm$5 & $\dots$ & $\dots$ & 0.57 &  no  \\
1592-52990-0418 & 2.47$\pm$0.04 & 355$\pm$6 & $\dots$ & $\dots$ & 2.49$\pm$0.03 & 443$\pm$6 & $\dots$ & $\dots$ & 0.62 &  no  \\
1594-52992-0566 & 3.34$\pm$0.09 & 865$\pm$30 & $\dots$ & $\dots$ & 3.38$\pm$0.10 & 987$\pm$36 & 21.2$\pm$0.67 & 1730$\pm$76 & 1.25 &  yes  \\
1691-53260-0603 & 3.52$\pm$0.33 & 363$\pm$97 & $\dots$ & $\dots$ & 3.64$\pm$0.78 & 295$\pm$15 & $\dots$ & $\dots$ & 0.50 &  no  \\
1692-53473-0151 & 2.94$\pm$0.03 & 400$\pm$4 & $\dots$ & $\dots$ & 2.90$\pm$0.03 & 507$\pm$5 & $\dots$ & $\dots$ & 1.19 &  no  \\
1722-53852-0289 & 3.16$\pm$0.34 & 91$\pm$20 & 8.54$\pm$1.58 & 101$\pm$19 & 3.69$\pm$0.30 & 75$\pm$7 & $\dots$ & $\dots$ & 1.52 &  no  \\
1722-53852-0625 & 3.20$\pm$0.04 & 287$\pm$4 & $\dots$ & $\dots$ & 2.62$\pm$0.03 & 303$\pm$4 & $\dots$ & $\dots$ & 1.94 &  no  \\
1732-53501-0363 & 2.71$\pm$0.50 & 19$\pm$3 & $\dots$ & $\dots$ & 1.87$\pm$0.52 & 12$\pm$3 & $\dots$ & $\dots$ & 1.04 &  no  \\
1752-53379-0416 & 3.76$\pm$0.29 & 131$\pm$16 & $\dots$ & $\dots$ & 3.46$\pm$0.23 & 121$\pm$7 & $\dots$ & $\dots$ & 0.54 &  no  \\
1754-53385-0485 & 2.28$\pm$0.05 & 382$\pm$18 & 5.53$\pm$0.98 & 64$\pm$17 & 2.48$\pm$0.02 & 592$\pm$6 & $\dots$ & $\dots$ & 1.05 &  no  \\
1758-53084-0178 & 3.45$\pm$0.15 & 176$\pm$13 & $\dots$ & $\dots$ & 1.73$\pm$0.19 & 85$\pm$29 & 2.34$\pm$0.29 & 143$\pm$29 & 1.20 &  no  \\
1775-53847-0226 & 4.20$\pm$0.20 & 197$\pm$8 & $\dots$ & $\dots$ & 4.48$\pm$0.19 & 248$\pm$9 & $\dots$ & $\dots$ & 0.72 &  no  \\
1794-54504-0415 & 3.19$\pm$0.32 & 37$\pm$3 & $\dots$ & $\dots$ & 3.15$\pm$0.23 & 52$\pm$3 & $\dots$ & $\dots$ & 0.92 &  no  \\
1821-53167-0528 & 3.20$\pm$0.08 & 175$\pm$4 & $\dots$ & $\dots$ & 2.65$\pm$0.04 & 263$\pm$4 & $\dots$ & $\dots$ & 1.87 &  no  \\
1827-53531-0008 & 2.44$\pm$0.03 & 606$\pm$10 & $\dots$ & $\dots$ & 2.19$\pm$0.02 & 1025$\pm$13 & 17.1$\pm$0.62 & 617$\pm$24 & 2.06 &  yes  \\
1829-53494-0062 & 4.05$\pm$0.35 & 101$\pm$22 & $\dots$ & $\dots$ & 2.26$\pm$0.44 & 60$\pm$26 & 4.68$\pm$0.72 & 76$\pm$27 & 1.50 &  no  \\
1833-54561-0586 & 3.63$\pm$0.35 & 83$\pm$7 & $\dots$ & $\dots$ & 3.43$\pm$0.80 & 34$\pm$7 & $\dots$ & $\dots$ & 0.65 &  no  \\
1842-53501-0073 & 2.50$\pm$0.12 & 134$\pm$6 & $\dots$ & $\dots$ & 2.97$\pm$0.20 & 105$\pm$6 & $\dots$ & $\dots$ & 0.64 &  no  \\
1854-53566-0011 & 3.03$\pm$0.63 & 59$\pm$17 & $\dots$ & $\dots$ & 3.88$\pm$0.31 & 76$\pm$5 & $\dots$ & $\dots$ & 1.02 &  no  \\
1864-53313-0466 & 4.16$\pm$0.32 & 95$\pm$9 & $\dots$ & $\dots$ & 3.40$\pm$0.40 & 57$\pm$9 & 18.5$\pm$3.63 & 68$\pm$23 & 0.58 &  yes  \\
1865-53312-0129 & 2.02$\pm$0.11 & 61$\pm$3 & $\dots$ & $\dots$ & 1.91$\pm$0.10 & 60$\pm$3 & 34.1$\pm$7.28 & 75$\pm$15 & 0.67 &  yes  \\
1870-53383-0466$^{**}$ & 1.47$\pm$0.01 & 800$\pm$7 & $\dots$ & $\dots$ & 1.56$\pm$0.02 & 283$\pm$4 & $\dots$ & $\dots$ & 2.28 &  no  \\
1920-53314-0489 & 4.28$\pm$0.04 & 1075$\pm$13 & $\dots$ & $\dots$ & 4.09$\pm$0.03 & 1362$\pm$11 & $\dots$ & $\dots$ & 4.06 &  no  \\
1927-53321-0261 & 2.40$\pm$0.01 & 1468$\pm$12 & $\dots$ & $\dots$ & 2.31$\pm$0.02 & 1511$\pm$14 & 16.0$\pm$0.66 & 488$\pm$23 & 2.30 &  yes  \\
1930-53347-0150 & 2.75$\pm$0.43 & 28$\pm$4 & $\dots$ & $\dots$ & 2.25$\pm$0.63 & 20$\pm$2 & $\dots$ & $\dots$ & 0.81 &  no  \\
1945-53387-0487 & 3.99$\pm$0.19 & 86$\pm$4 & $\dots$ & $\dots$ & 1.61$\pm$0.28 & 47$\pm$16 & 3.35$\pm$0.32 & 128$\pm$21 & 1.02 &  no  \\
1951-53389-0400 & 4.37$\pm$0.20 & 125$\pm$5 & $\dots$ & $\dots$ & 4.08$\pm$0.44 & 109$\pm$45 & $\dots$ & $\dots$ & 2.86 &  no  \\
1990-53472-0604 & 4.61$\pm$0.81 & 62$\pm$15 & $\dots$ & $\dots$ & 3.66$\pm$0.30 & 119$\pm$14 & 16.4$\pm$2.72 & 114$\pm$34 & 0.83 &  yes  \\
2004-53737-0520 & 3.99$\pm$0.08 & 283$\pm$5 & $\dots$ & $\dots$ & 3.82$\pm$0.05 & 389$\pm$5 & $\dots$ & $\dots$ & 1.36 &  no  \\
2007-53474-0631 & 1.81$\pm$0.38 & 32$\pm$11 & 5.12$\pm$0.34 & 183$\pm$12 & 1.61$\pm$0.19 & 53$\pm$10 & 4.90$\pm$0.60 & 87$\pm$11 & 1.10 &  no  \\
2025-53431-0173 & 3.59$\pm$0.17 & 151$\pm$7 & $\dots$ & $\dots$ & 2.88$\pm$0.84 & 154$\pm$16 & $\dots$ & $\dots$ & 0.52 &  no  \\
\hline
\end{tabular}
\end{center}
\end{table*}

\setcounter{table}{1}

\begin{table*}
\begin{center}
\caption{--to be continued.}
\begin{tabular}{ccccccccccc}
\hline\hline
pmf & \multicolumn{2}{c}{first [N~{\sc ii}]} & \multicolumn{2}{c}{second [N~{\sc ii}]} & \multicolumn{2}{c}{first H$\alpha$} &
	\multicolumn{2}{c}{second H$\alpha$} & $\chi^2$ & broad?\\
	& $\sigma$ & flux & $\sigma$ & flux & $\sigma$ & flux & $\sigma$ & flux & & \\
\hline
2090-53463-0487 & 5.13$\pm$0.21 & 263$\pm$12 & $\dots$ & $\dots$ & 3.54$\pm$0.23 & 102$\pm$8 & $\dots$ & $\dots$ & 0.85 &  no  \\
2139-53878-0330 & 1.58$\pm$0.21 & 391$\pm$149 & 2.85$\pm$0.09 & 1488$\pm$147 & 2.51$\pm$0.04 & 1278$\pm$24 & 16.9$\pm$0.62 & 1186$\pm$58 & 0.58 &  yes  \\
2168-53886-0218 & 3.21$\pm$0.38 & 86$\pm$21 & $\dots$ & $\dots$ & 2.87$\pm$0.17 & 80$\pm$4 & $\dots$ & $\dots$ & 0.86 &  no  \\
2196-53534-0565 & 2.87$\pm$0.07 & 256$\pm$6 & $\dots$ & $\dots$ & 2.64$\pm$0.06 & 294$\pm$6 & $\dots$  & $\dots$ & 0.52 &  no  \\
2227-53820-0518 & 3.02$\pm$0.39 & 212$\pm$84 & $\dots$ & $\dots$ & 2.12$\pm$0.28 & 208$\pm$78 &  $\dots$ & $\dots$ & 0.35 &  no  \\
2268-53682-0136 & 2.15$\pm$0.60 & 11$\pm$2 & $\dots$ & $\dots$ & 2.55$\pm$0.66 & 23$\pm$7 & $\dots$ & $\dots$ & 0.89 &  no  \\
2269-53711-0508 & 2.90$\pm$0.24 & 43$\pm$3 & $\dots$ & $\dots$ & 2.81$\pm$0.34 & 28$\pm$3 & $\dots$ & $\dots$ & 1.16 &  no  \\
2270-53714-0145 & 4.25$\pm$0.22 & 80$\pm$3 & $\dots$ & $\dots$ & 3.81$\pm$0.17 & 91$\pm$3 & $\dots$ & $\dots$ & 0.93 &  no  \\
2271-53726-0055 & 3.17$\pm$0.40 & 113$\pm$37 & $\dots$ & $\dots$ & 3.70$\pm$0.12 & 175$\pm$8 & $\dots$ & $\dots$ & 0.98 &  no  \\
2278-53711-0043 & 2.11$\pm$0.06 & 244$\pm$11 & 6.56$\pm$0.55 & 142$\pm$10 & 1.79$\pm$0.07 & 298$\pm$24 & 4.09$\pm$0.21 & 258$\pm$23 & 0.70 &  no  \\
2282-53683-0335 & 2.28$\pm$0.31 & 65$\pm$13 & 2.36$\pm$0.38 & 56$\pm$14 & 3.52$\pm$0.14 & 110$\pm$4 & $\dots$ & $\dots$ & 1.30 &  no  \\
2288-53699-0015 & 2.13$\pm$0.03 & 425$\pm$6 & $\dots$ & $\dots$ & 2.18$\pm$0.04 & 285$\pm$5 & $\dots$ & $\dots$ & 1.07 &  no  \\
2295-53734-0631 & 3.74$\pm$0.24 & 65$\pm$3 & $\dots$ & $\dots$ & 3.14$\pm$0.16 & 76$\pm$3 & $\dots$ & $\dots$ & 0.98 &  no  \\
2341-53738-0154 & 2.11$\pm$0.09 & 113$\pm$7 & 6.85$\pm$1.62 & 36$\pm$8 & 1.93$\pm$0.14 & 88$\pm$12 & $\dots$ & $\dots$ & 0.77 &  no  \\
2346-53734-0073 & 3.37$\pm$0.10 & 327$\pm$9 & $\dots$ & $\dots$ & 3.18$\pm$0.18 & 213$\pm$24 & $\dots$ & $\dots$ & 0.98 &  no  \\
2356-53786-0589 & 3.27$\pm$0.12 & 106$\pm$3 & $\dots$ & $\dots$ & 3.40$\pm$0.14 & 100$\pm$3 & $\dots$ & $\dots$ & 1.21 &  no  \\
2365-53739-0157 & 2.50$\pm$0.02 & 3085$\pm$32 & $\dots$ & $\dots$ & 2.25$\pm$0.02 & 3751$\pm$38 & $\dots$ & $\dots$ & 0.90 &  no  \\
2366-53741-0416 & 4.73$\pm$0.28 & 89$\pm$4 & $\dots$ & $\dots$ & 4.51$\pm$0.28 & 84$\pm$4 & $\dots$ & $\dots$ & 0.98 &  no  \\
2374-53765-0174 & 1.45$\pm$0.35 & 54$\pm$17 & 3.23$\pm$0.07 & 531$\pm$26 & 1.73$\pm$0.24 & 137$\pm$41 & 3.48$\pm$0.07 & 687$\pm$42 & 0.79 &  no  \\
2375-53770-0517 & 3.08$\pm$0.14 & 135$\pm$6 & $\dots$ & $\dots$ & 2.58$\pm$0.08 & 177$\pm$6 & 36.7$\pm$3.23 & 238$\pm$20 & 0.89 &  yes  \\
2418-53794-0079 & 2.85$\pm$0.07 & 135$\pm$3 & $\dots$ & $\dots$ & 2.94$\pm$0.05 & 211$\pm$3 & $\dots$ & $\dots$ & 1.75 &  no  \\
2419-54139-0036 & 3.64$\pm$0.14 & 300$\pm$10 & $\dots$ & $\dots$ & 3.97$\pm$0.19 & 262$\pm$11 & $\dots$ & $\dots$ & 0.69 &  no  \\
2419-54139-0050 & 3.60$\pm$0.21 & 138$\pm$8 & $\dots$ & $\dots$ & 3.28$\pm$0.39 & 69$\pm$7 & 103.$\pm$8.14 & 2372$\pm$391 & 0.80 &  yes  \\
2422-54096-0372 & 2.57$\pm$0.12 & 68$\pm$2 & $\dots$ & $\dots$ & 2.11$\pm$0.06 & 108$\pm$3 & $\dots$ & $\dots$ & 0.97 &  no  \\
2422-54096-0425 & 3.25$\pm$0.85 & 22$\pm$7 & $\dots$ & $\dots$ & 2.85$\pm$0.75 & 36$\pm$12 & $\dots$ & $\dots$ & 0.81 &  no  \\
2425-54139-0093 & 2.46$\pm$0.22 & 51$\pm$7 & $\dots$ & $\dots$ & 2.22$\pm$0.06 & 99$\pm$2 & $\dots$ & $\dots$ & 1.11 &  no  \\
2434-53826-0012 & 2.77$\pm$0.55 & 73$\pm$18 & $\dots$ & $\dots$ & 2.48$\pm$0.39 & 58$\pm$16 & 6.46$\pm$1.24 & 74$\pm$16 & 1.32 &  no  \\
2436-54054-0552 & 4.62$\pm$0.51 & 116$\pm$15 & $\dots$ & $\dots$ & 1.67$\pm$0.63 & 13$\pm$4 & 2.69$\pm$0.20 & 78$\pm$4 & 0.97 &  no  \\
2485-54176-0587 & 2.39$\pm$0.52 & 38$\pm$16 & $\dots$ & $\dots$ & 1.78$\pm$0.57 & 37$\pm$3 & $\dots$ & $\dots$ & 0.67 &  no  \\
2489-53857-0561 & 4.10$\pm$0.20 & 93$\pm$4 & $\dots$ & $\dots$ & 3.75$\pm$0.18 & 104$\pm$4 & $\dots$ & $\dots$ & 1.43 &  no  \\
2526-54582-0319 & 1.89$\pm$0.05 & 161$\pm$5 & 10.9$\pm$0.00 & 239$\pm$11 & 1.90$\pm$0.05 & 222$\pm$9 & 6.90$\pm$0.54 & 167$\pm$17 & 1.67 &  no  \\
2572-54056-0139 & 2.99$\pm$0.09 & 237$\pm$7 & $\dots$ & $\dots$ & 2.80$\pm$0.13 & 140$\pm$7 & 26.1$\pm$2.63 & 192$\pm$20 & 0.72 &  yes  \\
2593-54175-0025 & 1.59$\pm$0.27 & 53$\pm$13 & 5.65$\pm$1.99 & 50$\pm$15 & 2.19$\pm$0.11 & 167$\pm$7 & 89.5$\pm$5.91 & 1466$\pm$128 & 1.05 &  yes  \\
2644-54210-0229 & 1.94$\pm$0.10 & 76$\pm$3 & $\dots$ & $\dots$ & 1.55$\pm$0.48 & 125$\pm$3 & $\dots$ & $\dots$ & 0.81 &  no  \\
2744-54272-0490 & 2.25$\pm$0.78 & 36$\pm$10 & $\dots$ & $\dots$ & 3.25$\pm$0.44 & 61$\pm$11 & 17.9$\pm$2.58 & 92$\pm$26 & 1.22 &  yes  \\
2771-54527-0146 & 2.21$\pm$0.26 & 83$\pm$17 & 2.37$\pm$0.20 & 127$\pm$17 & 3.88$\pm$0.06 & 373$\pm$6 & $\dots$ & $\dots$ & 1.74 &  no  \\
2789-54555-0281 & 4.31$\pm$0.25 & 1101$\pm$191 & $\dots$ & $\dots$ & 4.58$\pm$0.70 & 630$\pm$125 & $\dots$ & $\dots$ & 0.28 &  no  \\
2790-54555-0124 & 1.83$\pm$0.14 & 49$\pm$3 & $\dots$ & $\dots$ & 1.82$\pm$0.11 & 55$\pm$3 & $\dots$ & $\dots$ & 0.69 &  no  \\
2886-54498-0629 & 2.83$\pm$1.28 & 72$\pm$21 & $\dots$ & $\dots$ & 2.02$\pm$0.31 & 49$\pm$12 & 2.34$\pm$0.25 & 90$\pm$12 & 0.75 &  no  \\
2948-54553-0408 & 4.14$\pm$0.24 & 90$\pm$5 & $\dots$ & $\dots$ & 2.18$\pm$0.46 & 30$\pm$12 & 6.32$\pm$1.55 & 49$\pm$11 & 1.08 &  no  \\
2968-54585-0061 & 3.19$\pm$0.03 & 683$\pm$7 & $\dots$ & $\dots$ & 3.09$\pm$0.03 & 718$\pm$8 & $\dots$ & $\dots$ & 2.67 &  no  \\
\hline
\end{tabular}
\end{center}
\end{table*}

\section*{Acknowledgements}
Zhang gratefully acknowledges the anonymous referee for giving us constructive comments and suggestions 
to greatly improve the paper. Zhang gratefully thanks the scientific research funds provided by GuangXi University 
and the kind grant support from NSFC-12173020. This manuscript has made use of the data from the SDSS projects, 
\url{http://www.sdss3.org/}, managed by the Astrophysical Research Consortium for the Participating Institutions 
of the SDSS-III Collaborations. The manuscript has made use of the data from the Catalina Sky Survey (CSS) 
\url{http://nesssi.cacr.caltech.edu/DataRelease/}, funded by the National Aeronautics and Space 
Administration under Grant No. NNG05GF22G issued through the Science Mission Directorate Near-Earth 
Objects Observations Program. The paper has made use of the public JAVELIN code 
\url{http://www.astronomy.ohio-state.edu/~yingzu/codes.html#javelin} as an approach to reverberation 
mapping that computes the lags between the AGN continuum and emission line light curves and their 
statistical confidence limits, and the MPFIT package \url{https://pages.physics.wisc.edu/~craigm/idl/cmpfit.html} 
to solve the least-squares problem through the Levenberg-Marquardt technique, and the MCMC code 
\url{https://emcee.readthedocs.io/en/stable/index.html}. 

\end{document}